\documentclass[twocolumn,twocolappendix]{aastex631}

\usepackage{amsmath}
\usepackage{palatino}
\usepackage{footnote}

\shorttitle{Radio study of nearby early-type galaxies}
\shortauthors{Grossová et al.}

\graphicspath{{./}{figures/}}

\begin{document}

\title{VLA Radio Study of a Sample of Nearby X-ray and Optically Bright Early-Type Galaxies}

\correspondingauthor{Romana Grossová}
\email{romana.grossova@gmail.com}

\author[0000-0003-3471-7459]{Romana Grossová}
\affiliation{Department of Theoretical Physics and Astrophysics, Faculty of Science, Masaryk University, Kotl\'a\v{r}sk\'a 2, Brno, 611 37, Czech Republic}
\affiliation{Dipartimento di Fisica, Universita degli Studi di Torino, via Pietro Giuria 1, I-10125 Torino, Italy}

\author[0000-0003-0392-0120]{Norbert Werner}
\affiliation{Department of Theoretical Physics and Astrophysics, Faculty of Science, Masaryk University, Kotl\'a\v{r}sk\'a 2, Brno, 611 37, Czech Republic}

\author[0000-0002-1704-9850]{Francesco Massaro}
\affiliation{Dipartimento di Fisica, Universita degli Studi di Torino, via Pietro Giuria 1, I-10125 Torino, Italy}

\author{Kiran Lakhchaura}
\affiliation{MTA-E\"otv\"os University Extragalactic Astrophysics Research Group, P\'azm\'any P\'eter s\'et\'any 1/A, Budapest, 1117, Hungary}

\author[0000-0001-6411-3651]{Tomáš Plšek}
\affiliation{Department of Theoretical Physics and Astrophysics, Faculty of Science, Masaryk University, Kotl\'a\v{r}sk\'a 2, Brno, 611 37, Czech Republic}

\author[0000-0003-1020-1597]{Krizstina Gab\'anyi}
\affiliation{ELKH-E\"otv\"os University Extragalactic Astrophysics Research Group, P\'azm\'any P\'eter s\'et\'any 1/A, Budapest, 1117, Hungary}
\affiliation{Konkoly Observatory, ELKH Research Center for Astronomy and Earth Sciences, Konkoly Thege Mikl\'os \'ut 15-17, H-1121 Budapest, Hungary}
\affiliation{ELTE Eötvös Loránd University, Institute of Geography and Earth Sciences, Department of Astronomy, Pázmány Péter sétány 1/A, Budapest, Hungary}

\author[0000-0001-7509-2972]{Kamlesh Rajpurohit}
\affiliation{Dipartimento di Fisica e Astronomia, Universit\'a di Bologna, Via Gobetti 93/2, 40131, Bologna, Italy}

\author{Rebecca E. A. Canning}
\affiliation{University of Portsmouth, Winston Churchill Ave, Portsmouth PO1 2UP, UK}

\author[0000-0003-0297-4493]{Paul Nulsen}
\affiliation{Harvard Smithsonian Center for Astrophysics, 60 Garden Street, Cambridge, MA 02138, USA}
\affiliation{ICRAR, University of Western Australia, 35 Stirling Hwy, Crawley,WA 6009, Australia}

\author[0000-0002-5671-6900]{Ewan O'Sullivan}
\affiliation{Harvard Smithsonian Center for Astrophysics, 60 Garden Street, Cambridge, MA 02138, USA}

\author[0000-0003-0667-5941]{Steven W. Allen}
\affiliation{Kavli Institute for Particle Astrophysics and Cosmology, Stanford University, 452 Lomita Mall, Stanford, CA 94305-4085, USA}

\author[0000-0002-9378-4072]{Andrew Fabian}
\affiliation{Institute of Astronomy, University of Cambridge, Madingley Road, Cambridge CB3 0HA, UK}

\begin{abstract}
Many massive early-type galaxies host central radio sources and hot X-ray atmospheres indicating the presence of radio-mechanical active galactic nucleus (AGN) feedback. The duty cycle and detailed physics of the radio-mode AGN feedback is still a matter of debate. To address these questions, we present 1--2\,GHz Karl G. Jansky Very Large Array (VLA) radio observations of a sample of the 42 nearest optically and X-ray brightest early-type galaxies. We detect radio emission in 41/42 galaxies. However, the galaxy without a radio source, NGC 499, has recently been detected at lower frequencies by the Low-Frequency Array (LOFAR). Furthermore, 27/42 galaxies in our sample host extended radio structures  and  34/42 sources show environmental interactions in the form of X-ray cavities. We find a significant correlation between the radio flux density and the largest linear size of the radio emission and between the radio power and the luminosity of the central X-ray point-source. The central radio spectral indices of the galaxies span a wide range of values, with the majority of the systems having steep spectra and the rest flat spectra. These results are consistent with AGN activity, where the central radio sources are mostly switched on, thus the duty cycle is very high. 7/14 galaxies with point-like radio emission (Fanaroff-Riley Class\,0; FR\,0) also show X-ray cavities indicating that, despite the lack of extended radio structures at 1--2\,GHz, these AGN do launch jets capable of inflating lobes and cavities.
\end{abstract}


\section{Introduction} \label{sec:intro}

The radio-mechanical feedback mode is thought to play a dominant role in the evolution of massive early-type galaxies, which host hot ($10^7$\,K) X-ray emitting atmospheres. In the absence of balancing heating, the atmospheric gas should cool radiatively and form stars, building much larger and bluer galaxies than are seen.
X-ray studies with {\it Chandra} and {\it XMM-Newton} as well as radio observations have shown that in these galaxies jet-inflated radio lobes displace the hot gas, creating `cavities' in the X-ray emitting plasma \citep[e.g.][]{fabian2003} and driving weak shocks and turbulence that heat the surrounding medium essentially isotropically \citep[for a recent review see][]{werner2019}. This feedback mode appears to be maintaining a remarkably long-lived delicate balance between heating and cooling in the hot X-ray emitting atmospheres of these systems \citep{mcnamara2007}. 

\citet{burns1990} showed that as much as 70$\%$ of the central dominant galaxies (CDGs) in clusters are radio loud. Later, \citet{mittal2009} found that all strong cool cores host a central radio source and \citet{sun2009} argued that all central brightest cluster and group galaxies with a radio emitting active galactic nucleus (AGN) are located in cool cores. \cite{kolokythas2018} used the Giant Metrewave Radio Telescope (GMRT) to study the Complete Local-volume Groups Sample (CLoGS), consisting of 53 local galaxy groups, at 235\,MHz and 610\,MHz and also found a high radio detection rate of 87$\%$. These results showed that the duty cycle of AGN in clusters and groups with short cooling times must be high.
Moreover, the duty cycle for the X-ray cavities in clusters of galaxies of around 70$\%$ was found, although lower mass systems like groups and giant elliptical galaxies showed also lower duty cycles (between 30-50$\%$)  \citep{dunn2005,dunn2006,nulsen2009,dong2010,birzan2012,birzan2020}. The ability to detect cavities depends on a number of factors including their location, size, age, the level of disturbance of the surrounding halo, the depth of the observation, and the instrument used, so estimates are likely subject to bias, and perhaps systematically underestimated as you go from the brightest clusters to fainter systems.

Results for the duty cycle in a population of massive galaxies are, however, somewhat less clear. \citet{best2005} studied galaxies in the redshift range of 0.03 $<$ z $<$ 0.1 and found that the fraction of galaxies with a radio-loud AGN increased with the stellar mass of the galaxy, reaching a maximum fraction of 30-40$\%$. For a volume limited sample of very local (up to 15\,Mpc) infrared luminous galaxies, \citet{goulding2009} found that only 27$\%$ of the galaxies host an AGN. Other studies found a higher fraction.
The K-band absolute magnitude limited ($M_{K}<-24$) sample of 396 early-type galaxies \citep{brown2011} showed the presence of radio continuum emission for all sources in the NRAO Very Large Array Sky Survey (NVSS) data combined with the single-dish data from the Green Bank Telescope (GBT), and Parkes Radio Telescope survey (PKS). \cite{sabater2019} used observations at lower frequencies (120--168\,MHz) from the LOw-Frequency ARray's (LOFAR's) Two-Metre Sky Survey (LoTSS) to investigate sources up to redshift 0.3 and confirmed previous findings of the high rate of radio source detection in the central region of massive galaxies. 

A different approach was taken by \citet{dunn2010}, who focused on a volume-limited study of the 18 nearest ($d<100$\,Mpc) optically and X-ray brightest early-type galaxies. Compared to other studies, their selection criteria also included X-ray luminosity, ensuring that the investigated galaxies really inhabit massive halos. Remarkably, their study revealed that nuclear radio emission is present in 17 out of the observed 18 galaxies. Furthermore, at least 10 of the galaxies with observed central radio emission also exhibited obvious spatially-extended jets in the Very Large Array (VLA) images. The authors concluded that the results present a severe challenge for models in which radio jets are considered a relatively rare and sporadic phenomenon \citep[e.g.][]{binney1995,kaiser2003} and the active `radio-mode' feedback most likely represents the default state for large elliptical galaxies.

Here, we extend the \citet{dunn2010} study to the 42 optically and X-ray brightest, nearest ($d<100$\,Mpc) early-type galaxies, with declination greater than -40 degrees (to ensure coverage by the VLA) to observe their radio properties, investigate the duty cycles, and search for correlations between the radio plasma and their hot atmospheres, as well as emission line nebulae. Our sample includes both galaxies at the centres of groups and clusters and field ellipticals with their own hot X-ray emitting atmosphere. Our focus on the nearest systems, as well as the relatively long baseline of the VLA A configuration (35\,km), ensures a good sensitivity and spatial resolution. \

Sections in this paper are arranged as follows: The selection criteria for our sample are stated in Section\,\ref{sec:sample_selection}. The radio morphology categories used in the paper are described in Section\,\ref{sec:categories} and followed by Section\,\ref{sec:observations}, which presents the observations and data reduction for radio (Section\,\ref{sec:vla_obs}) and X-ray (Section\,\ref{sec:xray_obs}) data.
The main results from the radio data reduction are summarized in Section\,\ref{sec:results} and Table\,\ref{tab:results} followed by the radio detection rates (Section\,\ref{sec:detec_rate}) and the X-ray cavity and central X-ray point-source rates (Section\,\ref{sec:comp_xrays}), the multifrequency correlations (Section\,\ref{sec:multif_correlations}, Table\,\ref{tab:corr}), and values of the central spectral indices (Section\,\ref{sec:spim}, Table\,\ref{tab:spim}). These results are discussed in Section\,\ref{sec:discussion}. 
In Section\,\ref{sec:summary}, we conclude with a summary of the results.\\
The observed radio morphologies and relevant multifrequency data for every source are described in Appendix\,\ref{app:individual}. The {\it Chandra} X-ray data overlaid by radio contours obtained in multiple VLA configurations at 1--2\,GHz are presented in Appendix\,\ref{app:xray_radio}, for both point-source-like (Appendix\,\ref{app:xray_radio_ps}) and extended (Appendix\,\ref{app:xray_radio_ext}) radio morphologies. The
Table\,\ref{tab:obs} in Appendix\,\ref{app:tab} includes information about the radio observations. Finally, Appendix\,\ref{app:add_sources} includes Table\,\ref{tab:add_sources} with additional sources excluded from the main sample.

Throughout the paper, the spectral indices, $\alpha$, are defined by flux density, $S_{\nu} \propto \nu^{\alpha}$ and radio powers as $P_{\nu} = 4\pi D_{l}^2 S_{\nu}$\footnote{The K correction was not applied given the redshift distribution of sources in our sample.}. The distances were determined through the redshift-independent surface brightness fluctuation method due to proximity of the sources in our sample (see Table\,\ref{tab:results}).
The following cosmological parameters were used in this paper: $\rm H_{0}=67.8\,\rm{km\,s}^{-1}\,\rm{Mpc}^{-1}$ \citep{planckcollaboration2016}, $\Omega_\mathrm{M}=0.308$ and $\Omega_{\Lambda}=0.692$.

\section{Sample Selection}
\label{sec:sample_selection}
We base our study on the extended sample of the nearest X-ray and optically brightest massive early-type galaxies discussed in \cite{dunn2010}. Their parent sample is the catalog of \citet{beuing1999}, which contains 530 elliptical and elliptical/lentilcular galaxies brighter than the total Johnson B-band magnitude of $B_{\mathrm T}=13.5$ and has a 90$\%$ completeness. \citet{beuing1999} also provide X-ray luminosities or upper limits for 293 galaxies, based on {\it ROSAT} All-Sky Survey (RASS) data. These measurements were updated by \citet{osullivan2001}, who also used data from {\it ROSAT} pointed observations. 

An extensive {\it Chandra} X-ray study of this sample was performed by \citet{lakhchaura2018}, who slightly modified the selection criteria and increased its completeness. Our paper is a radio counterpart to the X-ray study of \cite{lakhchaura2018}.   
Our final selection criteria combine the criteria of \citet{dunn2010} and \citet{lakhchaura2018} and are as follows: 
\begin{itemize}
    \item lower limit of the X-ray luminosity within 10\,kpc from the center of galaxy: \\$L_{\rm X}>10^{40}\,\rm{erg}\,\rm{s}^{-1}$ in the 0.5--7\,keV band
    \item upper limit of the optical absolute magnitudes: $B_{T}<{-20}$ (total Johnson B-band magnitudes)
    \item volume restriction: all sources are within the distance of 100\,Mpc
    \item declination cut: DEC $>$ -40 degrees to ensure radio coverage with the VLA
\end{itemize}
We note that while the X-ray luminosities of \citet{osullivan2001} were total luminosities, which also included the contribution from point sources, such as the central AGN and X-ray binaries, our X-ray luminosities based on the measurements of \citet{lakhchaura2018} only refer to the hot X-ray emitting atmospheres within 10\,kpc \footnote{According to \cite{goulding2016} most of our galaxies have their effective radius, $R_{\rm e}$, up to 10\,kpc from the core.} from the center of the galaxy. Additionally, the intrinsic 2--10\,keV central X-ray point source luminosities were estimated from the power-law components of the spectral models for a sub-sample of galaxies, for which central X-ray point source emission was detected \citep{lakhchaura2018}.

Given these selection criteria, our volume-limited sample of the X-ray and optically brightest nearby early-type galaxies (41 giant elliptical galaxies and 1 lenticular) has a high level of completeness. There are 6 sources which meet our volume and X-ray/optical brightness criteria, but have no dedicated VLA L-band observations or sufficient {\it Chandra} data. Furthermore, for additional 2 sources, IC\,310 and NGC\,4203, the X-ray emission is strongly dominated by the central point source and the contribution of the X-ray atmosphere cannot be determined reliably (see Appendix\,\ref{app:add_sources} and Table\,\ref{tab:add_sources}). The small incompleteness will not significantly affect our conclusions. NVSS total intensity maps, which were less sensitive than the dedicated observations presented in our work, reveal emission from a central radio source at 1.4 GHz in the majority of non-included galaxies (5 out of 8 galaxies).

The main goal of our study is to extend the radio sample of \cite{dunn2010} to the 42 nearest X-ray and optically brightest galaxies with the highest resolution VLA A configuration data in the L-band spanning the frequency range 1--2\,GHz and centered at $\sim$\,\,1.5\,GHz. 

We obtained new VLA A observations for 18 giant elliptical galaxies (proposal ID: 15A-305, PI: N. Werner) to complement the available archival radio data for the entire sample\footnote{ Unfortunately, NGC\,1521 was missed while proposing for the new VLA observations.}. The highest resolution array configuration was chosen to match the resolution of images from the {\it Chandra} X-ray observatory. Moreover, to be sensitive not only to the fine structure of the central region, but also to  more diffuse and extended radio emission, we decided to complement the high resolution VLA A configuration data with data from the more compact C and D VLA configurations.

\section{Radio morphology categories}
\label{sec:categories}
We define four categories of radio sources (see Figu-\newline\,re\,\ref{fig:categories}) depending on the total extent of radio emission: point sources, compact sources, diffuse sources, and sources with prominent jets and lobes. In some cases, the compact and diffuse sources reveal small-/large-scale jets and lobes, which we classify as a subcategory of the compact  (e.g. NGC\,5129) or diffuse (e.g. NGC\,741) sources with jets/lobes. 

The {\it point-source} category (PS) is defined for those sources with an extent smaller than twice the restoring beam size of their total intensity VLA map. Galaxies with radio emission larger than this threshold are indeed labeled as extended. In addition, {\it compact} sources (C) are defined when showing their  total extent smaller than 5\,kpc. On the other hand, {\it diffuse} (D) sources are those classified as having their radio emission extending to more than 5\,kpc from the nucleus with a rather dispersed morphology, without well-defined large-scale jets/lobes. Lastly, for the sources in the category of {\it prominent jets/lobes} (J/L), the radio jets and lobes are the most prominent features (e.g. IC\,4296) and well collimated and narrow (i.e., with opening angles less than a few tenths of degrees) extending more than 5\,kpc from their radio core.\\
\renewcommand\tablename{Figure}
\begin{table*}
\renewcommand{\arraystretch}{0}
\begin{tabular}{c@{}c@{}c}
\hline
\includegraphics[width=160pt,height=100pt]{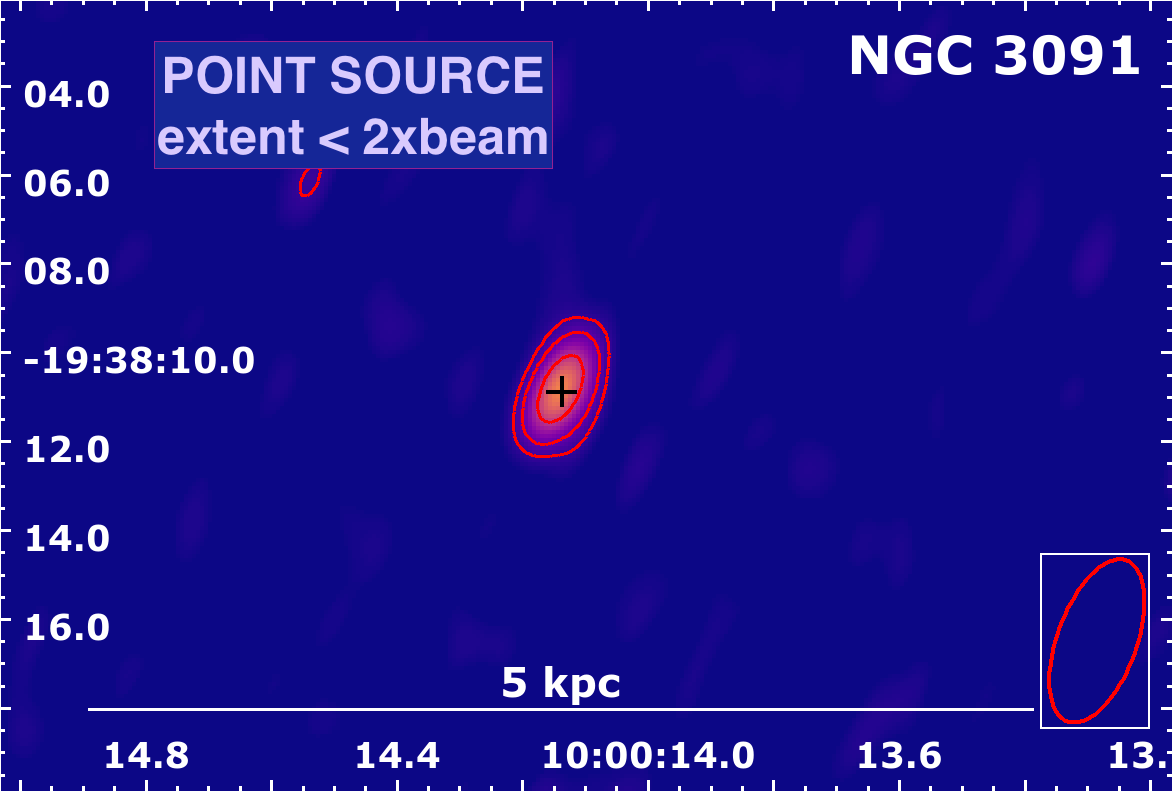}&
\includegraphics[width=160pt,height=100pt]{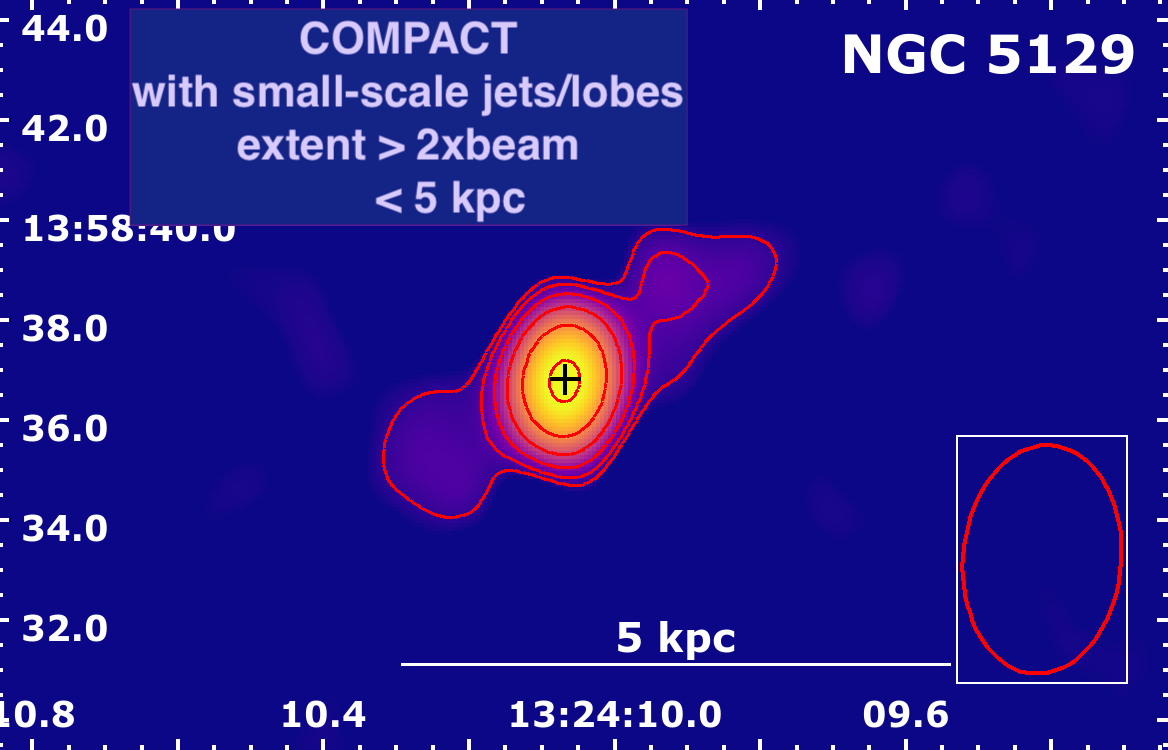}&
\includegraphics[width=160pt,height=100pt]{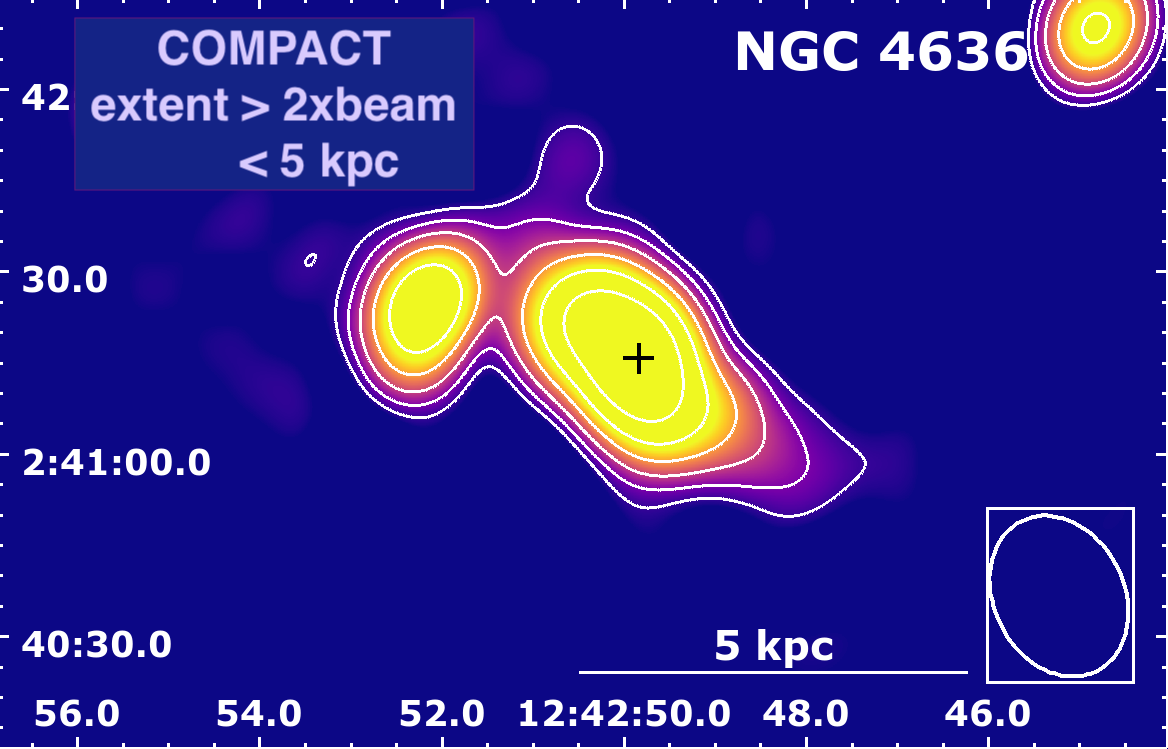}\\
\includegraphics[width=160pt,height=100pt]{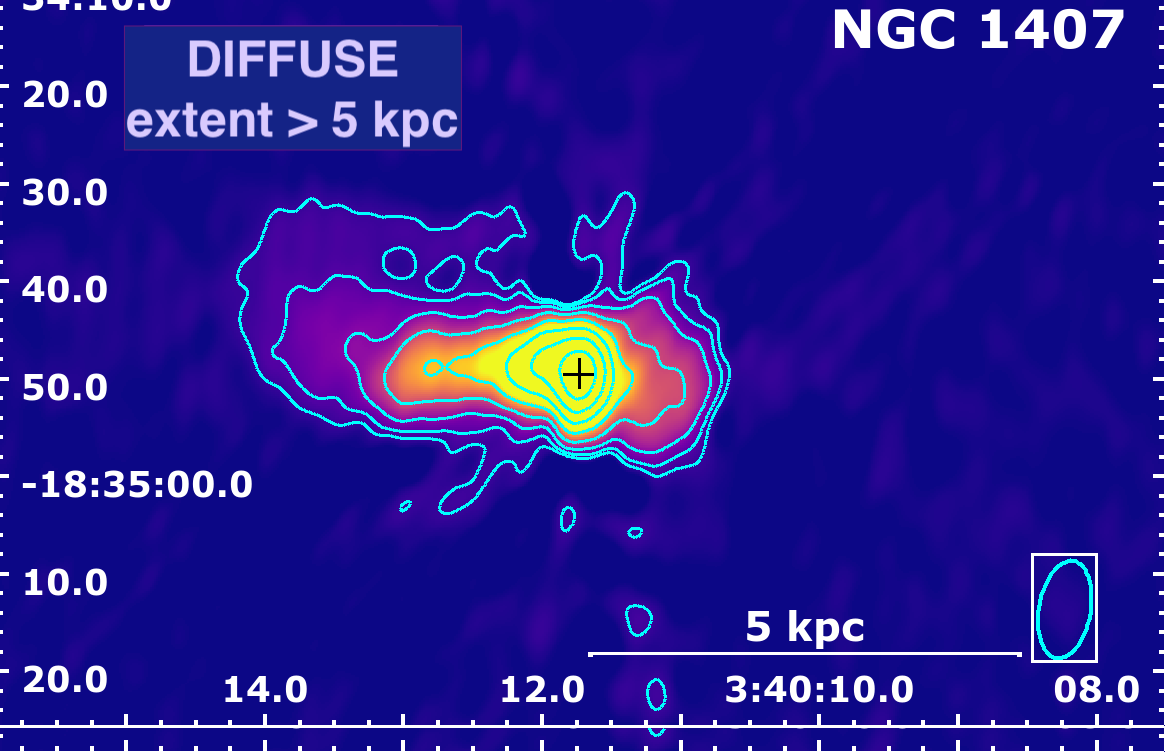}&
\includegraphics[width=160pt,height=100pt]{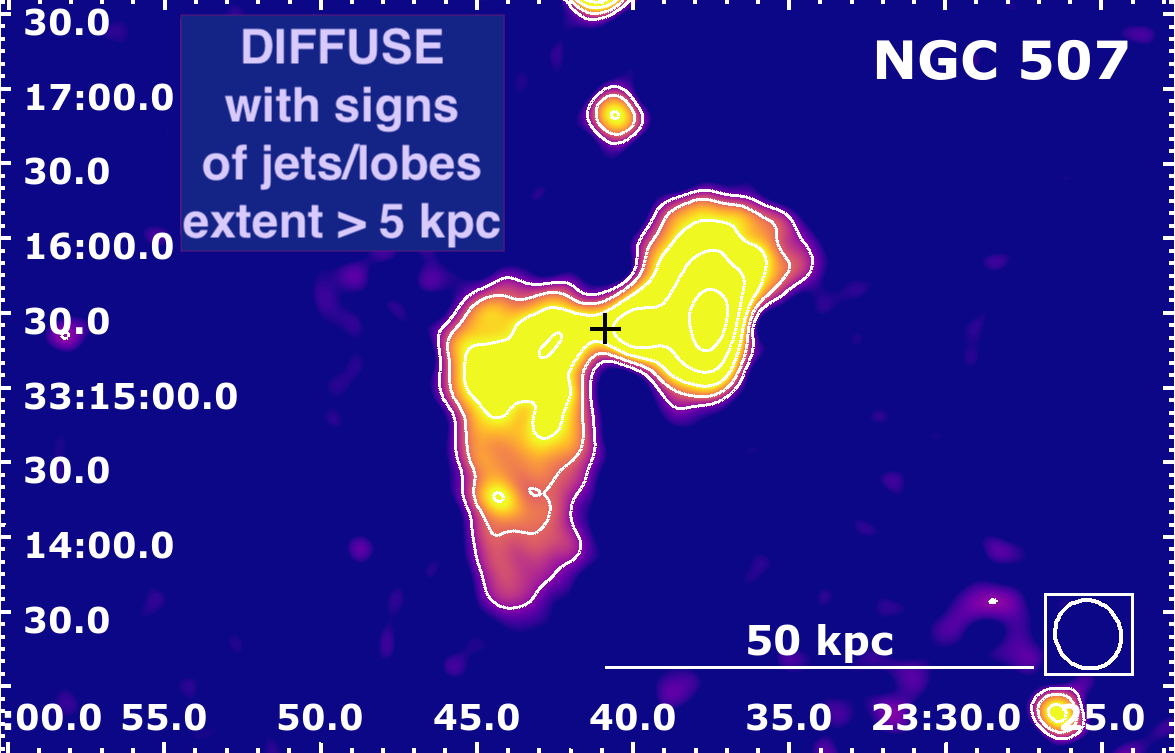}&
\includegraphics[width=160pt,height=100pt]{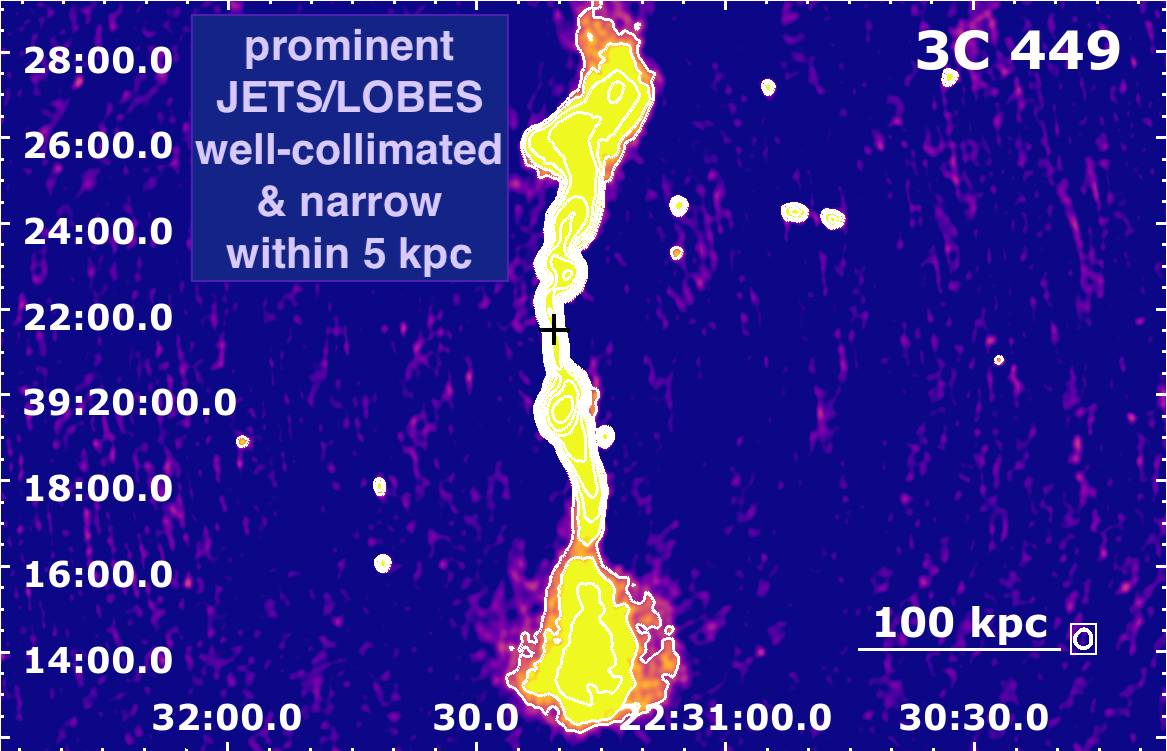}
\end{tabular}
\caption{Examples of radio morphology categories. The lowest level of the red/cyan/white (configuration A/ B/ C or D) radio contours corresponds to $5\times$\,RMS noise ($\sigma_{\rm RMS}$) in the image and the subsequent contours increase by factor of 2. The corresponding RMS noise and restoring beam size of the elliptical Gaussian for each source can be found in Table\,\ref{tab:results}. The black cross depicts the position of the radio center of the galaxy.}
\label{fig:categories}
\end{table*}

\section{Observations and data reduction}
\label{sec:observations}
\subsection{VLA Radio Observations and Analysis}
\label{sec:vla_obs}

We reduce and analyse VLA data obtained both before (historical VLA data) and after the major upgrade in 2011 \citep[Karl G. Jansky VLA/EVLA data;][]{perley2009,perley2011}. \

The major data size difference is in the increased bandwidth of the observations due to the upgrade to the Wideband Interferometer Digital ARchitecture (WIDAR) correlator system. The older historical data consisted of only one or two spectral windows with tens of channels with a channel width of 100--1000\,kHz, whereas the Karl G. Jansky VLA data have tens of spectral windows with up to hundreds of channels with a channel width of 3000\,kHz, which makes the size of the data set significantly larger. This affects our approach to the calibration methods we used for the two different data sets. For the Karl G. Jansky VLA data ($\sim$\,20-100\,GB), we used the more effective pipeline calibration method and in the case of historical data, a manual calibration approach was chosen, following the `Jupiter continuum calibration tutorial' available on the VLA NRAO website\footnote{
\href{https://casaguides.nrao.edu/index.php/Jupiter:\_continuum\_polarization\_calibration}{{https://casaguides.nrao.edu/index.php/Jupiter:\_continuum}\\ \href{https://casaguides.nrao.edu/index.php/Jupiter:\_continuum\_polarization\_calibration}{{\_polarization\_calibration}}}}. 

\subsubsection{Historical VLA data}
The pre-upgrade or historical VLA data sets were analysed for 33 galaxies in multiple configurations (details are given in Table\,\ref{tab:obs}) and observed for a large fraction of sources in our sample (i.e., more than $\sim$80\%) in two spectral windows between 1.4--1.7\,GHz. The important calibration steps for the historical data sets can be summarized as follows: The first step is to flag the imported data according to the suggestions from the NRAO observation log\footnote{\href{http://www.vla.nrao.edu/cgi-bin/oplogs.cgi}{http://www.vla.nrao.edu/cgi-bin/oplogs.cgi}} (if available), then we run {\tt tfcrop} (Time-Frequency Crop), the automatic flagging algorithm. The flux density for the corresponding VLA primary flux calibrators (Table\,\ref{tab:obs}) is set from a model \citep{perley2013} with the {\sc casa} task {\tt setjy}. Since there are only one or two single-channel spectral windows for most of the pre-upgrade historical data, we do not need to solve for the antenna delays and do the bandpass calibration. The next step is to determine the solution of the total gains for the flux density calibrator and finally, apply those gain solutions to the target.\\

\subsubsection{Karl G. Jansky VLA data}
The new Karl G. Jansky VLA observations at 1--2\,GHz presented here include 20 sources in A configuration from our project (ID: 15A-305, PI: Werner) and 10 archival observations in various VLA configurations (Table\,\ref{tab:obs}). The details of the data reduction and imaging were described in our recent paper on the giant elliptical galaxy IC\,4296 \citep{grossova2019}.

A similar approach is used to calibrate\footnote{{ {\sc casa} pipeline} v1.3.11 and CASA \citep[v4.7.2.;][]{mcmullin2007}} and image all sources from the 2015 project. The corresponding flux density calibrators for each observation are listed in Table\,\ref{tab:obs}.\\

The imaging was performed with the MultiTerm MultiFrequency synthesis ({\tt MTMFS}) {\tt clean} algorithm \citep{rau2011} with the {\tt briggs (robust=0)} weighting scheme \citep{briggs1995}. The second order Taylor polynomial \citep[{\tt nterms=2};][]{mcmullin2007} was used to account for the spectral behaviour of the sources.
The various spatial scales of radio emission require an individual approach to each source with different combinations of weightings, gridders, convolvers, and uv-tapers. When the dynamic range of the total intensity images reaches high enough values\footnote{We consider the dynamic range high enough when the ratio of the peak intensity to the RMS noise of the image is about 100.}, a few cycles of phase and possibly one cycle of amplitude and phase self-calibration are performed.\

An example of the final total intensity map is presented for the giant elliptical galaxy NGC\,4472 in Figu-\newline re\,\ref{fig:calib_nice_ngc4472}.

\setcounter{figure}{1}
\begin{figure}
    \centering
    \includegraphics[width=230pt]{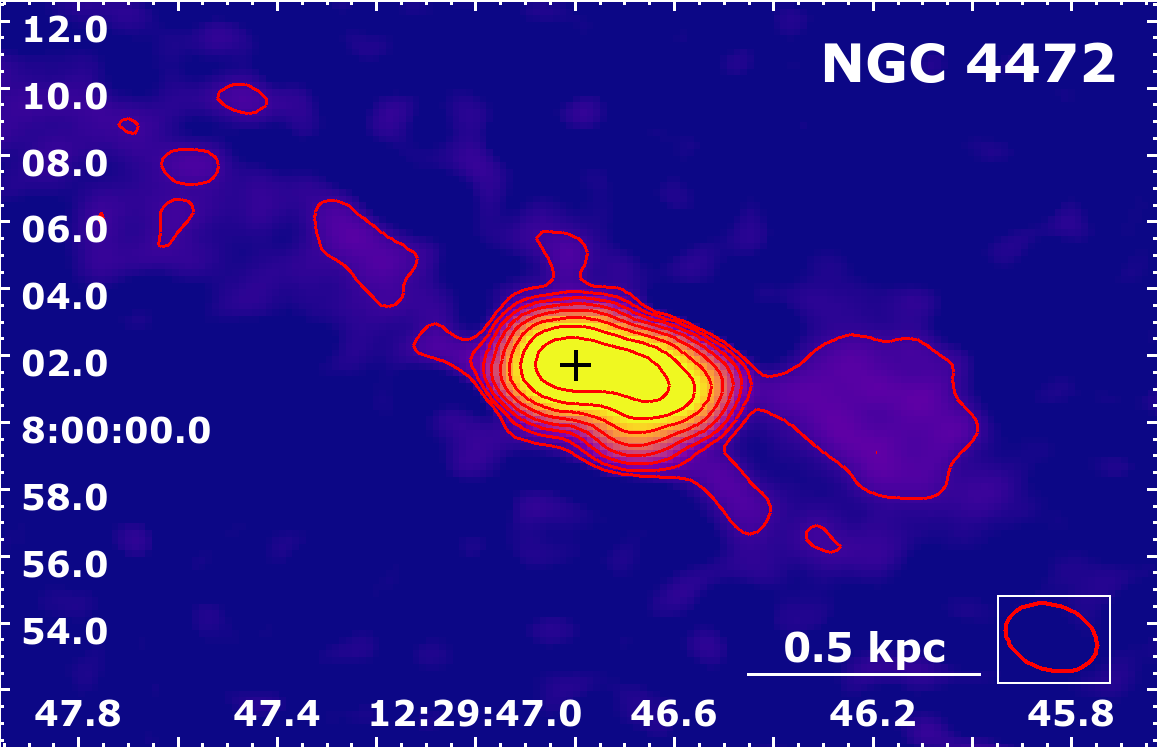}

    \caption{Example of the calibrated, final total intensity image of the giant elliptical galaxy NGC\,4472 in the upgraded VLA A configuration data observed in 2015. The red contour levels are created at [1,2,4,8,16,...]$\times\,5\,\sigma_{\rm RMS}$ up to peak intensity at 1.5\,GHz ($S_{\rm peak}$). The RMS noise and peak intensity values are $22\,\rm{\mu Jy}\,beam^{-1}$ and $\left(2.79 \pm 0.11\right)\times\,10^{-2}\,{\rm Jy}$  (Table\,\ref{tab:results}). The red ellipse in the white box represents the restoring beam of the final image. The black cross shows the position of the radio center of the giant elliptical galaxy.}
    \label{fig:calib_nice_ngc4472}
\end{figure}

\subsubsection{Total flux densities and uncertainties}
The final total flux densities centered at 1.5\,GHz ($S_{\rm 1.5GHz}$) for each source are derived using the {\tt imstat} {\sc casa} task. It is worth noting that, for sources in the sample observed only in A or B configuration, the final flux density might be lower than the `true' total flux density, because of the missing short baselines. On the other hand, we are mainly interested in central cores, where the flux density values are not affected by the missing short baselines.

The corresponding measurement uncertainties for the flux densities ($\Delta S_{\rm 1.5\,GHz}$) are determined as follows \citep[e.g.:][]{klein2003,rajpurohit2018}:

\begin{small}
\begin{equation}
\begin{split}
\Delta S_{\rm 1.5GHz} & = \sqrt{(S_{\rm 1.5\,GHz} \cdot S_{\rm cal})^2 + S_{\rm n}^2 + S_{\rm z}^2} \\ & =\sqrt{(S_{\rm 1.5GHz} \cdot S_{\rm cal})^2 + \left(\sigma_{\rm RMS} \cdot \sqrt{npts / A_{\rm beam}}\right)^2},
\end{split}
\end{equation}
\end{small}

 where $S_{\rm cal}$ is the specific calibration uncertainty, which is about 4$\%$ for VLA observations \citep{perley2013}. $S_{\rm n}$, the noise uncertainty, is defined as the off-source root-mean-square (RMS) noise ($\sigma_{\rm RMS}$) of the image multiplied by the square-root of {\it the number of points} ($npts$) in pixels per area, which covers the entire radio emission of the source divided by {\it the beam area} ($A_{\rm beam}$) in pixels. The term $S_{\rm z}$ accounts for a possibly wrong zero level, but may be neglected for interferometric observations \citep{klein2003}.   \\
 The RMS noise values were determined from four circular off-source regions in the image using {\tt casaviewer} and the median was taken as the final $\sigma_{\rm RMS}$. The {\sc casa} tool for image analysis, {\tt ia.beamarea}, gives an area that is covered by the beam or $A_{\rm beam}$ in pixels. Finally, we create a polygonal region enclosing the source's entire radio emission and save it into a {\sc casa} region file with the extension `.crtf'. The {\sc casa} task {\tt imstat} with a parameter {\tt region='*.crtf'} is used to determine the integrated flux density at 1.5\,GHz ($S_{\rm 1.5GHz}$) and the number of points ({\it npts}) or pixels within that specific region of the source emission. \\
 The value of the peak intensity at 1.5\,GHz ($S_{\rm peak}$) and the corresponding uncertainties are determined accordingly.

\subsubsection{Spectral index maps}
\label{sec:spindex}
To investigate the most recent activity of the AGN in our sample, we produce in-band spectral index maps from our 15A-305 VLA A configuration project for 18 giant elliptical galaxies. Each observation contains 16 spectral windows spanning the frequency range between 1--2\,GHz.\\

We adopted two different approaches to estimate the spectral index.
First, we used the MultiTerm MultiFrequency ({\tt MTMFS}) deconvolution algorithm in the {\sc tclean} {\sc casa} task. It produces two Taylor coefficient images: {\it tt0} as an equivalent of a Stokes\,I image, thus containing the information on the emission at a reference frequency, and {\it tt1} as a Taylor expansion term. The spectral index $\alpha$ is then defined as: $\alpha = {\rm tt}1/ {\rm tt}0$. The final output of the spectral index map is stored in the {\sc casa} product file, `.alpha.image.tt0' with the corresponding measurement uncertainty values saved in the {\sc casa} product, `.alpha.image.alpha.error.tt0'. To focus only on the emission above 5$\times\,\sigma_{\rm RMS}$, we applied this threshold value through the task {\tt widebandpbcor} to the final `alpha image' and with a parameter {\tt calcalpha}, we recalculated the spectral index map.\
In the {\tt casaviewer}, a circular region for each source with the diameter of 0.7\,kpc is defined to investigate the spectral index at the same physical scale in all 18 galaxies. The size of the central region (0.7\,kpc) was chosen to be applicable to all of our sources taking into account the different distances. This region was then used to extract the mean spectral index values from the threshold-adjusted {\sc casa} `alpha images'.\

For our second approach, we estimated the spectral index fitting flux densities measured in the central regions as described in the following. We split the 12 spectral windows\footnote{The 12 spectral windows were left after flagging the original 16 spectral windows.} into 4 chunks and `cleaned' them separately with specific reference frequencies (1.1, 1.3, 1.7 and 1.9\,GHz). Then, we smoothed the image chunks with the {\sc casa} task {\tt imsmooth} to gain the same restoring beam size (resolution). Lastly, in {\tt casaviewer}, we extracted the flux densities from a circle of 0.7\,kpc diameter centered at the radio core (consistent with the peak of radio intensity or the point from which the symmetrical jets/lobes are streaming out and are confirmed by the position of the center listed in the literature or NASA/IPAC Extragalactic Database (NED)\footnote{\href{https://ned.ipac.caltech.edu/}{https://ned.ipac.caltech.edu/}}) and calculated the spectral indices as a slope of the log-log fit with the extracted flux densities on the y-axis and the corresponding frequencies on the x-axis.

\begin{figure}
    \centering
\includegraphics[width=220pt]{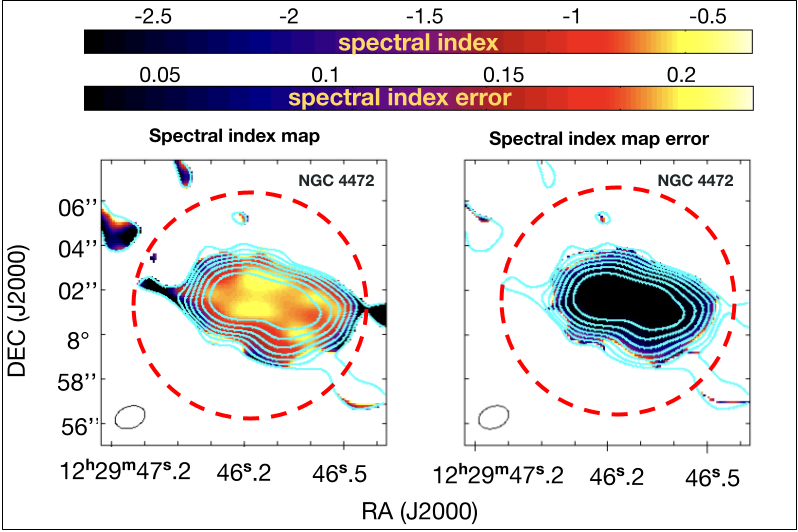}
\vspace{20pt}
\includegraphics[width=170pt]{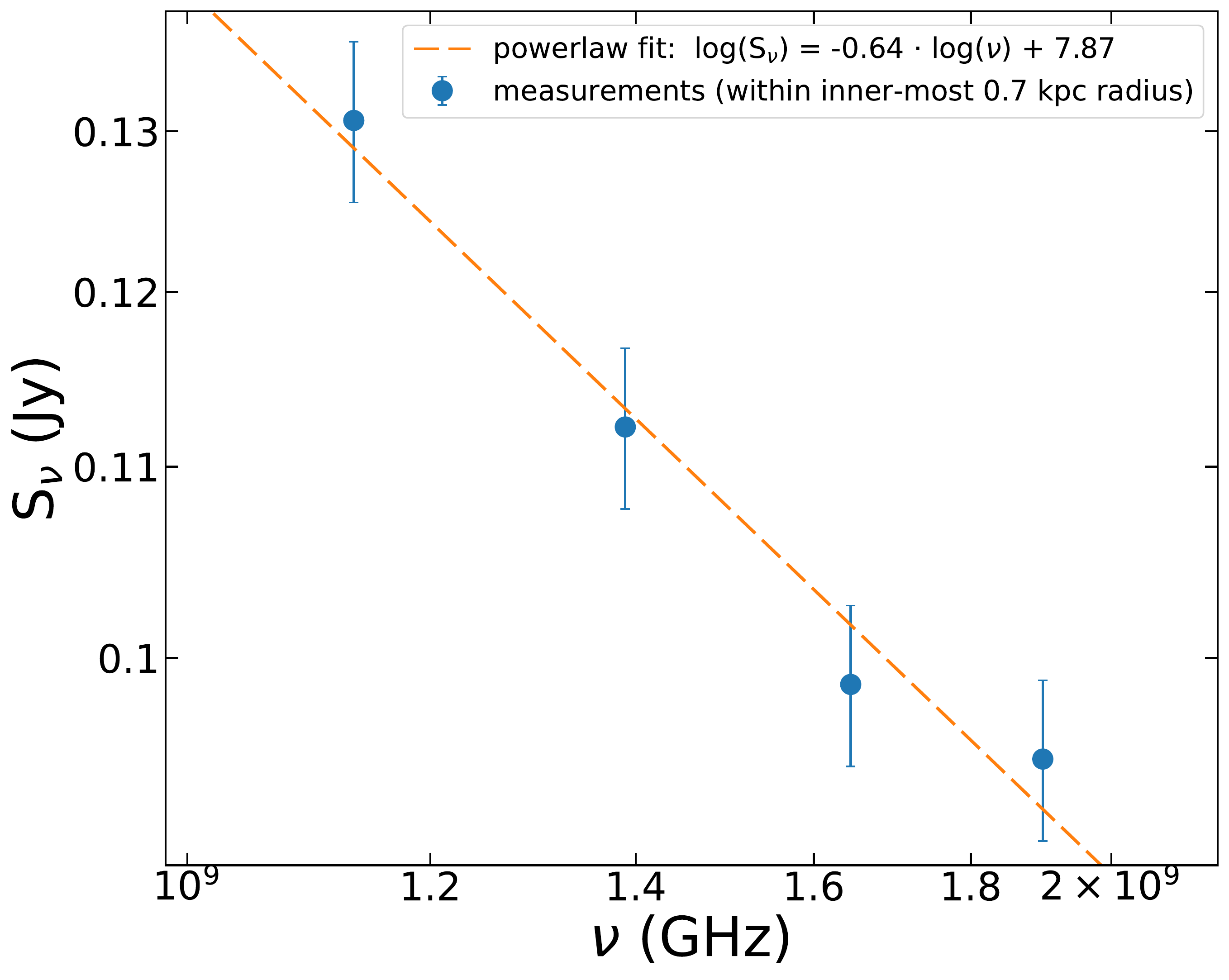}
\caption{Example of determining the spectral index for NGC\,4472 from a red circle with a diameter of 0.7\,kpc. {\bf Top:} The {\sc casa} output `alpha image' and its corresponding uncertainty values. In {\sc casaviewer}, we extracted the mean spectral index of (-0.55$\pm$0.05). {\bf Bottom:} The manual fitting of 4 frequency chunks and their corresponding flux densities gives a final spectral index value for NGC\,4472 of -0.64, which is comparable with the mean spectral index from the {\sc casa} `alpha image'.}
    \label{fig:spim_ngc4472}
\end{figure}

Comparison between the two methods adopted to estimate the spectral index are shown in Figure\,\ref{fig:spim_ngc4472} for the case of NGC\,4472 and results appear in good agreement.

Although, the latter method is simpler, because the spectral index maps (`alpha images') are automatically created during the {\sc tclean} task with the parameter {\tt nterms=2}. For further analysis of spectral indices, we concentrate only on this approach.

\subsection{X-ray  {\bfseries {\it Chandra}} Observations and Analysis}
\label{sec:xray_obs}
We use archival \textit{Chandra} Advanced CCD Imaging Spectrometer (ACIS) observations first reported and described in \cite{lakhchaura2018}.

The archival \textit{Chandra} observations were processed using the standard CIAO \citep[v4.13;][]{Fruscione2006}  procedures and a most recent calibration files (CALDB 4.9.4). For galaxies observed multiple times, the individual observations were reprojected and merged. Most galaxies were observed in the VFAINT mode using the ACIS-S chip, however, for a couple of galaxies we combined the ACIS-S and ACIS-I observations and for some of the galaxies there were only ACIS-I observations available.

The observations were deflared using the \texttt{lc\_clean} algorithm within the \texttt{deflare} routine. The images were generated from the merged observations using the \texttt{fluximage} procedure with a binsize of 1 pixel (0.492 arcsec). The images were background-subtracted using \texttt{blanksky} background files, which were scaled to match the observations in the $9-12\,$keV band, and exposure corrected using the corresponding exposure maps. The point sources were found using the \texttt{wavdetect} tool, inspected visually and filled with a mean surrounding surface brightness using the \texttt{dmfilth} procedure (CIAO 4.13).

The two-dimensional surface brightness distribution of each image obtained by this procedure was modelled using a projected version \citep{Ettori2000} of a classical beta model \citep{Cavaliere1967}. The fitting was performed in the SHERPA \citep[v4.13;][]{Freeman2001} package  using the Cash statistics \citep{Cash1979} and Monte Carlo optimization method \citep{Storn1995}. The fitted models were subtracted from the original images and the resulting residual images were used for further analysis.
Individual X-ray cavities were searched by eye using residual images and their reliability was checked using the Poisson statistics to be at least 4$\sigma$ under the surrounding background. The detection of these cavities is further supported by a novel machine learning method (a.k.a CADET; Plšek et al. in prep.).

\section{Results}
\label{sec:results}
Here, we report the results of radio observations for a sample of 42 early-type galaxies (41 giant elliptical galaxies and 1 lenticular galaxy). Our sample contains the most optically and X-ray luminous early-type galaxies within the luminosity distance of $\sim$\,100\,Mpc at declination that is accessible by the VLA (for more details see Section\,\ref{sec:sample_selection}). All details are reported in Table\,\ref{tab:results}. 

\renewcommand\tablename{Table}
\setcounter{table}{1}

\newpage
\begin{startlongtable}
\begin{longrotatetable}
\begin{deluxetable*}{lccccccccccr}
\tablenum{1}
\tablecaption{Radio imaging results for the sample of 42 early-type galaxies in several VLA configurations. The columns are listed as follows: (1) 3C, IC and NGC source names; (2) VLA configuration; (3) luminosity distance mostly derived from surface brightness fluctuations, a redshift-independent method (see Note); (4) the linear scale in kiloparsecs per arcseconds; (5) the radio morphology classification (Figure\,\ref{fig:categories} and Section\,\ref{sec:categories}); (6) the largest linear extent in kpc; (7) the restoring beam from the {\sc casa} {\tt clean} algorithm in arcseconds and corresponding (8) position angle; (9) the peak intensity in Jy/beam of the radio source in the total intensity image in the L-band (the frequency range between 1--2\,GHz, centered at 1.5\,GHz); (10) the integrated total flux density in Jy in L-band; (11) the total radio power in W/Hz; (12) the RMS noise ($\sigma_{\rm RMS}$) values of each total intensity image in Jy/beam. Details of each observation can be found in Table\,\ref{tab:obs} in the Appendix\,\ref{app:app_tab_radio}.}
 \label{tab:results}
\tablewidth{0pt}
\tabletypesize{\scriptsize}
\tablehead{
\colhead{Source} & \colhead{Config} & \colhead{Distance} & \colhead{Scale} &
\colhead{Radio} & \colhead{Extent} & \colhead{Beam} & \colhead{PA} & \colhead{$S_{ \rm peak}$ ${\rm \pm}$ ${eS_{\rm peak}}$} & \colhead{$S_{1.5\,{\rm GHz}}$ $\pm$ e$S_{1.5\,{\rm GHz}}$} & \colhead{$P_{1.5\,{\rm GHz}}$ $\pm$ e$P_{1.5\,{\rm GHz}}$} & RMS noise \\
\colhead{Name} & \colhead{VLA} & \colhead{(Mpc)} & \colhead{(kpc/$\arcsec$)} &
\colhead{Class} & \colhead{(kpc)} & \colhead{($\arcsec\times\arcsec$)} & \colhead{(deg)} & \colhead{(Jy/b)} & \colhead{(Jy)} & \colhead{(W/Hz)} & \colhead{(Jy/b)}
}
\decimalcolnumbers
\startdata
3C\,449 & A & $72.5^{L\#}$ & 0.352 & J/L & 36.6 & $1.9\times 1.4$ & 78.2 & $\left(1.46 \pm 0.06\right) \times 10^{-2}$ & $\left(2.55 \pm 0.10\right) \times 10^{-1}$ & $\left(1.60 \pm 0.06\right) \times 10^{23}$ & $1.0\times 10^{-4 } $ \\
3C\,449 & C & $72.5^{L\#}$ & 0.352 & J/L & 446.7 & $13.2\times 11.7$ & 5.5 & $\left(1.16 \pm 0.05\right) \times 10^{-1}$ & $\left(3.13 \pm 0.12\right) \times 10^{0}$ & $\left(1.97 \pm 0.08\right) \times 10^{24}$ & $1.1\times 10^{-4 } $ \\
IC1860 & A & $95.75^L$ & 0.465 & PS & - & $3.0\times 1.2$ &  8.9 & $\left(1.28 \pm 0.05\right) \times 10^{-2}$ & $\left(1.30 \pm 0.06\right) \times 10^{-2}$ & $\left(1.43 \pm 0.06\right) \times 10^{22}$ & $1.0\times 10^{-4 } $ \\
IC\,4296 & A & $47.31^L$ & 0.230 & J/L & 108.1 & $4.0\times 3.3$ & -12.5 & $\left(2.20 \pm 0.09\right) \times 10^{-1}$ & $\left(3.84 \pm 0.15\right) \times 10^{0}$ & $\left(1.03 \pm 0.04\right) \times 10^{24}$ & $6.0\times 10^{-5 } $ \\
IC\,4296 & D & $47.31^L$ & 0.230 & J/L & 437.0 & $84.3\times 22.6$ & -7.7 & $\left(5.31 \pm 0.21\right) \times 10^{-1}$ & $\left(8.56 \pm 0.34\right) \times 10^{0}$ & $\left(2.29 \pm 0.09\right) \times 10^{24}$ & $1.8\times 10^{-4 } $ \\
NGC\,57 & A & $77.15^L$ & 0.375 & PS & - & $1.8\times 1.0$ & 70.1 & $\left(3.26 \pm 0.13\right) \times 10^{-4}$ & $\left(5.41 \pm 0.29\right) \times 10^{-4}$ & $\left(3.85 \pm 0.21\right) \times 10^{20}$ & $1.0\times 10^{-5 } $ \\
NGC\,315 & A & $56.01^L$ & 0.272 & J/L & 13.4 & $1.6\times 1.4$ & 66.4 & $\left(4.40 \pm 0.18\right) \times 10^{-4}$ & $\left(3.39 \pm 0.37\right) \times 10^{-2}$ & $\left(1.27 \pm 0.14\right) \times 10^{22}$ & $8.9\times 10^{-5 } $ \\
NGC\,315 & B & $56.01^L$ & 0.272 & J/L & 154.2 & $4.2\times 3.7$ & 82.0 & $\left(4.61 \pm 0.18\right) \times 10^{-1}$ & $\left(1.27 \pm 0.05\right) \times 10^{0}$ & $\left(4.77 \pm 0.19\right) \times 10^{23}$ & $4.9\times 10^{-5 } $ \\
NGC\,315 & C & $56.01^L$ & 0.272 & J/L & 326.1 & $13.4\times 11.8$ & -13.2 & $\left(4.62 \pm 0.18\right) \times 10^{-1}$ & $\left(2.04 \pm 0.08\right) \times 10^{0}$ & $\left(7.66 \pm 0.31\right) \times 10^{23}$ & $7.3\times 10^{-5 } $ \\
NGC\,410 & A & $66.0^L$ & 0.320 & PS & - & $3.1\times 1.2$ & 66.6 & $\left(4.62 \pm 0.18\right) \times 10^{-3}$ & $\left(4.95 \pm 0.21\right) \times 10^{-3}$ & $\left(2.58 \pm 0.11\right) \times 10^{21}$ & $2.8\times 10^{-5 } $ \\
NGC\,499 & A & $60.74^L$ & 0.295 & NS & - & $1.4\times 1.2$ & -61.5 & - & - & - &  $1.2\times 10^{-4 } $  \\
NGC\,507 & A & $59.83^L$ & 0.290 & D & 13.2 & $2.7\times 1.1$ & 67.9 & $\left(1.70 \pm 0.07\right) \times 10^{-3}$ & $\left(7.20 \pm 0.40\right) \times 10^{-3}$ & $\left(3.08 \pm 0.17\right) \times 10^{21}$ & $2.8\times 10^{-5 } $ \\
NGC\,507 & C & $59.83^L$ & 0.290 & D$^{\dagger}$& 36.0 & $13.6\times 12.7$ & 62.8 & $\left(8.92 \pm 0.36\right) \times 10^{-3}$ & $\left(1.00 \pm 0.04\right) \times 10^{-1}$ & $\left(4.28 \pm 0.17\right) \times 10^{22}$ & $7.8\times 10^{-5 } $ \\
NGC\,533 & A & $61.58^L$ & 0.299 & C$^{\dagger}$& 1.1 & $1.2\times 1.1$ & -62.3 & $\left(1.02 \pm 0.04\right) \times 10^{-2}$ & $\left(2.41 \pm 0.10\right) \times 10^{-2}$ & $\left(1.09 \pm 0.04\right) \times 10^{22}$ & $1.9\times 10^{-5 } $ \\
NGC\,533 & B & $61.58^L$ & 0.299 & C & 2.8 & $4.8\times 3.7$ & -12.5 & $\left(4.26 \pm 0.17\right) \times 10^{-1}$ & $\left(6.95 \pm 0.30\right) \times 10^{-1}$ & $\left(3.15 \pm 0.13\right) \times 10^{23}$ & $4.4\times 10^{-3} $ \\
NGC\,708 & AB & $64.19^L$ & 0.312 & D$^{\dagger}$& 15.0 & $3.9\times 3.1$ & 89.6 & $\left(1.42 \pm 0.06\right) \times 10^{-2}$ & $\left(6.49 \pm 0.31\right) \times 10^{-2}$ & $\left(3.20 \pm 0.15\right) \times 10^{22}$ & $3.0\times 10^{-4 } $ \\
NGC\,708 & C & $64.19^L$ & 0.312 & D$^{\dagger}$& 38.6 & $8.3\times 5.0$ & -72.7 & $\left(1.29 \pm 0.05\right) \times 10^{-2}$ & $\left(4.49 \pm 0.18\right) \times 10^{-2}$ & $\left(2.21 \pm 0.09\right) \times 10^{22}$ & $9.3\times 10^{-6 } $ \\
NGC\,741 & A & $64.39^L$ & 0.313 & D$^{\dagger}$& 16.1 & $2.8\times 1.4$ & 50.9 & $\left(1.38 \pm 0.06\right) \times 10^{-2}$ & $\left(1.64 \pm 0.07\right) \times 10^{-1}$ & $\left(8.14 \pm 0.33\right) \times 10^{22}$ & $6.2\times 10^{-5 } $ \\
NGC\,741 & C & $64.39^L$ & 0.313 & C$^{\dagger}$& 137.7 & $15.8\times 14.1$ & 0.2 & $\left(6.80 \pm 0.27\right) \times 10^{-2}$ & $\left(9.12 \pm 0.37\right) \times 10^{-1}$ & $\left(4.52 \pm 0.18\right) \times 10^{23}$ & $7.0\times 10^{-4 } $ \\
NGC\,777 & A  & $58.08^L$ & 0.282 & C & 1.6 & $1.5\times 1.0$ & 73.7 & $\left(2.67 \pm 0.11\right) \times 10^{-3}$ & $\left(5.82 \pm 0.24\right) \times 10^{-3}$ & $\left(2.35 \pm 0.10\right) \times 10^{21}$ & $2.1\times 10^{-5 } $ \\
NGC\,777 & C & $58.08^L$ & 0.282 & NS & - & $3.2\times 3.0$ & 30.4 & - & - & - & $6.5\times 10^{-4 } $  \\
NGC\,1132 & C & $87.9^L$ & 0.427 & PS & - & $16.9\times 15.1$ & -7.4  & $\left(4.26 \pm 0.17\right) \times 10^{-3}$ & $\left(4.28 \pm 0.28\right) \times 10^{-3}$ & $\left(3.96 \pm 0.26\right) \times 10^{21}$ & $1.4\times 10^{-4 } $ \\
NGC\,1316 & BA  & $21.0^B$ & 0.102 & J/L & 11.8 & $4.5\times 3.5$ & 29.9 & $\left(9.15 \pm 0.37\right) \times 10^{-2}$ & $\left(2.62 \pm 0.10\right) \times 10^{-1}$ & $\left(1.38 \pm 0.06\right) \times 10^{22}$ & $8.5\times 10^{-5 } $ \\
NGC\,1316 & CD & $21.0^B$ & 0.102 & D & 257.7 & $60.3\times 34.4$ & -28.5 & $\left(2.04 \pm 0.08\right) \times 10^{-1}$ & $\left(3.30 \pm 0.18\right) \times 10^{1}$ & $\left(1.74 \pm 0.09\right) \times 10^{24}$ & $4.7\times 10^{-2 } $ \\
NGC\,1399 & A  & $20.09^B$ & 0.098 & J/L & 22.5 & $3.9\times 2.6$ & 13.6 & $\left(1.63 \pm 0.07\right) \times 10^{-2}$ & $\left(2.72 \pm 0.11\right) \times 10^{-1}$ & $\left(1.31 \pm 0.05\right) \times 10^{22}$ & $2.5\times 10^{-5 } $ \\
NGC\,1404 & A  & $20.02^B$ & 0.097 & PS & - & $3.9\times 2.6$ & 13.6 & $\left(2.04 \pm 0.08\right) \times 10^{-4}$ & $\left(2.15 \pm 0.32\right) \times 10^{-4}$ & $\left(1.03 \pm 0.15\right) \times 10^{19}$ & $2.5\times 10^{-5} $ \\
NGC\,1404 & CD & $20.02^B$ & 0.097 & NS & - & $47.8\times 42.8$ & 25.4 & - & - & - & $3.6\times 10^{-5 } $ \\
NGC\,1407 & A  & $23.27^L$ & 0.113 & D & 4.0 & $2.1\times 1.1$ & 36.1 & $\left(1.26 \pm 0.05\right) \times 10^{-2}$ & $\left(5.87 \pm 0.24\right) \times 10^{-2}$ & $\left(3.80 \pm 0.15\right) \times 10^{21}$ & $2.4\times 10^{-5 } $ \\
NGC\,1407 & B & $23.27^L$ & 0.113 & D & 6.0 & $5.1\times 2.7$ & -10.7 & $\left(2.24 \pm 0.09\right) \times 10^{-2}$ & $\left(7.12 \pm 0.29\right) \times 10^{-2}$ & $\left(4.61 \pm 0.18\right) \times 10^{21}$ & $1.4\times 10^{-5 } $ \\
NGC\,1407 & C & $23.27^L$ & 0.113 & D & 7.1 & $15.7\times 8.4$ & -3.0 & $\left(3.86 \pm 0.15\right) \times 10^{-2}$ & $\left(7.60 \pm 0.30\right) \times 10^{-2}$ & $\left(4.92 \pm 0.20\right) \times 10^{21}$ & $2.5\times 10^{-5 } $ \\
NGC\,1550 & A  & $67.30^L$ & 0.327 & C$^{\dagger}$& 10.1 & $1.8\times 1.0$ & 56.3 & $\left(8.52 \pm 0.34\right) \times 10^{-4}$ & $\left(2.35 \pm 0.13\right) \times 10^{-3}$ & $\left(1.27 \pm 0.07\right) \times 10^{21}$ & $2.1\times 10^{-5 } $ \\
NGC\,1550 & C & $67.30^L$ & 0.327 & D$^{\dagger}$& 36.2 & $17.3\times 15.3$ & -11.9 & $\left(3.13 \pm 0.12\right) \times 10^{-3}$ & $\left(1.70 \pm 0.11\right) \times 10^{-2}$ & $\left(9.21 \pm 0.60\right) \times 10^{21}$ & $2.3\times 10^{-4 } $ \\
NGC\,1600 & A  & $45.77^L$ & 0.222 & D$^{\dagger}$& 4.0 & $1.5\times 1.3$ & -3.5 & $\left(3.72 \pm 0.15\right) \times 10^{-3}$ & $\left(3.81 \pm 0.15\right) \times 10^{-2}$ & $\left(9.55 \pm 0.38\right) \times 10^{21}$ & $1.1\times 10^{-5 } $ \\
NGC\,2300 & A  & $41.45^L$ & 0.201 & PS & - & $1.7\times 1.0$ & 89.4 & $\left(1.67 \pm 0.07\right) \times 10^{-3}$ & $\left(1.70 \pm 0.08\right) \times 10^{-3}$ & $\left(3.49 \pm 0.16\right) \times 10^{20}$ & $2.1\times 10^{-5 } $ \\
NGC\,2300 & D & $41.45^L$ & 0.201 & PS & - & $66.4\times 36.8$ & 15.2 & $\left(2.32 \pm 0.09\right) \times 10^{-3}$ & $\left(1.99 \pm 0.19\right) \times 10^{-3}$ & $\left(4.09 \pm 0.40\right) \times 10^{20}$ & $1.4\times 10^{-4 } $ \\
NGC\,3091 & A  & $48.32^L$ & 0.235 & PS & - & $1.9\times 0.9$ & -24.3 & $\left(7.53 \pm 0.30\right) \times 10^{-4}$ & $\left(7.39 \pm 0.51\right) \times 10^{-4}$ & $\left(2.06 \pm 0.14\right) \times 10^{20}$ & $2.8\times 10^{-5 } $ \\
NGC\,3923 & A  & $20.97^L$ & 0.102 & PS & - & $2.2\times 0.8$ & 5.0 & $\left(3.76 \pm 0.15\right) \times 10^{-4}$ & $\left(4.90 \pm 0.40\right) \times 10^{-4}$ & $\left(2.58 \pm 0.21\right) \times 10^{19}$ & $2.3\times 10^{-5 } $ \\
NGC\,3923 & CD & $20.97^L$ & 0.102 & NS & - &  $40.7\times 32.1$ & -88.2 & - & - & - & $7.9\times 10^{-5 } $ \\
NGC\,4073 & A  & $60.08^L$ & 0.292 & PS & - & $1.5\times 1.1$ & -51.5 & $\left(7.23 \pm 0.29\right) \times 10^{-4}$ & $\left(7.15 \pm 0.43\right) \times 10^{-4}$ & $\left(3.09 \pm 0.19\right) \times 10^{20}$ & $2.0\times 10^{-5 } $ \\
NGC\,4125 & D & $21.41^L$ & 0.104 & PS & - & $44.0\times 34.2$ &  17.7 & - & - & - & $6.7\times 10^{-4 } $ \\
NGC\,4261 & A  & $29.0^M$ & 0.141 & J/L & 7.9 & $1.4\times 1.3$ & -22.9 & $\left(1.13 \pm 0.05\right) \times 10^{-2}$ & $\left(1.72 \pm 0.07\right) \times 10^{-2}$ & $\left(1.73 \pm 0.07\right) \times 10^{21}$ & $8.2\times 10^{-6 } $ \\
NGC\,4261 & C & $29.0^M$ & 0.141 & J/L & 55.5 & $18.8\times 13.5$ & 51.8 & $\left(2.11 \pm 0.08\right) \times 10^{-1}$ & $\left(1.27 \pm 0.05\right) \times 10^{-1}$ & $\left(1.28 \pm 0.05\right) \times 10^{22}$ & $1.4\times 10^{-3 } $ \\
NGC\,4374 & A & $18.5^B$ & 0.090 & J/L & 4.3 & $1.5\times 1.3$ & 47.6 & $\left(1.18 \pm 0.05\right) \times 10^{-1}$ & $\left(4.78 \pm 0.23\right) \times 10^{-1}$ & $\left(1.96 \pm 0.09\right) \times 10^{22}$ & $2.0\times 10^{-3}$ \\
NGC\,4374 & B & $18.5^B$ & 0.090 & J/L & 13.3 & $4.6\times 4.4$ & -4.5 & $\left(1.30 \pm 0.05\right) \times 10^{-1}$ & $\left(4.18 \pm 0.17\right) \times 10^{0}$ & $\left(1.71 \pm 0.07\right) \times 10^{23}$ & $1.4\times 10^{-3}$ \\
NGC\,4374 & C & $18.5^B$ & 0.090 & J/L & 18.7 & $38.8\times 32.4$ & -45.2 & $\left(1.28 \pm 0.05\right) \times 10^{0}$ & $\left(5.94 \pm 0.24\right) \times 10^{0}$ & $\left(2.43 \pm 0.10\right) \times 10^{23}$ & $5.8\times 10^{-3}$ \\
NGC\,4406 & A  & $17.9^B$ & 0.087 & PS & - & $1.1\times 1.0$ & -1.4 & $\left(3.57 \pm 0.14\right) \times 10^{-4}$ & $\left(2.77 \pm 0.53\right) \times 10^{-4}$ & $\left(1.06 \pm 0.20\right) \times 10^{19}$ & $4.5\times 10^{-5} $ \\
NGC\,4406 & D & $17.9^B$ & 0.087 & NS & - & $45.0\times  43.2$ & -6.0 & - & - & - & $2.6 10^{-4 } $ \\
NGC\,4472 & A  & $16.7^B$ & 0.081 & C$^{\dagger}$& 2.6 & $1.4\times 1.0$ & -72.7 & $\left(2.79 \pm 0.11\right) \times 10^{-2}$ & $\left(1.20 \pm 0.05\right) \times 10^{-1}$ & $\left(4.00 \pm 0.16\right) \times 10^{21}$ & $2.2\times 10^{-5 } $ \\
NGC\,4472 & C & $16.7^B$ & 0.081 & D$^{\dagger}$& 12.4 & $16.2\times 13.8$ & -32.6 & $\left(1.21 \pm 0.05\right) \times 10^{-1}$ & $\left(2.28 \pm 0.09\right) \times 10^{-1}$ & $\left(7.61 \pm 0.31\right) \times 10^{21}$ & $3.1\times 10^{-4 } $ \\
NGC\,4486 & A  & $16.7^B$ & 0.081 & J/L & 4.4 & $1.4\times 1.2$ & 49.1 & $\left(3.31 \pm 0.13\right) \times 10^{0}$ & $\left(2.03 \pm 0.09\right) \times 10^{1}$ & $\left(6.77 \pm 0.29\right) \times 10^{23}$ & $2.8\times 10^{-2 } $ \\
NGC\,4486 & B & $16.7^B$ & 0.081 & D$^{\dagger}$& 9.7 & $6.5\times 3.6$ & -51.1 & $\left(8.14 \pm 0.33\right) \times 10^{0}$ & $\left(1.23 \pm 0.05\right) \times 10^{2}$ & $\left(4.10 \pm 0.16\right) \times 10^{24}$ & $1.3\times 10^{-2 } $ \\
NGC\,4486 & C & $16.7^B$ & 0.081 & D$^{\dagger}$& 46.5 & $13.4\times 12.2$ & 0.8 & $\left(2.03 \pm 0.08\right) \times 10^{1}$ & $\left(1.52 \pm 0.06\right) \times 10^{2}$ & $\left(5.07 \pm 0.20\right) \times 10^{24}$ & $1.8\times 10^{-2 } $ \\
NGC\,4552 & A& $16.0^B$ & 0.078 & PS & - & $1.4\times 1.1$ & 13.6 & $\left(5.89 \pm 0.24\right) \times 10^{-2}$ & $\left(5.97 \pm 0.25\right) \times 10^{-2}$ & $\left(1.83 \pm 0.08\right) \times 10^{21}$ & $2.7\times 10^{-4 } $ \\
NGC\,4552 & C & $16.0^B$ & 0.078 & C$^{\dagger}$& 7.8 & $11.4\times 10.4$ & -26.5 & $\left(1.09 \pm 0.04\right) \times 10^{-1}$ & $\left(1.65 \pm 0.07\right) \times 10^{-1}$ & $\left(5.05 \pm 0.21\right) \times 10^{21}$ & $2.6\times 10^{-4} $ \\
NGC\,4636 & A & $15.96^L$ & 0.077 & C & 2.9 & $2.7\times 1.6$ & -74.7 & $\left(7.84 \pm 0.31\right) \times 10^{-3}$ & $\left(5.95 \pm 0.24\right) \times 10^{-2}$ & $\left(1.81 \pm 0.07\right) \times 10^{21}$ & $4.4\times 10^{-6} $ \\
NGC\,4636 & C & $15.96^L$ & 0.077 & C & 3.7 & $13.8\times 10.7$ & 116.7 & $\left(2.99 \pm 0.12\right) \times 10^{-2}$ & $\left(6.91 \pm 0.28\right) \times 10^{-2}$ & $\left(2.11 \pm 0.08\right) \times 10^{21}$ & $7.4\times 10^{-5 } $ \\
NGC\,4649 & A & $16.5^B$ & 0.080 & C & 0.3 & $1.4\times 1.3$ & -8.5 & $\left(9.68 \pm 0.39\right) \times 10^{-3}$ & $\left(1.44 \pm 0.06\right) \times 10^{-2}$ & $\left(4.69 \pm 0.19\right) \times 10^{20}$ & $2.2\times 10^{-5}$ \\
NGC\,4649 & D & $16.5^B$ & 0.080 & D$^{\dagger}$& 6.4 & $11.5\times 9.1$ & -6.8 & $\left(2.02 \pm 0.08\right) \times 10^{-2}$ & $\left(2.83 \pm 0.11\right) \times 10^{-2}$ & $\left(9.22 \pm 0.37\right) \times 10^{20}$ & $4.2\times 10^{-5} $ \\
NGC\,4696 & A  & $37.48^L$ & 0.182 & D$^{\dagger}$& 7.9 & $5.0\times 1.0$ & 8.5 & $\left(2.42 \pm 0.10\right) \times 10^{-1}$ & $\left(2.80 \pm 0.11\right) \times 10^{0}$ & $\left(4.71 \pm 0.19\right) \times 10^{23}$ & $3.8\times 10^{-4 } $ \\
NGC\,4696 & BC & $37.48^L$ & 0.182 & D$^{\dagger}$& 10.9 & $31.3\times 10.1$ & -34.6 & $\left(1.36 \pm 0.05\right) \times 10^{-1}$ & $\left(2.43 \pm 0.10\right) \times 10^{-1}$ & $\left(4.08 \pm 0.16\right) \times 10^{22}$ & $4.5\times 10^{-5 } $ \\
NGC\,4778 & A B & $59.29^{L\#}$ & 0.288 & PS & - & $6.7\times 5.6$ & -41.5 & $\left(5.57 \pm 0.22\right) \times 10^{-3}$ & $\left(5.15 \pm 0.53\right) \times 10^{-3}$ & $\left(2.17 \pm 0.22\right) \times 10^{21}$ & $3.5\times 10^{-4 } $ \\
NGC\,4778 & C & $59.29^{L\#}$ & 0.288 & PS & - & $18.1\times 11.5$ & -2.0 & $\left(3.81 \pm 0.15\right) \times 10^{-3}$ & $\left(3.95 \pm 0.73\right) \times 10^{-3}$ & $\left(1.66 \pm 0.31\right) \times 10^{21}$ & $4.4\times 10^{-4 } $ \\
NGC\,4782 & A  & $48.63^L$ & 0.236 & J/L & 30.2 & $0.5\times 0.4$ & -177.1 & $\left(7.55 \pm 0.30\right) \times 10^{-2}$ & $\left(1.24 \pm 0.05\right) \times 10^{0}$ & $\left(3.51 \pm 0.14\right) \times 10^{23}$ & $3.6\times 10^{-5 } $ \\
NGC\,4782 & BA & $48.63^L$ & 0.236 & D$^{\dagger}$& 55.9 & $5.8\times 3.9$ & 16.0 & $\left(7.48 \pm 0.30\right) \times 10^{-2}$ & $\left(5.43 \pm 0.22\right) \times 10^{0}$ & $\left(1.54 \pm 0.06\right) \times 10^{24}$ & $2.7\times 10^{-4 } $ \\
NGC\,4936 & A  & $31.36^L$ & 0.152 & PS & - & $3.3\times 1.2$ & 12.9 & $\left(3.33 \pm 0.13\right) \times 10^{-3}$ & $\left(3.34 \pm 0.14\right) \times 10^{-3}$ & $\left(3.93 \pm 0.17\right) \times 10^{20}$ & $2.6\times 10^{-5 } $ \\
NGC\,5044 & A  & $35.75^L$ & 0.174 & C$^{\dagger}$& 1.4 & $2.1\times 1.3$ & -17.6 & $\left(2.72 \pm 0.11\right) \times 10^{-2}$ & $\left(2.95 \pm 0.12\right) \times 10^{-2}$ & $\left(4.51 \pm 0.18\right) \times 10^{21}$ & $3.0\times 10^{-5 } $ \\
NGC\,5044 & BA  & $35.75^L$ & 0.174 & D$^{\dagger}$& 23.1 & $5.3\times 3.6$ & -54.2 & $\left(2.78 \pm 0.11\right) \times 10^{-2}$ & $\left(3.25 \pm 0.13\right) \times 10^{-2}$ & $\left(4.97 \pm 0.20\right) \times 10^{21}$ & $2.7\times 10^{-5 } $ \\
NGC\,5044 & D & $35.75^L$ & 0.174 & D & 24.3 & $62.1\times 37.2$ & 2.1 & $\left(3.34 \pm 0.13\right) \times 10^{-2}$ & $\left(3.40 \pm 0.14\right) \times 10^{-2}$ & $\left(5.20 \pm 0.21\right) \times 10^{21}$ & $1.7\times 10^{-4 } $ \\
NGC\,5129 & A  & $86.85^L$ & 0.422 & C$^{\dagger}$& 4.0 & $2.3\times 1.6$ & -5.5 & $\left(5.50 \pm 0.22\right) \times 10^{-3}$ & $\left(5.05 \pm 0.21\right) \times 10^{-3}$ & $\left(4.56 \pm 0.19\right) \times 10^{21}$ & $3.0\times 10^{-5 } $ \\
NGC\,5419 & A  & $50.87^L$ & 0.247 & D & 6.1 & $3.7\times 1.1$ & 16.4 & $\left(3.84 \pm 0.15\right) \times 10^{-3}$ & $\left(2.11 \pm 0.08\right) \times 10^{-2}$ & $\left(6.53 \pm 0.26\right) \times 10^{21}$ & $2.3\times 10^{-6 } $ \\
NGC\,5419 & B & $50.87^L$ & 0.247 & D & 13.5 & $10.7\times 3.8$ & 0.6 & $\left(1.24 \pm 0.05\right) \times 10^{-1}$ & $\left(3.23 \pm 0.13\right) \times 10^{-1}$ & $\left(1.00 \pm 0.04\right) \times 10^{23}$ & $2.5\times 10^{-4 } $ \\
NGC\,5419 & CD & $50.87^L$ & 0.247 & D & 181.8 & $33.6\times 28.3$ & 19.8 & $\left(2.97 \pm 0.12\right) \times 10^{-1}$ & $\left(5.54 \pm 0.22\right) \times 10^{-1}$ & $\left(1.72 \pm 0.07\right) \times 10^{23}$ & $1.3\times 10^{-4 } $ \\
NGC\,5813 & A B & $29.23^L$ & 0.142 & D & 2.7 & $5.3\times 4.5$ & -33.3 & $\left(4.04 \pm 0.16\right) \times 10^{-3}$ & $\left(7.66 \pm 0.36\right) \times 10^{-3}$ & $\left(7.83 \pm 0.37\right) \times 10^{20}$ & $6.9\times 10^{-5 } $ \\
NGC\,5813 & B & $29.23^L$ & 0.142 & D & 4.1 & $6.2\times 4.7$ & -21.9 & $\left(3.60 \pm 0.14\right) \times 10^{-3}$ & $\left(5.34 \pm 0.38\right) \times 10^{-3}$ & $\left(5.46 \pm 0.38\right) \times 10^{20}$ & $1.7\times 10^{-4 } $ \\
NGC\,5813 & D & $29.23^L$ & 0.142 & D & 19.5 & $45.2\times 36.5$ & 20.2 & $\left(1.27 \pm 0.05\right) \times 10^{-2}$ & $\left(1.46 \pm 0.06\right) \times 10^{-2}$ & $\left(1.49 \pm 0.06\right) \times 10^{21}$ & $6.5\times 10^{-5 } $ \\
NGC\,5846 & A  & $27.13^L$ & 0.132 & C & 4.2 & $1.5\times 1.3$ & 8.3 & $\left(9.25 \pm 0.37\right) \times 10^{-3}$ & $\left(2.07 \pm 0.09\right) \times 10^{-2}$ & $\left(1.82 \pm 0.08\right) \times 10^{21}$ & $2.6\times 10^{-5 } $ \\
NGC\,5846 & B & $27.13^L$ & 0.132 & C & 2.7 & $8.0\times 5.7$ & -20.3 & $\left(1.02 \pm 0.04\right) \times 10^{-2}$ & $\left(1.57 \pm 0.09\right) \times 10^{-2}$ & $\left(1.38 \pm 0.08\right) \times 10^{21}$ & $2.2\times 10^{-4 } $ \\
NGC\,5846 & CD & $27.13^L$ & 0.132 & D & 19.4 & $58.7\times 25.6$ & -61.7 & $\left(1.82 \pm 0.07\right) \times 10^{-2}$ & $\left(2.00 \pm 0.10\right) \times 10^{-2}$ & $\left(1.76 \pm 0.08\right) \times 10^{21}$ & $3.0\times 10^{-4 } $ \\
NGC\,7619 & A  & $50.53^L$ & 0.245 & C & 2.2 & $2.2\times 1.1$ & 58.8 & $\left(1.20 \pm 0.05\right) \times 10^{-2}$ & $\left(1.91 \pm 0.08\right) \times 10^{-2}$ & $\left(5.84 \pm 0.24\right) \times 10^{21}$ & $2.8\times 10^{-5 } $ \\
NGC\,7619 & C & $50.53^L$ & 0.245 & PS & 104.2 & $15.6\times 15.2$ & -47.0 & $\left(2.1 \pm 0.1\right) \times 10^{-2}$ & $\left(2.5 \pm 0.1\right) \times 10^{-2}$ & $\left(7.64 \pm 0.34\right) \times 10^{21}$ & $2.4\times 10^{-4 } $ \\
\enddata
\tablecomments{(1) sources marked with `*' are our new VLA A observations obtained within the project 15A-305; (3) distances marked with $\#$ are based on redshift measurements; {\bf References:} L:\cite{lakhchaura2018}; B:\cite{blakeslee2009}; M: \cite{mieske2005}; (5) radio morphological categories: J/L: jets/lobes; D: diffuse and D$^{\dagger}$: diffuse with signs of jet/lobe-like morphology; C: compact and C$^{\dagger}$ with signs of small-scale jet/lobe-like morphology; PS: point source-like radio emission; NS: no radio source detected.}
\end{deluxetable*}
\end{longrotatetable}
\end{startlongtable}

\normalsize

\subsection{Multi-scale radio emission}
\label{sec:detec_rate}
\subsubsection{Central radio emission}
\label{sec:radio_central}
Our results show a high radio core\footnote{A radio core is point-like radio emission within \\ twice the beam size located at the center of host galaxy.} detection rate of $98 \%$ (41/42) for the early-type galaxies in our sample within the frequency range between 1--2\,GHz, centered at 1.5\,GHz (Table\,\ref{tab:results}).

For the remaining galaxy, NGC\,499, no central radio emission has been detected in the available VLA observations in the L-band (1--2\,GHz). However, an archival single-dish Arecibo observation at 2.38\,GHz \citep{dressel1978}, an observation by the NVSS survey at 1.4\,GHz \citep[although, only with a 2$\sigma$ detection]{brown2011} and most recently a  low-frequency observation with LOFAR \citep{birzan2020} detect a faint nuclear point-like radio source in the centre of NGC\,499. Thus, due to the previously confirmed presence of radio emission from NGC\,499 in the above-mentioned studies, the detection rate in our sample could be considered to be 100$\%$.

The non-detection of the radio emission from NGC\,499 in the highest resolution VLA A configuration data could be due to the shallow observations with a resulting RMS noise of $\sim$\,12.0$\times 10^{-5}$\,Jy/beam, which is twice the sensitivity limit for our faintest detected radio source in our sample: NGC\,4406 (Table\,\ref{tab:results}). Observations with LOFAR at $\sim$\,150\,MHz detected a radio core with a total integrated flux density of 0.046$\pm$0.009\,Jy. Assuming a spectral index of about -0.7, we estimate that NGC\,499 would have a radio source with the total flux density of 0.009\,Jy at 1.5\,GHz, which is still above the threshold of the archival VLA A observation, potentially indicating a steeper spectrum. We note that the much higher flux density of 0.26\,Jy reported by \cite{condon1988} at  1.4\,GHz frequency does not agree with all other available radio measurements.

\subsubsection{Extended radio emission}
\label{sec:radio_extended}
For 67$\%$ (28/42) of the galaxies in our sample, we detect a diffuse, extended morphology.

It is worth noting that we present only a lower limit for the number of galaxies with extended radio emission. Due to our main interest in detecting the radio cores, the highest resolution A configuration data are prioritized. Sensitivity to the diffuse, extended emission was sacrificed at the expense of the higher resolution of the fine structures in the central region. The more compact configuration observations (C or D) are missing for a fraction of sources (11 sources out of 42) in our sample.

\subsection{Radio morphology}
The morphology of the radio emission at 1--2\,GHz varies widely within our sample of early-type galaxies.

On the one hand, the total intensity radio maps reveal well-collimated large-scale radio jets and lobes extending up to hundreds of kiloparsecs (e.g. NGC\,315 ; Figure\,\ref{fig:chandra_vla_a}c) or small-scale jets residing within a few kpc from the radio nucleus (e.g. NGC\,5129; Figure\,\ref{fig:chandra_vla_c}h). On the other hand, we also observe many galaxies, where the jets and lobes are disturbed by the influence of the surrounding hot gas or interaction with other galaxies: narrow or wide angle tails (e.g. NGC\,507; Figure\,\ref{fig:chandra_vla_a}d; and IC\,4296; Figure\,\ref{fig:chandra_vla_a}b), tails tracing the path of the interaction with another galaxy (e.g. NGC\,741; Figure\,\ref{fig:chandra_vla_a}g), S-shaped radio emission (e.g. NGC\,1316; Figure\,\ref{fig:chandra_vla_b}a), diffuse morphology with no clear jets or lobes (e.g. NGC\,1407; Figure\,\ref{fig:chandra_vla_b}c) and relic-like large-scale diffuse emission (e.g. NGC\,5419; Figure\,\ref{fig:chandra_vla_d}a). A more detailed definition of our categories is given in Section\,\ref{sec:categories}. 

For IC\,4296 (Figure\,\ref{fig:chandra_vla_a}b), NGC\,57 (Figure\,\ref{fig:chandra_vla_ps_a}b), NGC\,533 (Figure\,\ref{fig:chandra_vla_a}e), NGC\,1550 (Figure\,\ref{fig:chandra_vla_b}d), NGC\,4261 (Figure\,\ref{fig:chandra_vla_b}f), NGC\,4374 ( Figure\,\ref{fig:chandra_vla_b}g), NGC\,4472 (Figure\,\ref{fig:chandra_vla_b}h), NGC\,4552 (Figure\,\ref{fig:chandra_vla_c}b), NGC\,5129 (Figure\,\ref{fig:chandra_vla_c}h) and NGC\,5419 (Figure\,\ref{fig:chandra_vla_d}a), we found new, previously unobserved and undescribed radio morphology at the high resolution of about 1--2\,arcseconds in the frequency range of 1--2\,GHz (red contours in Figures in Appendix\,\ref{app:xray_radio_ext}).\

Details of radio morphologies together with relevant information on the multifrequency data for each source individually are given in Appendix\,\ref{app:individual}.

\subsection{Comparison with X-ray data}
\label{sec:comp_xrays}
\subsubsection{X-ray cavity rate}
The hot X-ray emitting atmospheres embedded in galaxies are closely related to the activity of their central radio-emitting AGN \citep[e.g.:][]{mcnamara2005,werner2019}.

In our sample, we compared {\it Chandra} X-ray images with the total intensity radio images searching for signatures of interaction as X-ray cavities. We reported also all known X-ray cavities from the literature.\

We also detected X-ray cavities for 7 sources with only point-like radio morphology and for NGC\,499 \citep{panagoulia2014b,kim2019}. Cavities in  NGC\,1132 and NGC\,4778, which host point-like radio sources were previously detected by \citet{dong2010} and \citet{morita2006}. Additionally, Plšek et al. (in prep.) found potential cavities for another five point-like radio sources: NGC\,2300, NGC\,3091, NGC\,3923, NGC\,4073, NGC\,4125 (see Discussion \ref{sec:ghost_cavities} for more details). 

In our sample we can summarize that 34 out of 42 (81$\%$) early-type galaxies show detectable X-ray cavities in their hot X-ray atmospheres. \

\subsubsection{Central X-ray point source rate}
In the parent X-ray study, \cite{lakhchaura2018} found nuclear X-ray point sources for 32$\%$ of early-type galaxies (16 out of 49 in their study), from which 14 X-ray point sources are relevant to our study.

It is worth noting that while radio emission is detected in 98(--100)$\%$ of the sources from our sample, only about 33$\%$ (14/42) of the systems also host a detectable central X-ray point source.

\subsection{Multifrequency correlations}
\label{sec:multif_correlations}
\begin{figure}
    \centering
    \includegraphics[scale=5,width=230pt]{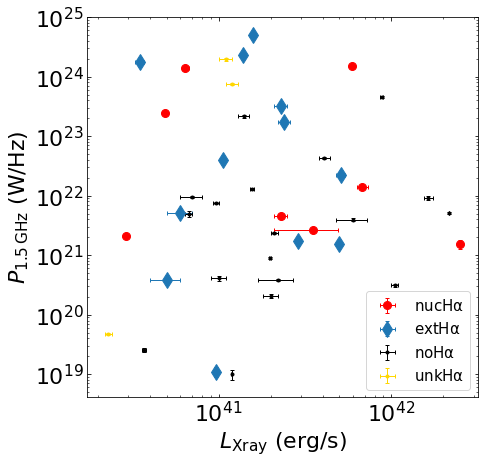}
    \includegraphics[scale=5,width=238pt]{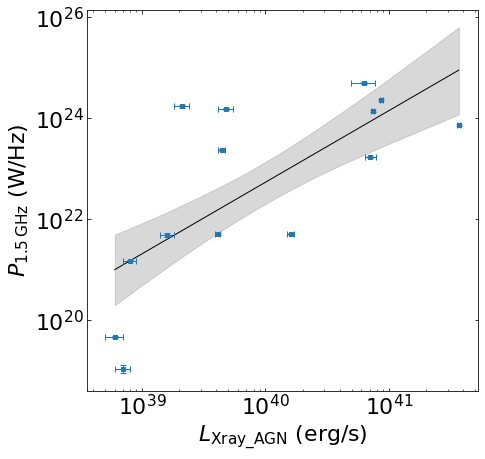}

    \caption{{\bf Top:} Relation between radio power at $\sim$\,1.5\,GHz ($P_{\rm 1.5GHz}$) vs. the X-ray luminosity ($L_{\mathrm{Xray}}$) within the 10\,kpc from the center of the galaxy shows no correlation. {\bf Bottom:} Significant correlation (Spearman and Pearson coefficient of 0.71 with $p$-value of 0.004)) is seen, when we compare the radio power at $\sim$\,1.5\,GHz ($P_{\rm 1.5GHz}$) with the luminosity of the central X-ray point sources ($L_{\mathrm{\rm{Xray\_AGN}}}$) \citep{lakhchaura2018} detected only in a subset of our main sample of early-type galaxies. Error bars on the radio power are mostly smaller than the plotted symbol size. The correlation coefficients and corresponding $p$-values are given in Table\,\ref{tab:corr}. The black line represents the linear fit of the distribution, while the shaded area shows the one-sigma confidence region for the correlation.
   } 
    \label{fig:p_lx}
\end{figure}

\begin{figure}
    \centering
    \includegraphics[width=230pt]{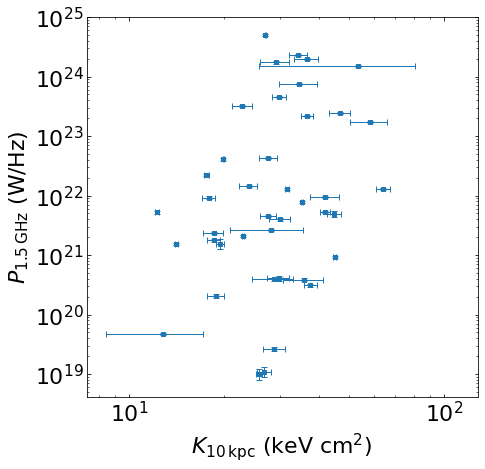} \includegraphics[width=230pt]{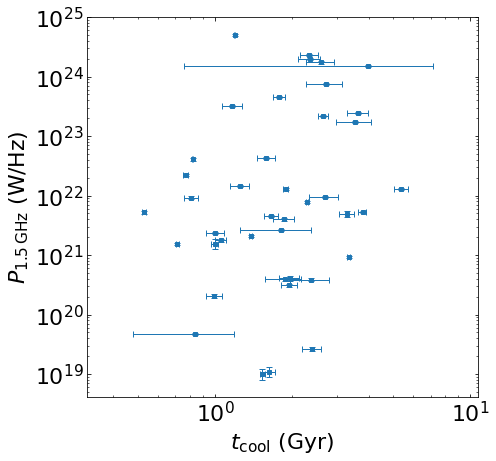}
    \caption{The radio power ($P_{\rm 1.5GHz}$) plotted against the entropy ({\bf top}) and cooling time ({\bf bottom}) of the hot X-ray emitting gas measured within $r<10$ kpc shows no correlation. Error bars on the radio power are mostly smaller than the plotted symbol size.}
    \label{fig:thermo_pv}
\end{figure}

\begin{figure}
    \centering
    \includegraphics[scale=5,width=205pt]{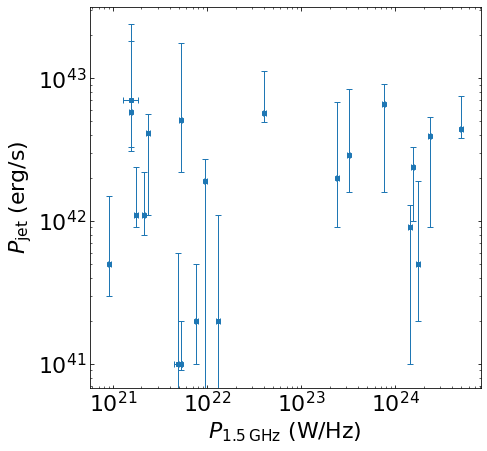}\\
    
    \includegraphics[scale=5,width=205pt]{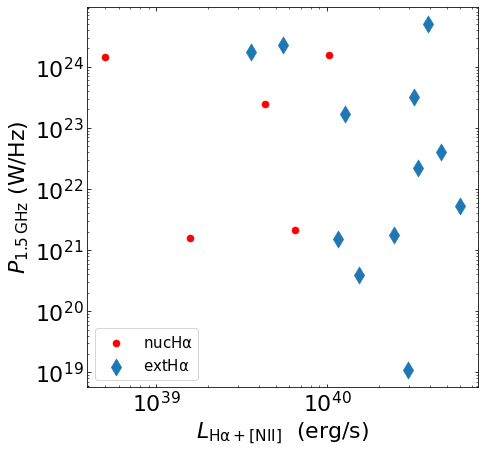}\\
    
    \includegraphics[scale=5,width=205pt]{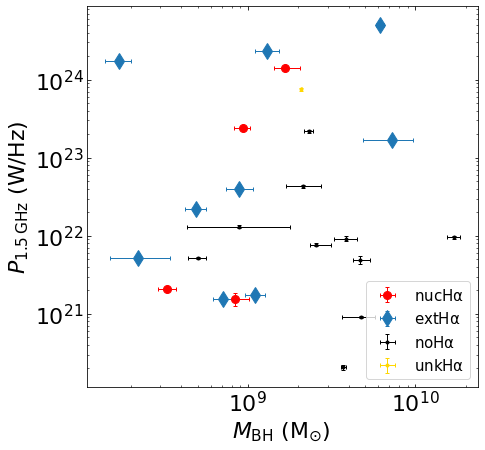}
    \caption{ \small {\bf Top:} No correlation is found between the radio power ($P_{1.5\,{\rm GHz}}$) and the jet power ($P_{\rm jet}$; {\bf top}), the H$\alpha +$[N II] luminosity ($L_{\rm H_{alpha}+[N_{II}]}$; {\bf middle}) as well as the supermassive black hole mass ($M_{\rm BH}$; {\bf bottom}). The nuclear (nucH$\alpha$), extended (extH$\alpha$), undetected (noH${\alpha}$) and unknown (unkH${\alpha}$) H${\alpha}+\mathrm{[NII]}$ emission are distinguished. The correlation coefficients and corresponding $p$-values are given in Table\,\ref{tab:corr}. Error bars on the radio power are mostly smaller than the plotted symbol size.
    \label{fig:pv_pjet_ha_mbh}}
\end{figure}

\begin{figure*}[h!]
\centering
\includegraphics[width=480pt]{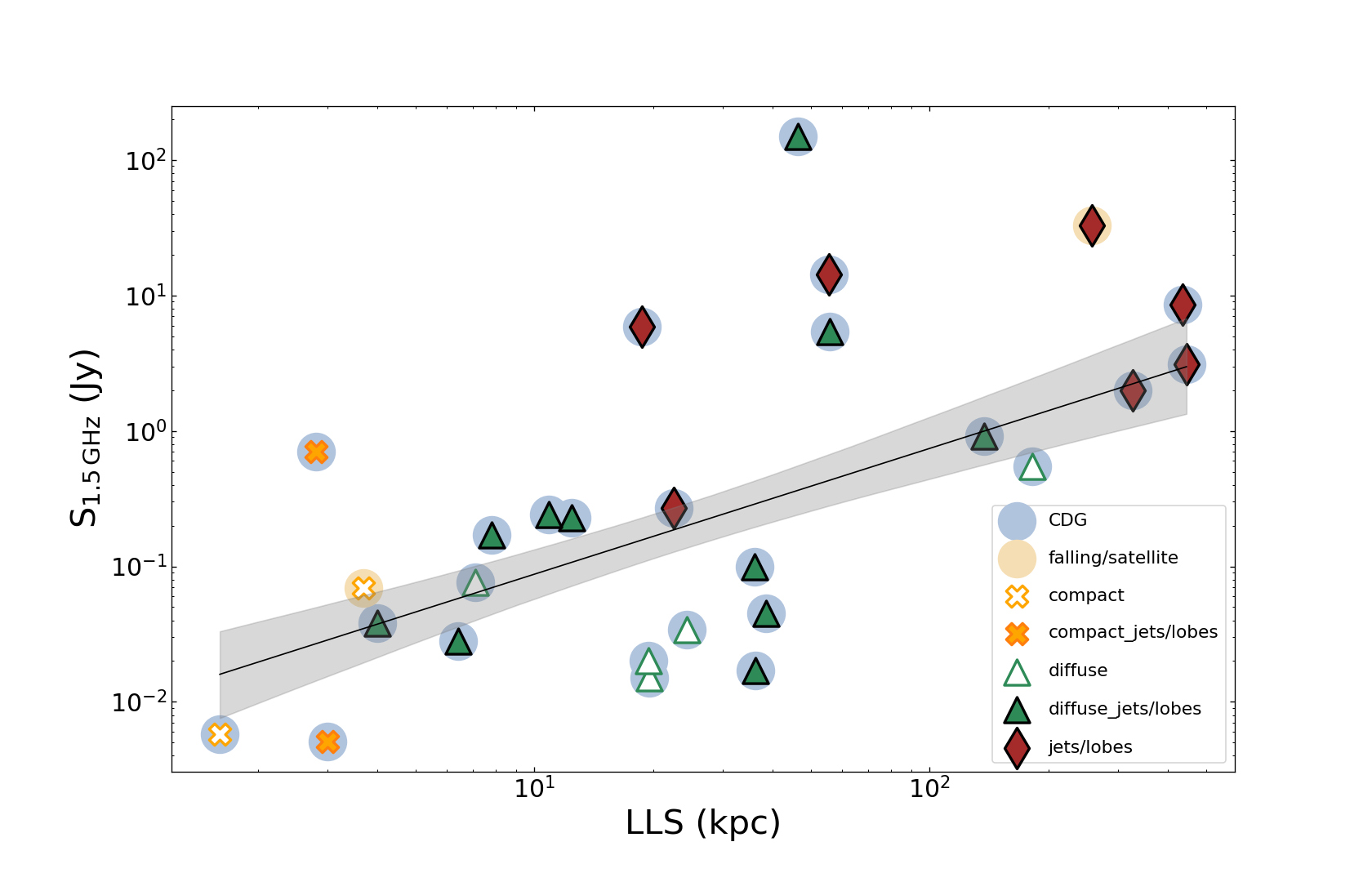}
    \caption{The total flux density at 1.5\,GHz ($S_{1.5\,{\rm GHz}}$) is plotted against the largest linear size ($LLS$) of the radio source. A positive correlation with a {\tt linmix} correlation coefficient of $0.67^{+0.08}_{-0.09}$ is determined. The brighter the radio source, the larger its linear extent. The correlation coefficients and corresponding $p$-values are given in Table\,\ref{tab:corr}. The black line represents the linear fit of the distribution, while the shaded area shows the 68$\%$ confidence region for the correlation.}
        \label{fig:llsVSp1.5}
\end{figure*}

\begin{figure*}[h!]
    \centering
    \includegraphics[width=320pt]{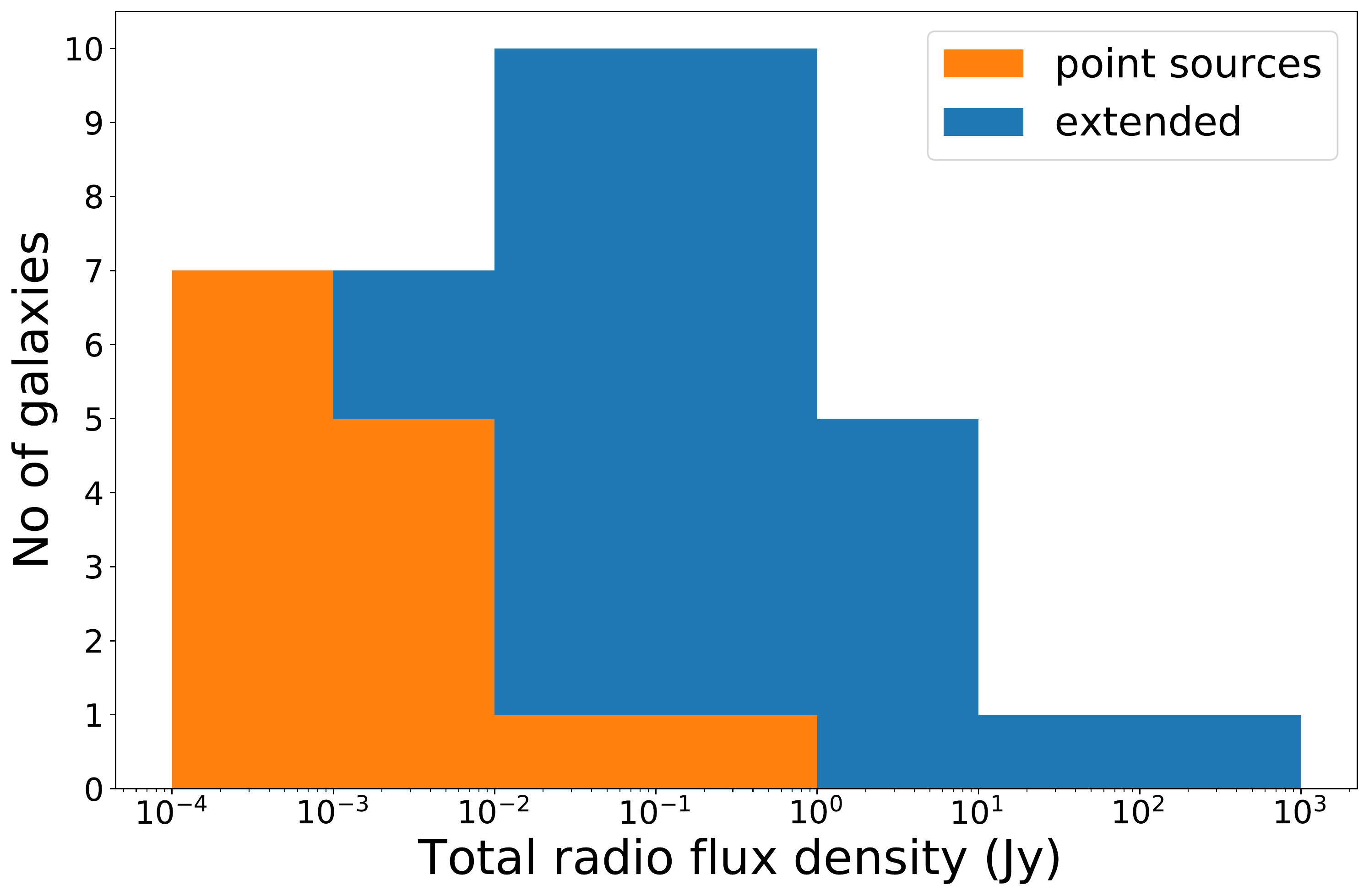}
    \caption{The total integrated radio flux density histogram with two morphologically different categories: point sources and extended sources. The fainter sources fall into the point sources category, whereas the brighter galaxies are the extended ones.}
    \label{fig:hist_numVSflux}
\end{figure*}

We investigate the correlations and trends between the radio power at $\sim$\,1.5\,GHz ($P_{\rm 1.5GHz}$) and the X-ray luminosity within 10\,kpc from the center of the galaxy (Figure\,\ref{fig:p_lx}; top), the luminosity of the central X-ray point source or AGN (Figure\,\ref{fig:p_lx}; bottom), entropy and cooling time of the hot X-ray emitting gas (Figure\,\ref{fig:thermo_pv}; top and bottom, respectively), the power of the jet estimated from X-ray cavities (Figure\,\ref{fig:pv_pjet_ha_mbh}; top), the luminosity of H$\alpha +$[N II] nebulae (Figure\,\ref{fig:pv_pjet_ha_mbh}; middle), the mass of the central supermassive black hole (Figure\,\ref{fig:pv_pjet_ha_mbh}; bottom), and the largest linear size of the radio emission (Figure\,\ref{fig:llsVSp1.5}).
Moreover, the flux density distribution for point-like and extended radio sources is presented in Figure\,\ref{fig:hist_numVSflux}. 

The details of the H$\alpha +$[N II] nebulae emission extent and other information about multiphase gas is given in Appendix\,\ref{app:app_tab_multi-fr} and Table\,\ref{tab:multi-f}.\\

The Spearman and Pearson correlation coefficients are used to derive the trends between two investigated quantities without accounting for the measurement uncertainties. 
Moreover, a Bayesian approach to linear regression taking into account the corresponding measurement uncertainties for both $x$ and $y$ variables, as well as the upper limits was used within the {\tt linmix} package\footnote{\href{https://github.com/jmeyers314/linmix}{https://github.com/jmeyers314/linmix}} \citep{Kelly2007}. 

The statistical significance of the trends is represented by the probability derived from the null hypothesis test (i.e., the $p$-value).

\begin{table*}
	\centering
	\tablenum{2}
	\begin{tabular}{lcccccccc}
		 \hline\hline
		Investigated & & \multicolumn{2}{c}{Spearman}& \multicolumn{2}{c}{Pearson} & {\tt linmix} & & \\
		Relations & Fitted points & $r$ &  $p$-value & $r$ & $p$-value & $\rho$ & $\alpha$ & $\beta$ \\ 
        $P_{1.5\,{\rm GHz}}$ vs. & & & &  &  & & & \\
		\hline
$L_{\mathrm{Xray}}$ & all &  $0.06 $ & $ 0.71$  & $0.06 $ & $ 0.71$ & $0.06^{+0.17}_{-0.16}$ & $40.9 \pm 1.2$ & $0.02 \pm 0.06$ \\
 & nuc/extH${\alpha}$ & $-0.12 $ & $ 0.62$ & $-0.10 $ & $ 0.66$  & $-0.10^{+0.20}_{-0.20}$ &  $42.1 \pm 2.0$ & $-0.4 \pm 0.9$ \\
 & no H${\alpha}$ & $0.19$ & $ 0.046$ & $0.33 $ & $ 0.19$ & $0.32^{+0.22}_{-0.26}$ &$38.5 \pm 2.4$ & $0.1 \pm 0.11$   \\
$L_{\mathrm{Xray\_AGN}}$ & all & $0.71$ & $ <0.004$ & $0.71$ & $ <0.004$  & $0.68^{+0.14}_{-0.21}$  & $31.9\pm2.9$ & $0.4 \pm 0.13$ \\
$K_{10{\rm\,kpc}}$ & all &  $ 0.03$ & $0.37$ & $ 0.02$ & $0.34$ & $0.34^{+0.14}_{-0.16}$  & $0.7 \pm 0.4$ & $0.04 \pm 0.02$ \\
$t_{{\rm\,cool}}$ & all & $0.29$ & $ 0.06$ & $0.31$ & $ 0.05$ & $0.17^{+0.17}_{-0.15}$ & $-0.7\pm 0.6$ & $0.04 \pm 0.03$  \\
$P_{\mathrm{jet}}$ & all & $0.05$ & $ 0.83$ & $0.22$ & $ 0.33$ & $0.33^{+0.23}_{-0.27}$ &  $39.4 \pm 2.4$ & $0.13 \pm 0.1$  \\
$L_{\mathrm{H}{\alpha}+\mathrm{[NII]}}$ & all & $-0.23$ & $ 0.34$ & $-0.31$ & $ 0.19$ & $-0.30^{+0.24}_{-0.22}$  & $-0.1 \pm 0.1$ & $42.3 \pm 2.0$ \\ \vspace{-0.6mm}
$M_{\mathrm{BH}}$ & all & $0.07$ & $ 0.75$ & $0.02$ & $ 0.93$ & $0.01^{+0.22}_{-0.22}$  & $9.1 \pm 2.1$ & $0.01 \pm 0.09$ \\ \vspace{-0.6mm}
 & nuc/extH${\alpha}$ & $0.41$ & $ 0.16$ & $0.39$ & $ 0.18$ & $0.38^{+0.25}_{-0.31}$ & $5.8 \pm 2.8$ & $0.1 \pm 0.1$ \\
 & no H${\alpha}$ & $-0.37$ & $ 0.29$ & $-0.18$ & $0.62$ & $-0.16^{+0.35}_{-0.38}$  & $11.4\pm 5.4$ & $-0.1 \pm 0.3$ \\
 \hline       
$S_{1.5{\rm\,GHz}}$ vs. $LLS$ & all & $0.64$ & $<0.0002$ &  $0.64$ & $<0.0003$ & $0.63^{+0.11}_{-0.14}$ & $1.62 \pm 0.1$ & $0.4 \pm 0.1$ \\
\hline
\end{tabular}
	\caption{The {\tt linmix} correlation coefficients between the radio power at 1.5\,GHz and the properties of the ambient hot and warm gas and the mass of the central supermassive black hole. The corresponding parameters of the fit are: the Spearman correlation coefficient and Pearson correlation coefficient and the corresponding $p$-values, the {\tt linmix} correlation coefficient ($\rho$), the log normalisation ($\alpha$) and the power-law index ($\beta$).}
	\label{tab:corr}
\end{table*}

Significant trends, with Spearman and Pearson correlation factor of 0.71 and corresponding $p$-value of 0.004 are obtained when comparing the radio power with the luminosity of the central X-ray point source  (Figure\,\ref{fig:p_lx}; bottom) and the radio flux density with the largest linear size of the detected radio emission (Figure\,\ref{fig:llsVSp1.5}).

Negligible correlations are found when comparing the radio power with the jet power required to inflate the cavities; the luminosity of the X-ray atmosphere (Figure\,\ref{fig:p_lx}; top), and the luminosity of the warm ionized nebulae, traced by H${\alpha} +$[N II] line emission (Figure\,\ref{fig:pv_pjet_ha_mbh}; middle), the mass of the supermassive black hole (Figure\,\ref{fig:pv_pjet_ha_mbh}; bottom) as well as various thermodynamical properties of the X-ray emitting hot gas like entropy (Figure\,\ref{fig:thermo_pv}; top) and cooling time (Figure\,\ref{fig:thermo_pv}; bottom) \citep{lakhchaura2018}.

\clearpage
\newpage
 \begin{figure}[h!]
    \centering
\includegraphics[width=250pt]{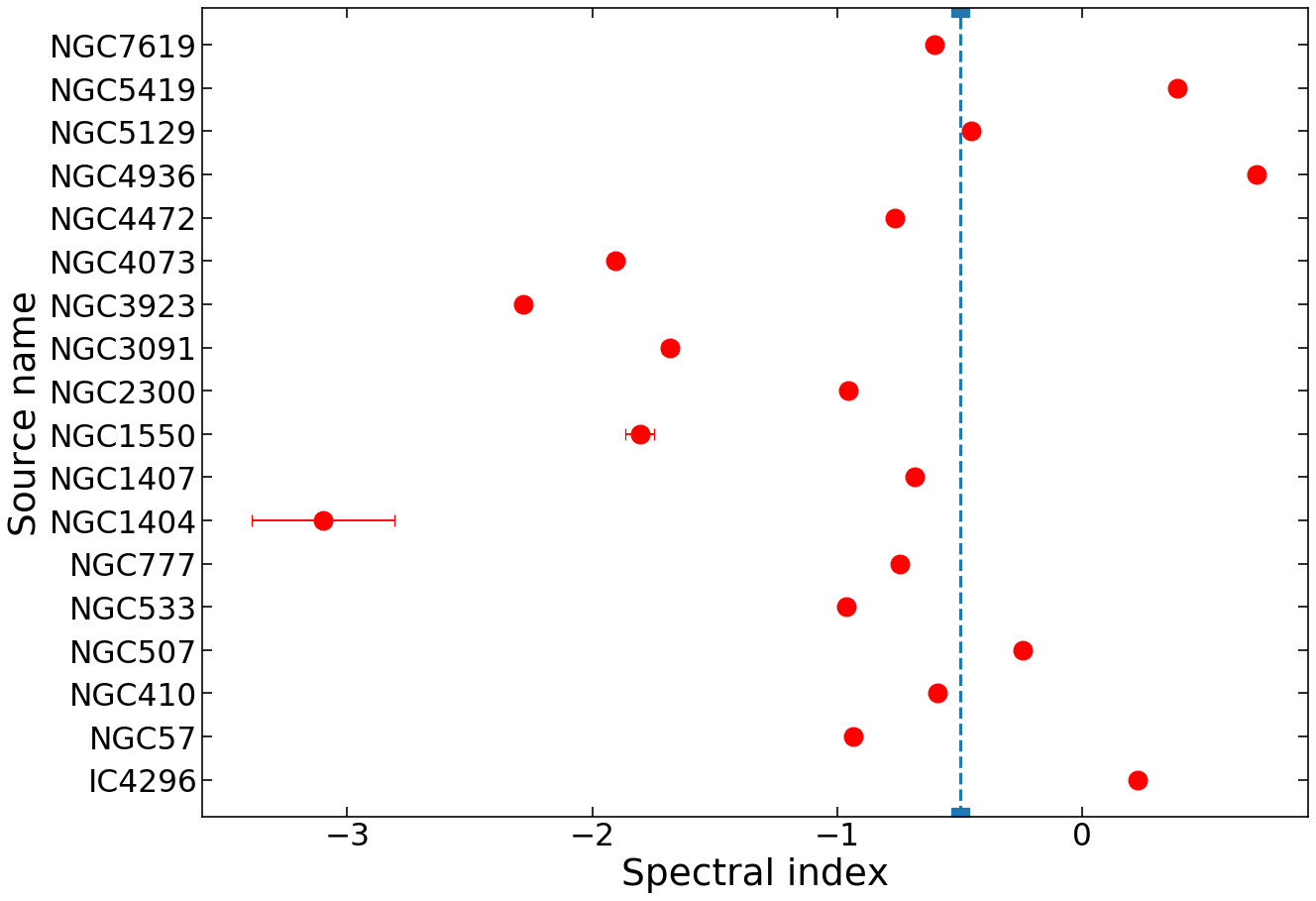}
    \caption{Radio spectral indices for a subset of our sample. Approximately, 3/4 of the sources have steep spectra ($\alpha > -0.5$) and the rest have flat and inverted radio spectra ($\alpha \leq -0.5$). The vertical blue dashed line indicates the spectral index of -0.5.}
    \label{fig:spim_galaxies}
\end{figure}

\subsection{Nuclear spectral indices}
\label{sec:spim}
To estimate the age of the radio emitting plasma in the innermost central regions (within 0.7\,kpc radius of the radio core) of the early-type galaxies, we determined in-band spectral indices for 18 out of 42 galaxies observed with the upgraded VLA A configuration.  
The description of the approach we used to determine the spectral indices for the subsample of our galaxies can be found in Section\,\ref{sec:spindex}.

The final central spectral indices are presented in Table\,\ref{tab:spim} and Figure\,\ref{fig:spim_galaxies}. 
Interestingly, about 1/3 of our sources have inverted or flat radio spectra ($\alpha\geq-0.5$), while the majority showed steep spectral indices ($\alpha<-0.5$). This result is consistent with a scenario where some of the sources were recently active, while for some, the spectral shape is in agreement with synchrotron cooling.

\begin{table}
	\centering
\tablenum{3}
	\begin{tabular}{lc}
		\hline\hline
		Source & Spectral index \\
		\hline
IC\,4296 & 0.224$\pm$0.005 \\
NGC\,57 & -0.934$\pm$0.014 \\
NGC\,410 & -0.698$\pm$0.055 \\
NGC\,507 & -0.289$\pm$0.058 \\
NGC\,533 & -0.978$\pm$0.068 \\
NGC\,777 & -0.757$\pm$0.080 \\
NGC\,1404 & -2.554$\pm$0.621 \\
NGC\,1407 & -0.727$\pm$0.025 \\
NGC\,1550 & -1.628$\pm$0.387 \\
NGC\,2300 & -0.954$\pm$0.004 \\
NGC\,3091 & -1.259$\pm$0.227 \\ 
NGC\,3923 & -2.074$\pm$0.851 \\
NGC\,4073 & -0.647$\pm$0.053 \\
NGC\,4472 & -0.548$\pm$0.228 \\
NGC\,4936 & 0.491$\pm$0.042 \\
NGC\,5129 & -0.506$\pm$0.006 \\
NGC\,5419 & 0.342$\pm$0.002 \\
NGC\,7619 & -0.614$\pm$0.014 \\
\hline
	\end{tabular}
	\caption{The mean values of the spectral indices and corresponding uncertainties determined from the central region with diameter of 0.7\,kpc for a subset of our early-type galaxies.}
	\label{tab:spim}
\end{table}

\section{Discussion}
\label{sec:discussion}
\subsection{High detection rate}
The early-type galaxies in our sample show a very high radio detection rate at frequencies between 1--2\,GHz. For 41 out of the sample of 42 galaxies, we detected at least point-like emission in the central region. A significant fraction, 27/42 galaxies, shows extended radio emission from jets and lobes. For 26 out of these 42 galaxies, the presence of X-ray cavities was detected. Moreover, X-ray cavities were also detected for 8 sources without jets/lobes. Altogether, 34 sources show signatures of interaction with the X-ray emitting medium. This suggests that most of these systems are operating in the radio-mechanical or maintenance mode AGN feedback. 

Our high detection rate of radio sources appears to be consistent with previous findings \citep[e.g.:][]{dunn2010,brown2011, sabater2019}.
At first glance, these results indicate a very high AGN duty cycle. It is, however, worth noting that the detected extended radio emission does most likely not always represent the current state of AGN activity and is often a remnant from a previous cycle.\

\subsection{Origin of radio emission}
\label{sec:sf}
To investigate whether the origin of the observed radio emission is solely related to AGN activity or could also be linked to star formation\footnote{Especially for low luminosity galaxies with no significant extended emission known as FR\,0s \citep{baldi2015}, the synchrotron emission at GHz frequencies could also be due to particles accelerated in star forming regions \cite[see e.g.,][]{condon1992,lacki2013}}, we follow the analysis of \cite{kolokythas2018}. We determine the far-ultraviolet (FUV) fluxes of the galaxies with point-source-like radio emission using the Galaxy Evolution Explorer (GALEX) Survey GR6 catalog\footnote{\href{https://galex.stsci.edu/GR6/?page=mastform}{https://galex.stsci.edu/GR6/?page=mastform}} to estimate the star formation rate (SFR) \citep{bell2003}, from which we determine the expected radio power \citep{salim2007}. If the expected radio power from the FUV estimated SFR is at least half of the observed radio power, star formation could potentially be the dominant radio emission mechanism.

Our analysis shows that for 5 out of the 14 galaxies with point-like radio morphology, namely NGC\,1404, NGC\,3091, NGC\,3923, NGC\,4073 and NGC\,4406, star formation could dominate the observed radio emission (see Table\,\ref{tab:fuvflux}). However, 3/5 sources where the radio emission could potentially be dominated by star-formation activity, also host ghost cavities, indicating that radio-mode AGN activity is also present in these galaxies (see Section\,\ref{sec:ghost_cavities}). Importantly, 7/14 point-like galaxies display cavities indicating that despite the lack of extended radio structures at 1--2\,GHz, these galaxies host a radio AGN capable of inflating lobes and cavities. The two galaxies, which lack observable signatures of radio mode AGN feedback are NGC\,1404 and NGC\,4406, which are falling through and being stripped by the ICM of the Fornax and Virgo clusters, respectively.

Using FUV fluxes as an indicator of star formation, \cite{kolokythas2018} found that for 5 of the 26 galaxies in their high richness subsample of the Complete Local-volume Groups Sample (CLoGS), the star formation could have a significant contribution to the radio emission.

\begin{startlongtable}
\begin{deluxetable*}{lcccc}
\tablenum{4}
\tablecaption{The star formation contribution to the radio emission in the case of early-type galaxies with point-source radio morphologies. The columns: (1) the source name; ` * ': sources observed within our project 15A-305; (2) the far-ultraviolet flux from the GALEX-DR5 (GR5) catalogue \citep{bianchi2011}; (3) the estimated star formation rate (SFR) according to the relation of \cite{bell2003}; (4) the expected radio power from the SFR \citep{salim2007}; (5) the half value of the radio power; if the expected power is greater than 1/2 of the observed power, star formation dominates.}
\label{tab:fuvflux}
\tablewidth{0pt}
\tablehead{
\colhead{Source} & \colhead{FUV$_{\rm flux}$ $^{(a)}$} & \colhead{SFR$_{\rm FUV}$} & \colhead{$P_{\rm 1.5GHz\_exp}$} & 50$\%$ of $P_{\rm 1.5GHz\_obs}$ \\
\colhead{Name} & \colhead{($\mu$Jy)} & \colhead{(10$^{-2}$\,$M_{\odot}$/yr)} & \colhead{(W/Hz)} & (W/Hz) 
}
\decimalcolnumbers
\startdata
IC1860 & $\left( 60.3\pm 6.3\right)$ & $\left( 7.10\pm 0.70\right)$ & $\left( 1.29\pm 0.13\right) \times 10^{20}$ & $\left( 7.13\pm 0.33\right) \times 10^{21}$ \\
NGC\,57 & $\left( 94.5\pm 7.2\right)$ & $\left( 7.30\pm 0.60\right)$ & $\left( 1.32\pm 0.10\right) \times 10^{20}$ & $\left( 1.93\pm 0.10\right) \times 10^{20}$ \\
NGC\,410 & $\left( 77.8\pm 9.3\right)$ & $\left( 4.40\pm 0.50\right)$ & $\left( 7.92\pm 0.90\right) \times 10^{19}$ & $\left( 1.29\pm 0.05\right) \times 10^{21}$ \\
NGC\,410 & $\left( 97.6\pm 2.7\right)$ & $\left( 5.49\pm 0.15\right)$ & $\left( 9.95\pm 0.28\right) \times 10^{19}$ & $\left( 1.29\pm 0.05\right) \times 10^{21}$ \\
NGC\,1132 & $\left( 59.7\pm 9.4\right)$ & $\left( 6.00\pm 0.90\right)$ & $\left( 1.08\pm 0.17\right) \times 10^{20}$ & $\left( 1.98\pm 0.13\right) \times 10^{21}$ \\
NGC\,1404 & $\left( 550.1\pm 27.2\right)$ & $\left( 2.85\pm 0.14\right)$ & $\left( 5.16\pm 0.26\right) \times 10^{19}$ & $\left( 5.20\pm 0.80\right) \times 10^{18}$ \\
NGC\,1404 & $\left( 702.9\pm 1.5\right)$ & $\left( 3.64\pm 0.01\right)$ & $\left( 6.60\pm 0.01\right) \times 10^{19}$ & $\left( 5.20\pm 0.80\right) \times 10^{18}$ \\
NGC\,1404 & $\left( 778.2\pm 3.9\right)$ & $\left( 4.03\pm 0.02\right)$ & $\left( 7.30\pm 0.04\right) \times 10^{19}$ & $\left( 5.20\pm 0.80\right) \times 10^{18}$ \\
NGC\,1404 & $\left( 732.3\pm 5.7\right)$ & $\left( 3.79\pm 0.03\right)$ & $\left( 6.90\pm 0.06\right) \times 10^{19}$ & $\left( 5.20\pm 0.80\right) \times 10^{18}$ \\
NGC\,2300 & $\left( 150.8\pm 6.7\right)$ & $\left( 3.35\pm 0.2\right)$ & $\left( 6.07\pm 0.27\right)\times 10^{19}$ & $\left( 2.05\pm 0.20\right) \times 10^{20}$ \\
NGC\,2300 & $\left( 163.0\pm 4.2\right)$ & $\left( 3.62\pm 0.01\right)$ & $\left( 6.56\pm 0.17\right) \times 10^{19}$ & $\left( 2.05\pm 0.20\right) \times 10^{20}$ \\
NGC\,3091 & $\left( 186.4\pm 3.4\right)$ & $\left( 5.62\pm 0.1\right)$ & $\left( 1.02\pm 0.02\right) \times 10^{20}$ & $\left( 7.60\pm 0.40\right) \times 10^{19}$ \\
NGC\,3091 & $\left( 163.5\pm 4.2\right)$ & $\left( 4.93\pm 0.34\right)$ & $\left( 8.90\pm 0.60\right) \times 10^{19}$ & $\left( 7.60\pm 0.40\right) \times 10^{19}$ \\
NGC\,3923 & $\left( 438.4\pm 6.1\right)$ & $\left( 2.49\pm 0.03\right)$ & $\left( 4.51\pm 0.06\right) \times 10^{19}$ & $\left( 1.29\pm 0.11\right) \times 10^{19}$ \\
NGC\,3923 & $\left( 357.5\pm 21.5\right)$ & $\left( 2.03\pm 0.12\right)$ & $\left( 3.68\pm 0.22\right) \times 10^{19}$ & $\left( 1.29\pm 0.11\right) \times 10^{19}$ \\
NGC\,4073 & $\left( 191.0\pm 3.9\right)$ & $\left( 8.91\pm 0.18\right)$ & $\left( 1.61\pm 0.03\right) \times 10^{20}$ & $\left( 1.54\pm 0.09\right) \times 10^{20}$ \\
NGC\,4073 & $\left( 130.0\pm 10.2\right)$ & $\left( 6.1\pm 0.05\right)$ & $\left( 1.10\pm \right) 0.09\times 10^{20}$ & $\left( 1.54\pm \right) 0.09\times 10^{20}$ \\
NGC\,4125 & $\left( 206.80\pm 16.6\right)$ & $\left( 1.22\pm 0.10\right)$ & $\left( 2.22\pm 0.18\right) \times 10^{19}$ & $\left( 2.36\pm 0.11\right) \times 10^{19}$ \\
NGC\,4406 & $\left( 1004.0\pm 9.0\right)$ & $\left( 4.16\pm 0.04\right)$ & $\left( 7.53\pm 0.07\right) \times 10^{19}$ & $\left( 5.30\pm 1.00\right) \times 10^{18}$ \\
NGC\,4406 & $\left( 1029.6\pm 12.6\right)$ & $\left( 4.26\pm 0.05\right)$ & $\left( 7.72\pm 0.09\right) \times 10^{19}$ & $\left( 5.30\pm 1.00\right) \times 10^{18}$ \\
NGC\,4406 & $\left( 800.7\pm 17.2\right)$ & $\left( 3.32\pm 0.07\right)$ & $\left( 6.01\pm 0.13\right) \times 10^{19}$ & $\left( 5.30\pm 1.00\right) \times 10^{18}$ \\
NGC\,4936 & $\left( 70.0\pm 9.2\right)$ & $\left( 0.89\pm 0.12\right)$ & $\left( 1.61\pm 0.21\right) \times 10^{19}$ & $\left( 1.96\pm 0.08\right) \times 10^{20}$ \\
NGC\,7619 & $\left( 81.2\pm 6.6\right)$ & $\left( 2.31\pm 0.22\right)$ & $\left( 4.20\pm 0.40\right) \times 10^{19}$ & $\left( 3.82\pm 0.15\right) \times 10^{21}$ \\
NGC\,7619 & $\left( 95.4\pm 3.3\right)$ & $\left( 2.31\pm 0.11\right)$ & $\left( 4.18\pm 0.20\right) \times 10^{19}$ & $\left( 3.82\pm 0.15\right) \times 10^{21}$ \\
NGC\,7619 & $\left( 81.6\pm 7.0\right)$ & $\left( 2.31\pm 0.23\right)$ & $\left( 4.20\pm 0.40\right) \times 10^{19}$ & $\left( 3.82\pm 0.15\right) \times 10^{21}$ \\
\hline
\enddata
\tablecomments{ (a) Multiple measurements of the FUV flux were found in the catalogue for some sources.}
\end{deluxetable*}
\end{startlongtable}

\subsection{AGN duty cycle}
To study the duty cycle of radio mode AGN activity in more detail, we estimate the spectral indices in the nuclear regions for a subsample of our early-type galaxies (Table\,\ref{tab:spim}). The distribution of the central spectral indices could help us place better constraints on the duty cycle of the AGN.

From in-band VLA analysis of the mean spectral indices of the radio emission within 0.7\,kpc from the central region, we determined that approximately 1/3 of the sources have a relatively flat ($\alpha \geq -0.5$) spectral index. The rest of the galaxies have steep spectra. This result indicates that the age of the radio emitting plasma in the centres of these galaxies spans a range of values and while the radio mode AGN activity is variable, it has a relatively high duty cycle.

This analysis could also motivate a more in-depth investigation of the nuclear activity of these AGN. For example, higher resolution VLBI observations would offer a closer look at the central regions and their most recent state of activity. From our sample, an example is provided by the recently active source NGC\,5044, where \cite{schellenberger2020a} confirm ongoing jet activity using VLBA and summarize the presence of multiple generations of X-ray cavities.\

Another example in the sample is NGC 1407, where we see a small-scale young jet in the core \citep{giacintucci2012}.\
A recent study of NGC\,1316 \citep{maccagni2021b} using the upgraded Karoo Array Telescope (MeerKAT), as well as VLA and ALMA data, found a highly variable central radio source in the last three cycles of its activity. One cycle, forming large diffuse radio lobes, could have started around 20\,Myr ago. The second cycle started possibly around 3\,Myr ago, forming a flattened S-shaped structure. The current activity appears to be 1\,Myr old.

\subsection{Correlations with radio power}

We investigated the possible correlations between the radio power and the X-ray luminosity of the hot halo as well as the central point source, the thermodynamical properties of the hot gas, the H$\alpha +$[N II] luminosity, and the jet power calculated from the X-ray cavities and between the radio flux density and the largest linear size of the radio emission (Table\,\ref{tab:corr} and Figure\, \ref{fig:p_lx}, \ref{fig:thermo_pv}, \ref{fig:pv_pjet_ha_mbh}, \ref{fig:llsVSp1.5}). 

We see a significant correlation with Spearman and Pearson coefficient of 0.64 with $p$-value of 0.004) between the total radio flux density and the largest linear size of the radio emission. The larger the total extent of the radio emission, the more powerful the source is \citep{singal1993,lara2001,shabala2013,tang2020}. While the distances of our sources span a factor of 6, the radio fluxes span four orders of magnitude. The range of radio power is thus too large to be accounted for by the more powerful sources being more distant.

Similarly, a significant correlation with Spearman and Pearson coefficient of 0.71 with the $p$-value of 0.004) is found between the radio power and the X-ray luminosity of the central X-ray point source\footnote{To take into account biasing due to the same D$_l^2$ dependence, we compared X-ray luminosity of the atmosphere ($L_{\rm Xray}$) within 10\,kpc radius from the core with the radio flux ($S_{1.5GHz}$) instead of the radio power ($P_{1.5GHz}$), resulting in a slightly lower correlation coefficient of 0.61 and $p$-value of 0.01}. The correlation is consistent with a scenario where the X-ray emission comes from an unresolved X-ray jet or the base of the jet.

The relation between the luminosity of the hot X-ray atmospheres within 10\,kpc radius from the central region and the radio power at 1.5\,GHz shows no correlation.
The lack of trends could, at the first glance, appear surprising. It is worth noting that the lack of any trend detected here could be affected by the limited range of X-ray luminosities in the innermost regions up to radius of 10\,kpc spanning over approximately one order of magnitude in comparison with a large span of radio power values ($\sim$\,6 orders of magnitude; Figure\,\ref{fig:p_lx}; left). Moreover, for a stable continuous accretion of the hot atmospheric gas, at a given black hole mass, one would expect higher accretion rates and thus more powerful jets for lower entropies and shorter cooling times. However, the gas with typical cooling times seen in early-type galaxies is expected to be thermally unstable \citep[e.g.][]{fabiannulsen1977,nulsen1986}. Assuming that the ambient medium cools and forms dense clouds with very small volume filling fractions that fall toward the central black hole, the accretion rate can for short moments rise by orders of magnitude, triggering a feedback response \citep{gaspari2013,voit2015,mcnamara2016}. Since the infall of dense clouds with a range of masses is essentially chaotic, the power of a given triggered outbursts can also have a range of values. Thus, we are not expecting any trend between the thermodynamic properties of the atmospheric gas and the observed radio emission.\

The radio power showed no correlation with the radio power and the jet power computed using X-ray cavities \citep[and references therein]{lakhchaura2018} and H$\alpha +$[N II] luminosity \citep{lakhchaura2018} and the mass of the central supermassive black hole\footnote{The supermassive black hole masses were adapted from \cite{kormendy2013} and \cite{saglia2016} using direct measurements and from \cite{lauer2007} and \cite{makarov2014}, who derived the masses from the $M_{BH} - \sigma$ scaling relations and taken from \citep{lakhchaura2018}.}.\\
The H$\alpha +$[N II] emitting gas represents only a fraction of the cold/warm gas mass in the centres of these galaxies and the observed nebulae are not necessarily located in the vicinity of the supermassive black hole. Therefore, the lack of correlation is not entirely surprising. Results of the statistical analysis are consistent with previous ones found in the literature \citep{franceschini1998,liu2006}. Weak correlations with CO emitting cold molecular gas have previously been observed by \cite{babyk2019}.

\subsection{Interaction with the X-ray gas}
\subsubsection{Widening radio jets}
One of the signatures of the interaction between the radio plasma and the hot X-ray gas could be seen in the widening of the jets after they pierce through the relatively dense galactic atmosphere.\

 \begin{figure*}[h!]
    \centering
\includegraphics[width=220pt]{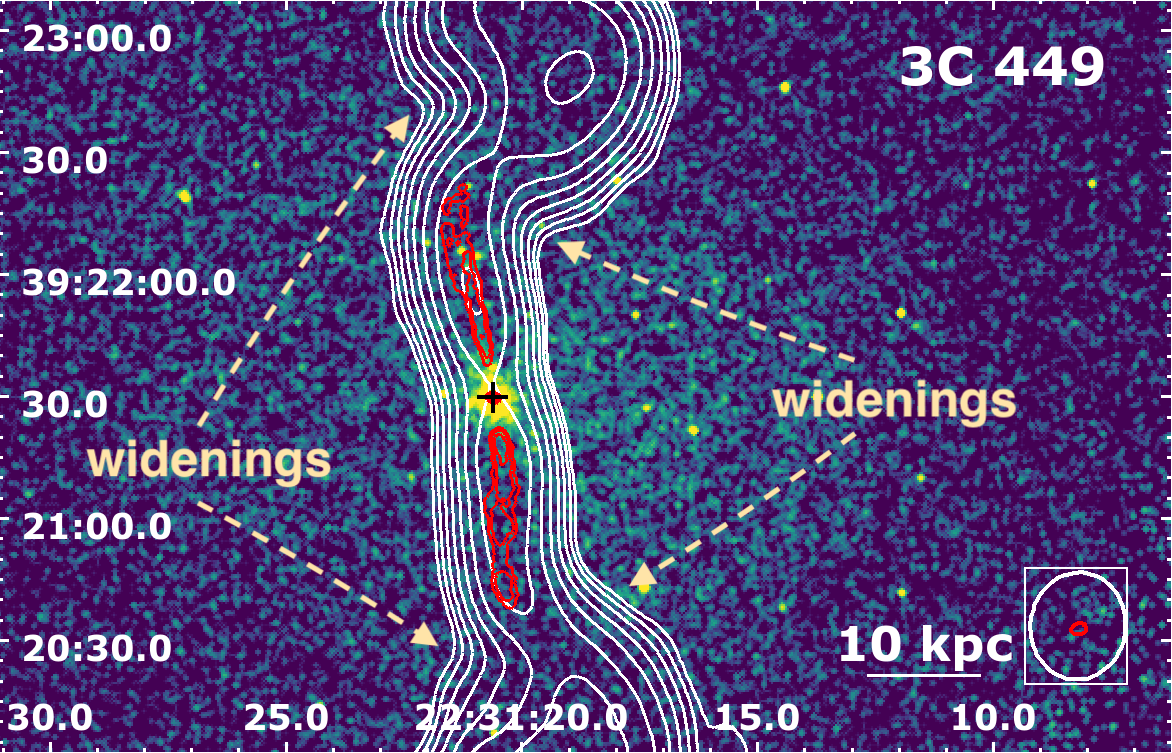}
\includegraphics[width=220pt]{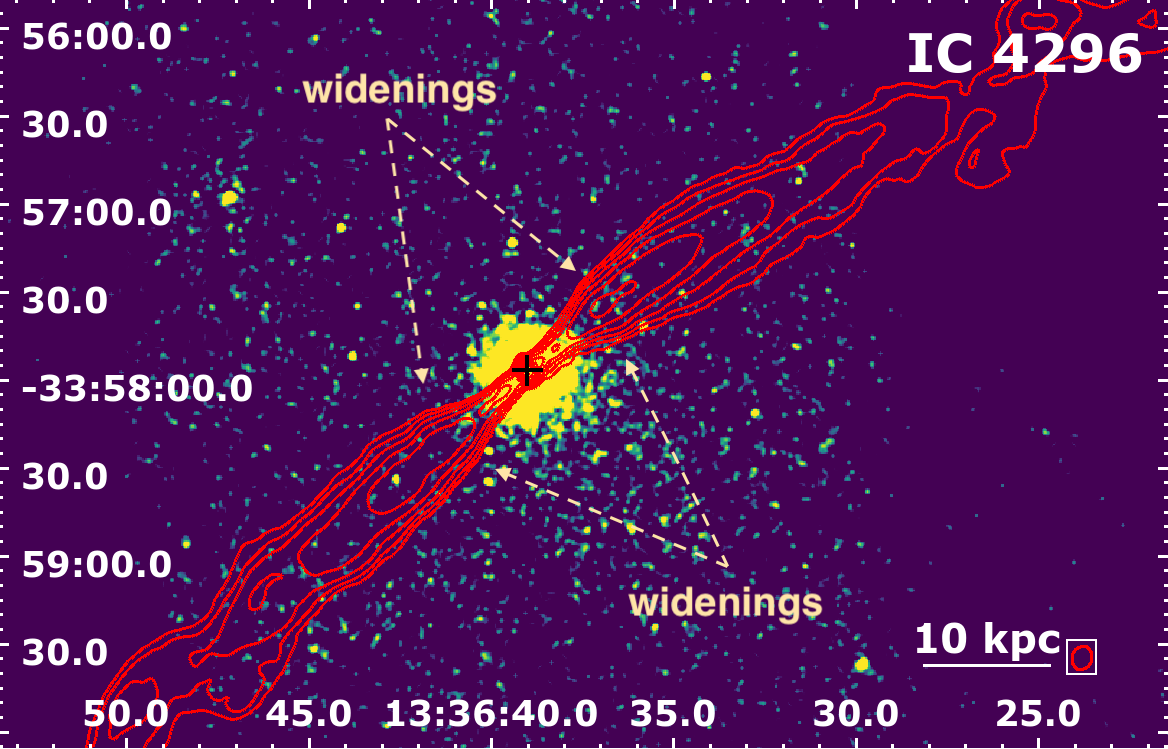}

\includegraphics[width=220pt,height=144pt]{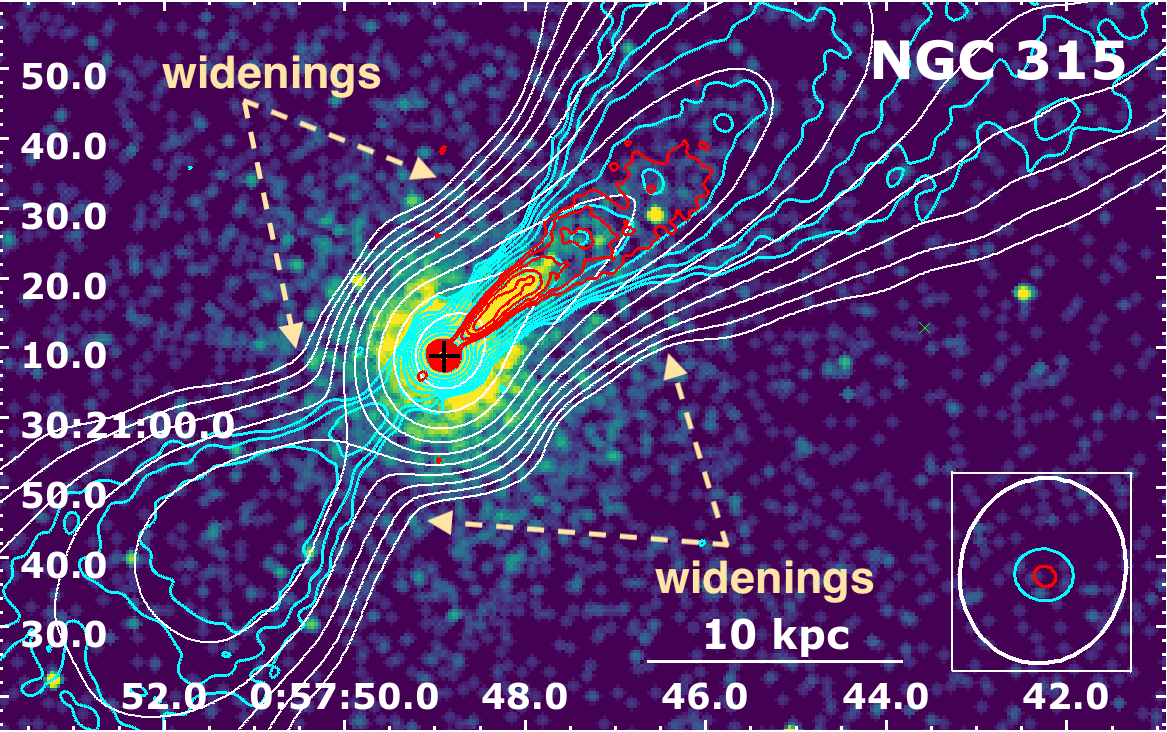}
\includegraphics[width=220pt]{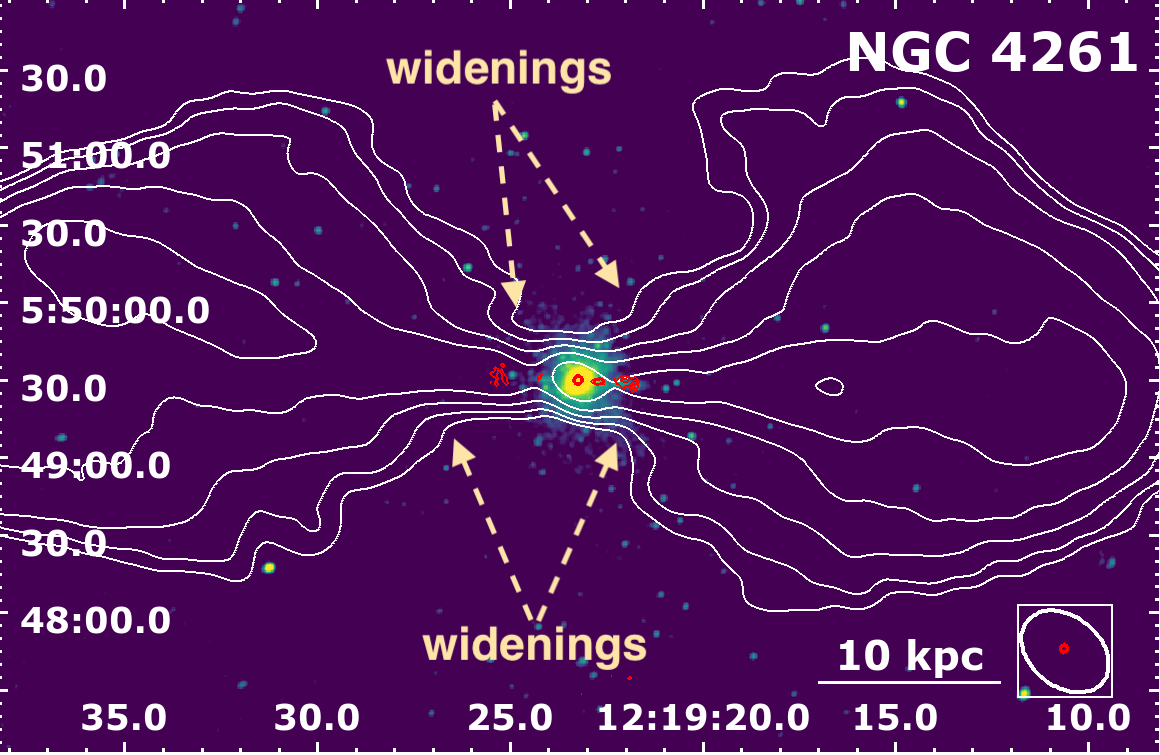}
\caption{Widening of the radio contours visible for 3C\,449, IC\,4296, NGC\,315 and NGC\,4261. The radio contour levels are drawn at 5$\times$RMS noise, increasing by factors of 2 up to the peak intensity. The RMS noise and peak intensity values can be found in Table\,\ref{tab:results}. The black cross depicts the radio center of the galaxy.}
\label{fig:widenings}
\end{figure*}

The radio contours in 3C\,449, NGC\,315, NGC\,4261 and IC\,4296\footnote{The mechanism was described for IC\,4296 } by e.g.: \cite{killeen1988,grossova2019} (Figure\,\ref{fig:widenings}) widen significantly at about 10-20\,kpc as the relativistic plasma in jets is released from the higher pressure environment of the hot X-ray gas. In other words, if the ambient gas pressure drops below a certain value, then the jets can widen significantly.\\
The widening and the corresponding brightening of the jets could be connected to the lower ambient ICM pressure and possibly to additional thermal energy from the interaction of the radio and X-ray plasma that will increase the pressure of the radio jet causing it to expand \citep[e.g.][]{gizani1999}.

\subsubsection{Ghost X-ray cavities}
\label{sec:ghost_cavities}
Evidence for the past interaction between the radio and X-ray plasma, in the form of `ghost' cavities\footnote{In some systems the relativistic radio emitting plasma filling the cavities has aged and stopped produce detectable 1--2 GHz radio emission, giving rise to `ghost' cavities.}, was also observed in the giant elliptical galaxy NGC\,499 \citep{panagoulia2014b,kim2019}, which is the only galaxy in our sample with a non-detection of a radio source at 1--2\,GHz (Section\,\ref{sec:radio_central}) using our VLA data.
Similarly, X-ray cavities have been observed in seven galaxies with only a central point source in radio. Those previously known are: NGC\,1132 \citep{dong2010} and NGC\,4778 \citep{morita2006}. Additionally, Plšek et al. (in prep.) found potential cavities for another five sources: NGC\,2300, NGC\,3091, NGC\,3923, NGC\,4073, and NGC\,4125.\

The lack of radio emission filling the volume of X-ray cavities is most likely due to the old, aged plasma present in lobes. Assuming the lobes rose and expanded at the sound speed $c_{s}\approx 500\,{\rm km/s}$ of 1\,keV gas, then the radio plasma in the ghost cavities has aged at the time scales $\sim\,10$\,Myr thus barely visible at the observed radio frequencies unless other mechanisms occur.

\subsubsection{Possible X-ray tail}
\begin{figure}
    \centering
\includegraphics[width=220pt]{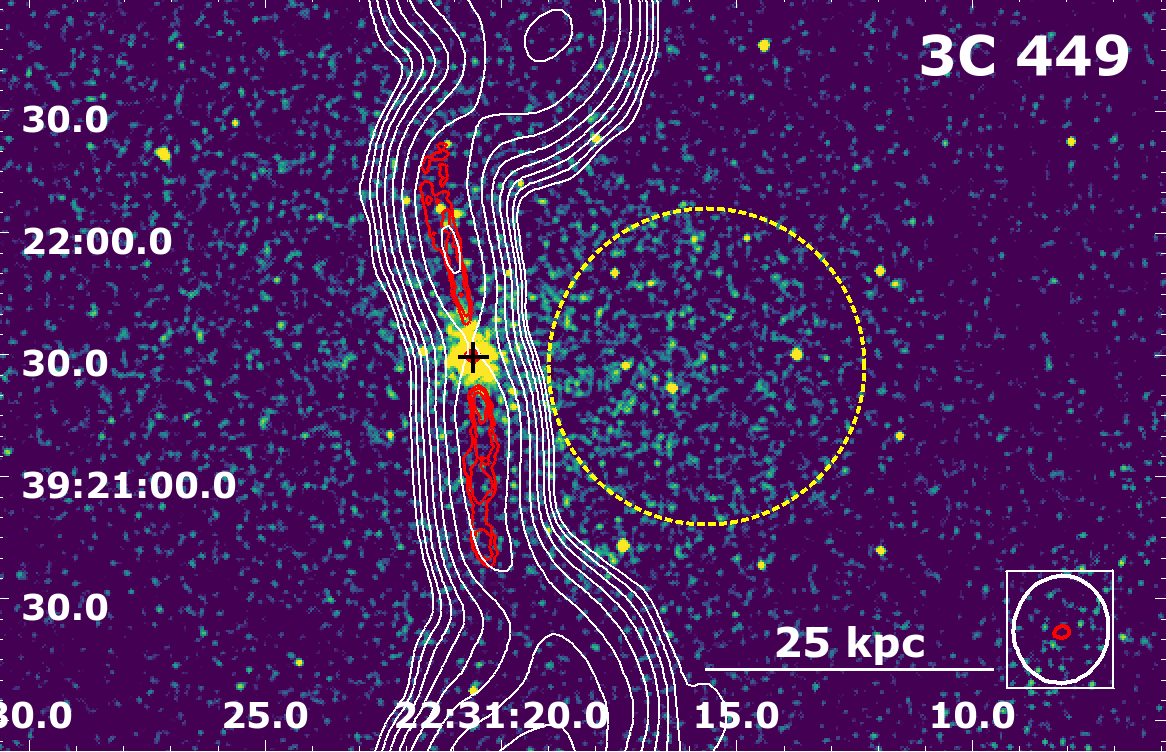}
\includegraphics[width=220pt]{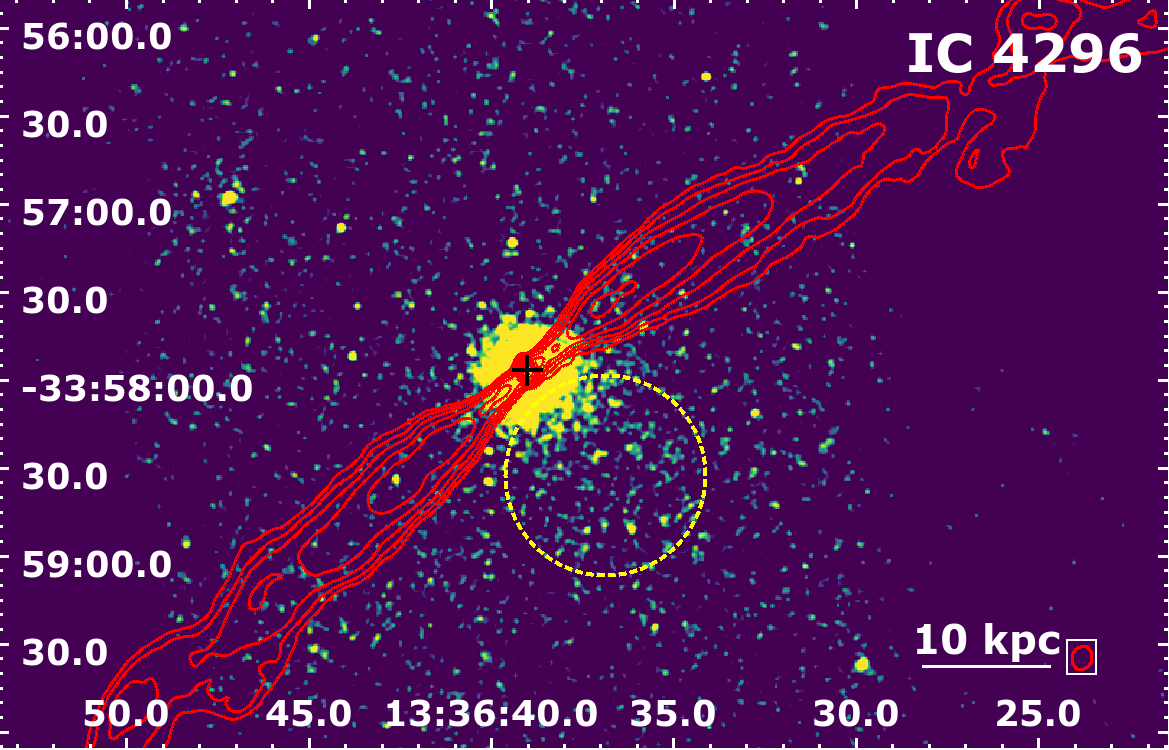}
    \caption{The {\it Chandra} X-ray images smoothed to enhance the X-ray tail enclosed by the radio jets of 3C\,449 ({\bf top}) and IC\,4296 (bottom). The black cross indicates the position of the radio center of the galaxy and yellow circles show the cross-cone region between the jets with the X-ray tail}.
    \label{fig:excess}
\end{figure}

We observe an X-ray tail in the so-called cross-cone region between the jets (yellow circles in the images of IC\,4296 and 3C\,449; Figure\,\ref{fig:excess}) due to the relative motion of the intracluster/group medium and the centrally located jetted galaxy in a system that experiences ICM sloshing. With a large opening angle of the jets and the presence of potential 'warm` spots, the galaxies can be categorised as Wide Angled Tail (WAT) galaxies \citep[e.g.:][]{owen1976,o'donoghue1990,leahy1993,missaglia2019}.\

\subsection{Offset radio sources}
\label{sec:recoil}
In some cases, the radio emission is observed to be offset from the optical/X-ray emission of the host galaxy. Several mechanisms can be responsible for such misalignment. One possibility is, that the offset radio emission might belong to a merging galaxy where only one nucleus is active \citep[a so-called `offset AGN'][]{steinborn2016} or to a recoiled black hole kicked out of the system at large velocities (ranging from hundreds to thousands of km s$^{-1}$) when the central black holes of merging galaxies coalesce.

To date, there have been tens of observed candidates for such black holes ejected by gravitational wave recoil \citep[also supported by simulations, e.g.][]{campanelli2007,blecha2016,blecha2013,chiaberge2017,condon2017,komossa2012}.  The broad line region can be carried away by the recoiling black hole and the AGN can continue to be active \citep[e.g.:][]{komossa2012}. Moreover, due to previously published candidates for an offset black hole within our sample, namely NGC\,4486 and NGC\,5813, we investigate this mechanism for three more potential candidates (NGC\,3923, NGC\,5129 and NGC\,4125). 

\subsubsection{Small offsets: NGC\,3923, NGC\,4486, NGC\,5846, NGC\,5129}
Multifrequency studies were carried out to search for such recoiling black holes \citep{lena2014,barrows2016,skipper2018,ward2021}. Even in the cases of the giant elliptical galaxies M\,87 (NGC\,4486) and NGC\,5846, a pc-scale displacement of the SMBH from the center of the host galaxy was observed and ascribed to the recoil mechanism \citep{batcheldor2010,lena2014}. Although, the apparent supermassive black hole offset in M\,87 could be just due to the variation of the flux as suggested by \cite{lopez-navas2018}. 

In our analysis, we present a previously unpublished candidate for the radio/X-ray offset from the optical core of 1.2\,arcseconds (0.5\,kpc), which may be the result of a displaced supermassive black hole from the central potential of the host giant elliptical galaxy NGC\,5129. And we confirm the previous findings for NGC\,4486 and NGC\,5846, where we also find small offsets of the central radio emission relative to the optical/X-ray emission. Moreover, we found a previously unpublished small offset of 0.21\,kpc for the radio emission in the central regions of NGC\,3923.\

\begin{figure*}
    \centering
    \includegraphics[width=220pt]{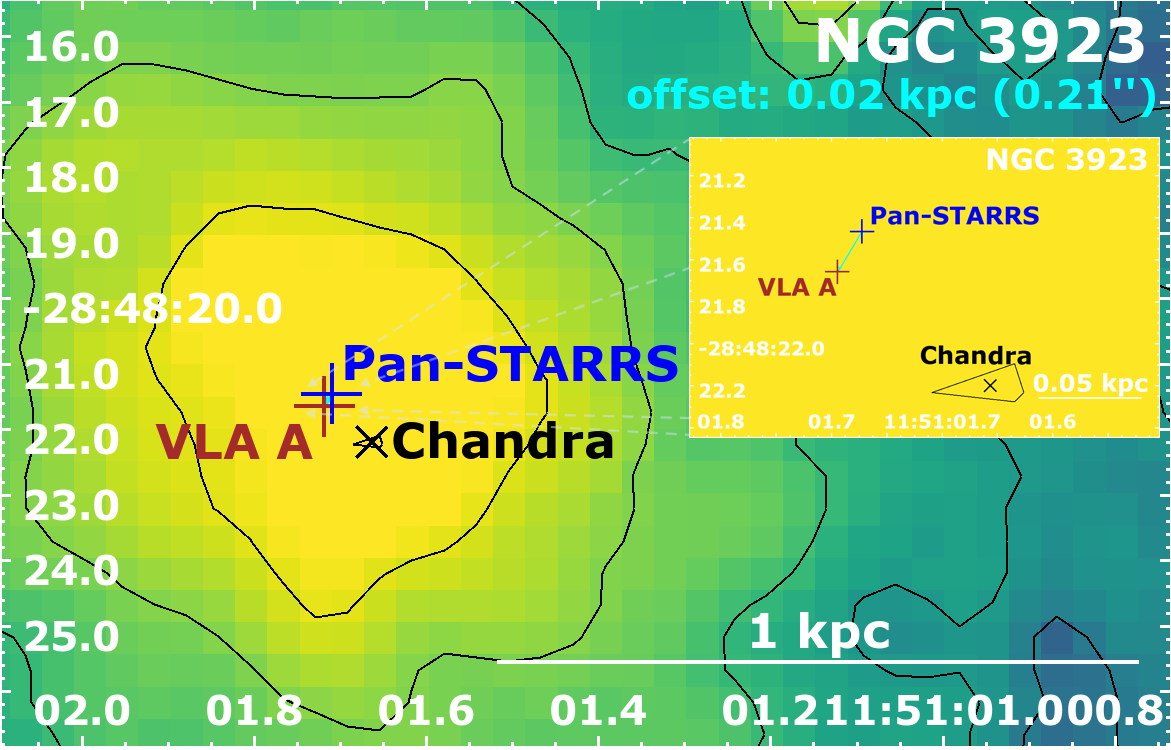}
    \includegraphics[width=220pt]{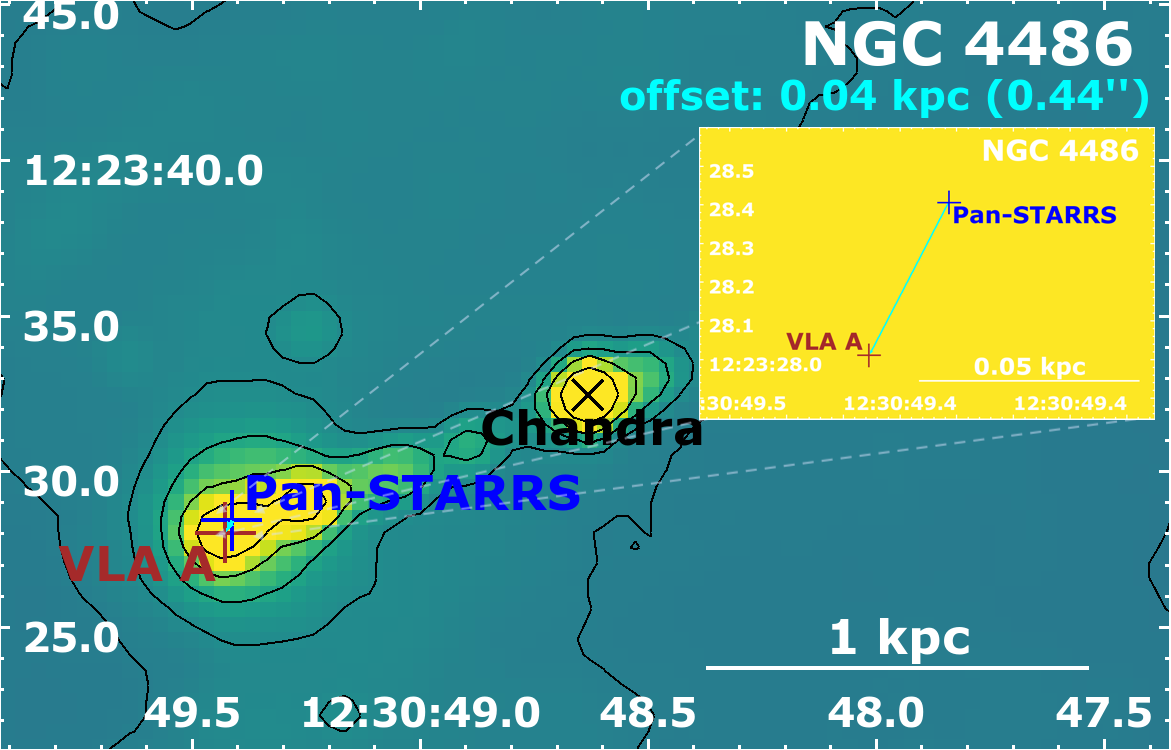}
        
    \includegraphics[width=220pt]{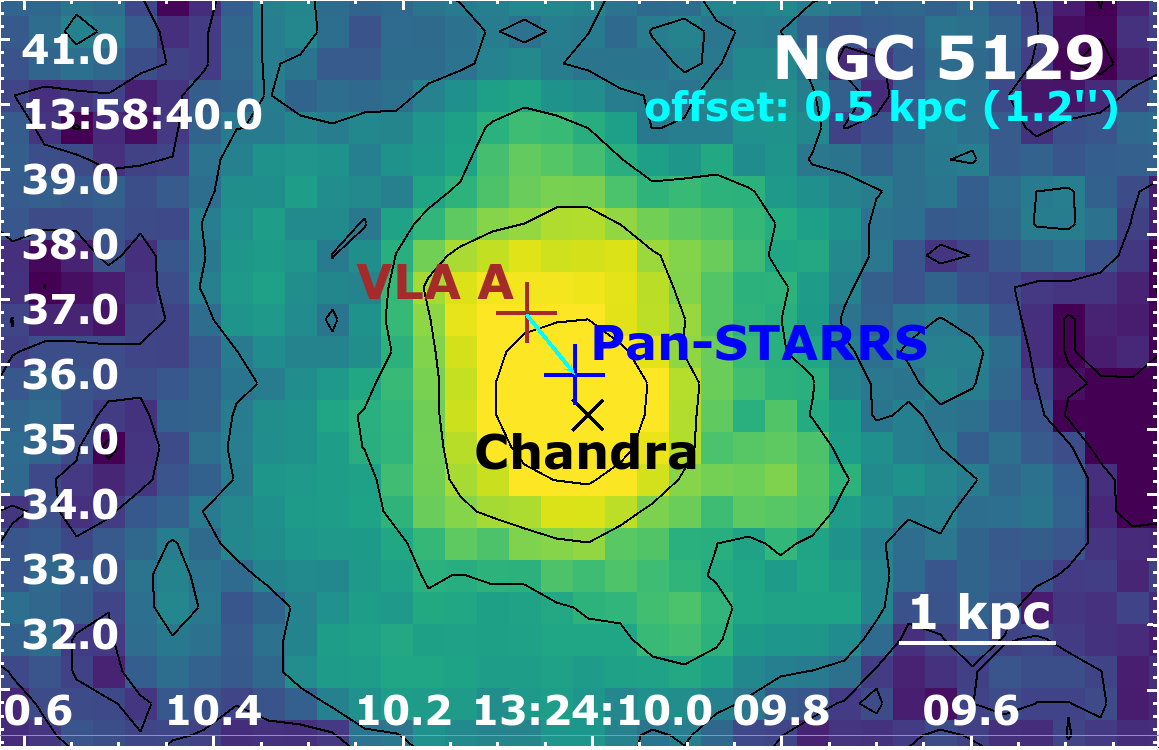}
    \includegraphics[width=220pt]{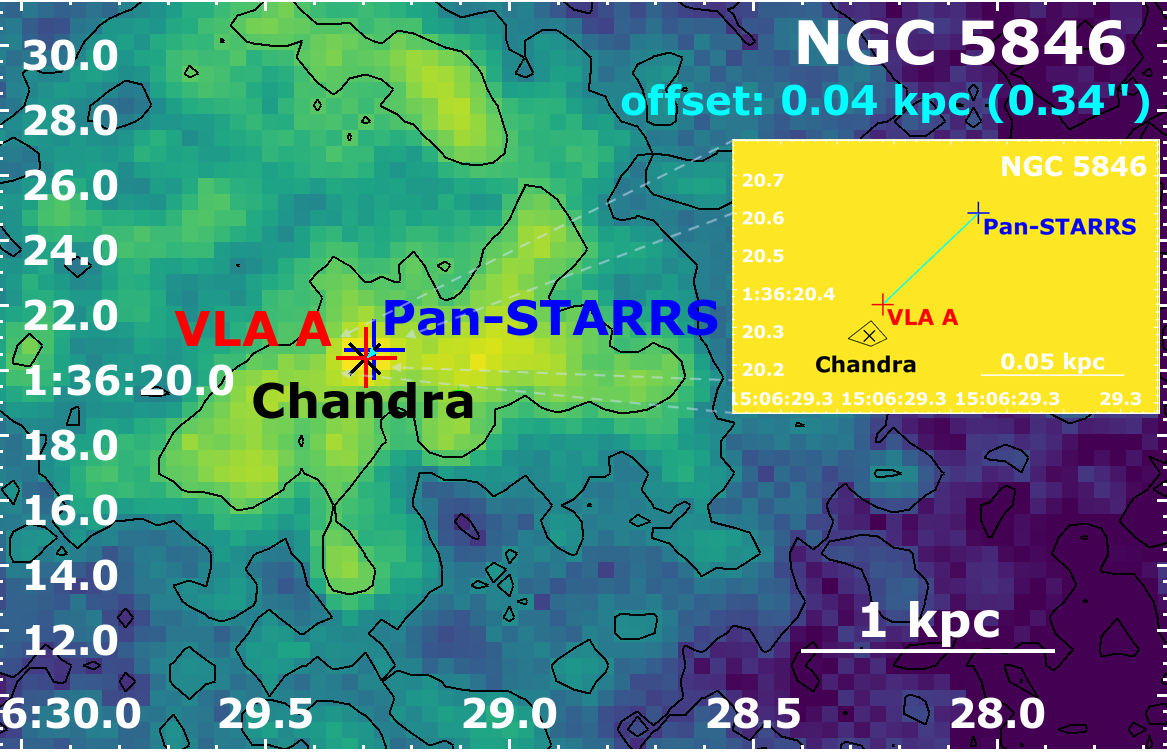}
    \caption{{\it Chandra} images with overlaid black X-ray contours and a black `X’ to mark the central X-ray brightness peak. The innermost X-ray contour is drawn by estimating the peak of the average count value of the {\it Chandra} image with the DS9 \citep{joye2003} {\tt projection region} tool. The subsequent contours levels are decreasing as a power of two. The brown cross shows the core of the radio emission from VLA A configuration data and the blue cross represents the position of the optical center from Pan-STARRS in each image (estimated from the innermost optical contour). The offset of the observed radio core emission with respect to the central peak of the optical emission is shown by cyan lines. Small offsets (within 0.5\,kpc) are seen for NGC\,3923, NGC\,4486, NGC\,5129 and NGC\,5846.}
    \label{fig:off_less}
\end{figure*}

All small-scale offsets are shown in Figure\,\ref{fig:off_less}.

\subsubsection{Intriguing radio lobe emission in NGC\,4125}
\label{disc:offset_ngc4125}
\begin{figure*}
    \centering
    \includegraphics[scale=5,width=400pt]{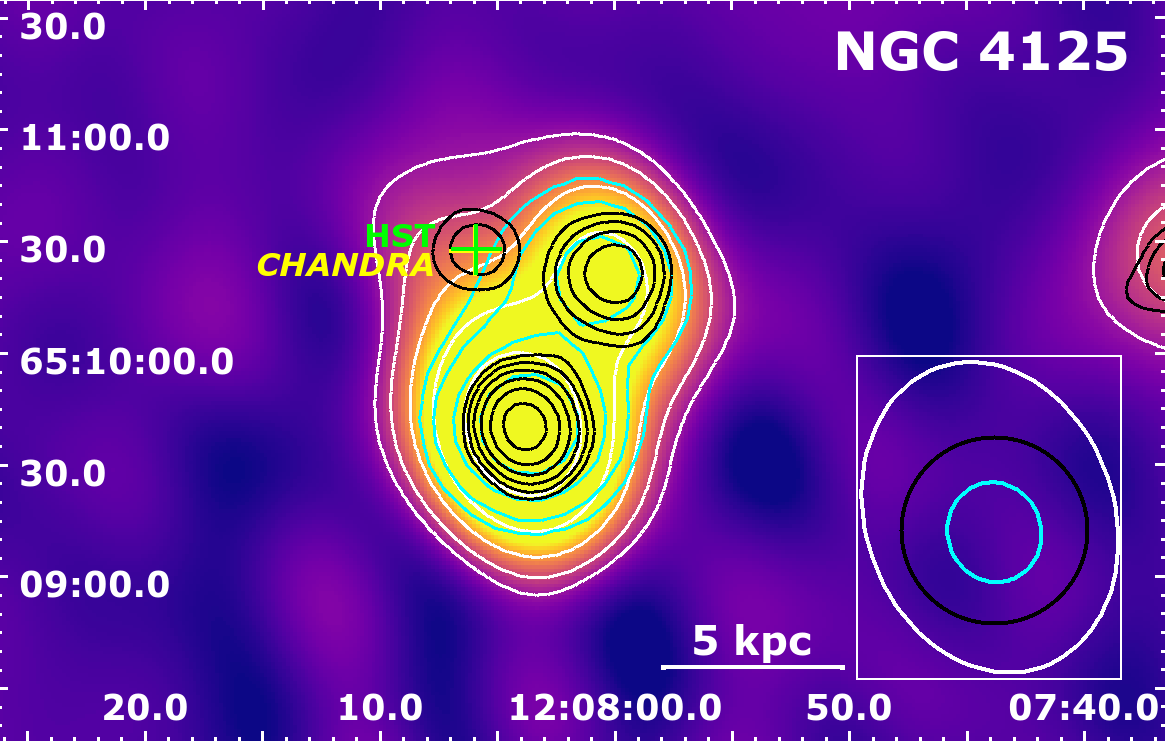}
    \includegraphics[scale=5,width=290pt]{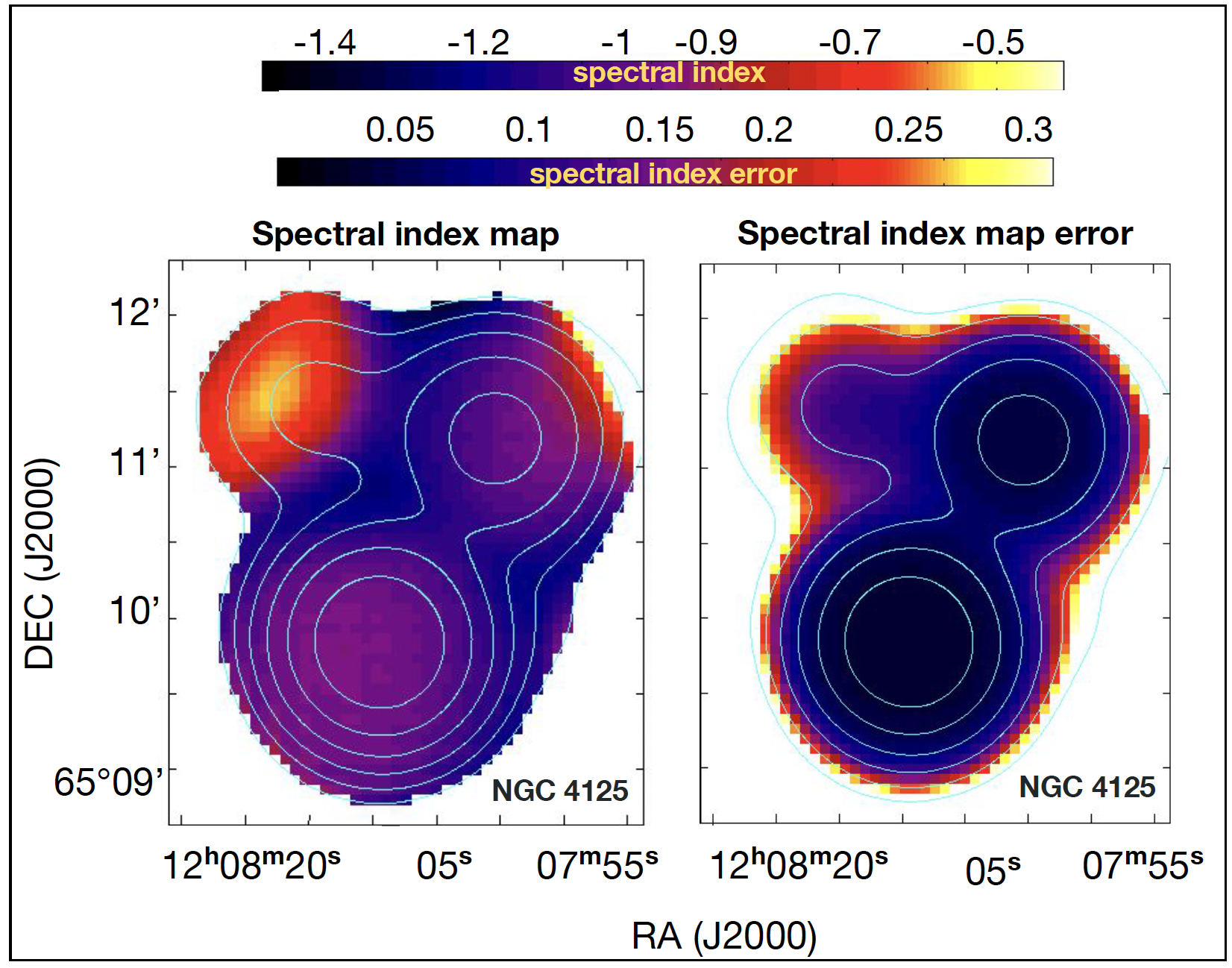}
    
    \caption{{\bf Top: } VLA D configuration total intensity image of NGC\,4125 with overlaid radio isophots in white. The contours in cyan and black represent the emission from TGSS at 148\,MHz and WSRT at 1.4\,GHz, respectively. In all cases the contour levels are at [1, 2, 4, 8, 16,...] $\times\,5\cdot\,\sigma_{\rm RMS}$. The corresponding RMS noise levels ($\sigma_{\rm RMS}$) for VLA D configuration, TGSS and WSRT data is $67\,{\rm\mu Jy}$, $1.7\,{\rm mJy}$ and $60\,{\rm\mu Jy}$, respectively. The restoring beam (resolution) for each observation can be found in the bottom right corner of the image in the white box. The optical (HST) center of the galaxy is shown as a green cross and is consistent with the X-ray center ({\it Chandra}; a yellow cross) and the radio center, most promint in the WSRT image at 1.4\,GHz. {\bf Bottom: } The spectral index map between 148\,MHz (TGSS) and 1.4\,GHz (WSRT).}
    \label{fig:off_ngc4125}
\end{figure*}

The most intriguing is the offset radio emission located 46\,arcseconds (4\,kpc) from the radio core of NGC\,4125 (Figure\,\ref{fig:off_ngc4125}; top). The unusual position of the radio emission has been previously reported by \cite{krajnovic2002}, who claimed that due to the missing optical counterpart, it is probably a background source. However, our analysis, where we estimated how distant a potential background source would need to be if its optical counterpart is not seen by the Panoramic Survey Telescope and Rapid Response System (Pan-STARRS)\footnote{\href{https://panstarrs.stsci.edu/}{https://panstarrs.stsci.edu/}} showed that it is quite unlikely that the observed radio emission is connected to a background source.

Our analysis takes into account the limiting apparent magnitude of the Pan-STARRS survey of 23\,mag. We derived that a galaxy similar to M\,87, with an absolute magnitude of -21.5, would need to be at least at the luminosity distance of $\sim$\,6300\,Mpc (corresponding to a redshift of 0.95) to be missed by the optical survey. At this luminosity distance, the potential background radio source would have a size of $\sim$\,1\,Mpc (assuming a linear scale 8\,kpc/arcseconds) and large radio power of 10$\times$10$^{25}$\,W/Hz, corresponding to twice the power of the giant radio halo in the cluster merger `El Gordo' \citep{lindner2014} at the redshift of $z=0.87$

Moreover, we examined the archival data from TIFR GMRT Sky Survey (TGSS) at 148\,MHz and Westerbork Synthesis Radio Telescope (WSRT) data at 1.4\,GHz and created a spectral index map between these two frequencies (Figure\,\ref{fig:off_ngc4125}; bottom). We found a flat spectral index at the position of the radio/X-ray/optical core of NGC\,4125 and steeper spectral indices in a clear double morphology of the radio emission resembling a FR\,II radio galaxy, which would potentially disfavor the recoiled black hole scenario. A `kicked-out' black hole, which recoils and moves with a high velocity relative to the galaxy, would be expected to form narrow tail structures along its path \citep{blecha2011}, whereas we observe an apparently untouched radio emission from potential hotspots (especially in the north-western side).\ 

The analysis of the X-ray {\it Chandra} data reveals possible X-ray cavities (Plšek et al.; in prep.) perpendicular to the core of NGC\,4125, thus they do not correspond to the offset radio emission in the form of radio lobes.\

\cite{burke-spolaor2017} found a similar offset radio source in the central dominant galaxy (CDG) of the galaxy cluster Abell\,2261 with a missing optical and X-ray counterpart \citep{gultekin2021}. Interestingly, the radio morphology of the emission resembles the one in NGC\,4125 with its pear-like shape.\

Further analysis needs to be carried out to investigate the origin and source of this unusual double-lobe radio emission.

\subsection{Fanaroff-Riley Class\,II radio sources?}
\label{sec:frII}
Most of the early-type galaxies in our sample are low power Fanaroff-Riley Class\,I \citep[FR\,I,][]{capetti2017a} radio sources. Although two giant ellipticals, NGC\,533 and NGC\,1600, show morphological similarities to Fanaroff-Riley Class\,II \citep[FR\,II,][]{capetti2017b} radio sources, they still have low power more typical for FR\,Is (see Section\,\ref{sec:frII}).

\subsubsection*{NGC\,533}
\begin{figure*}[h!]
    \centering
\includegraphics[width=220pt]{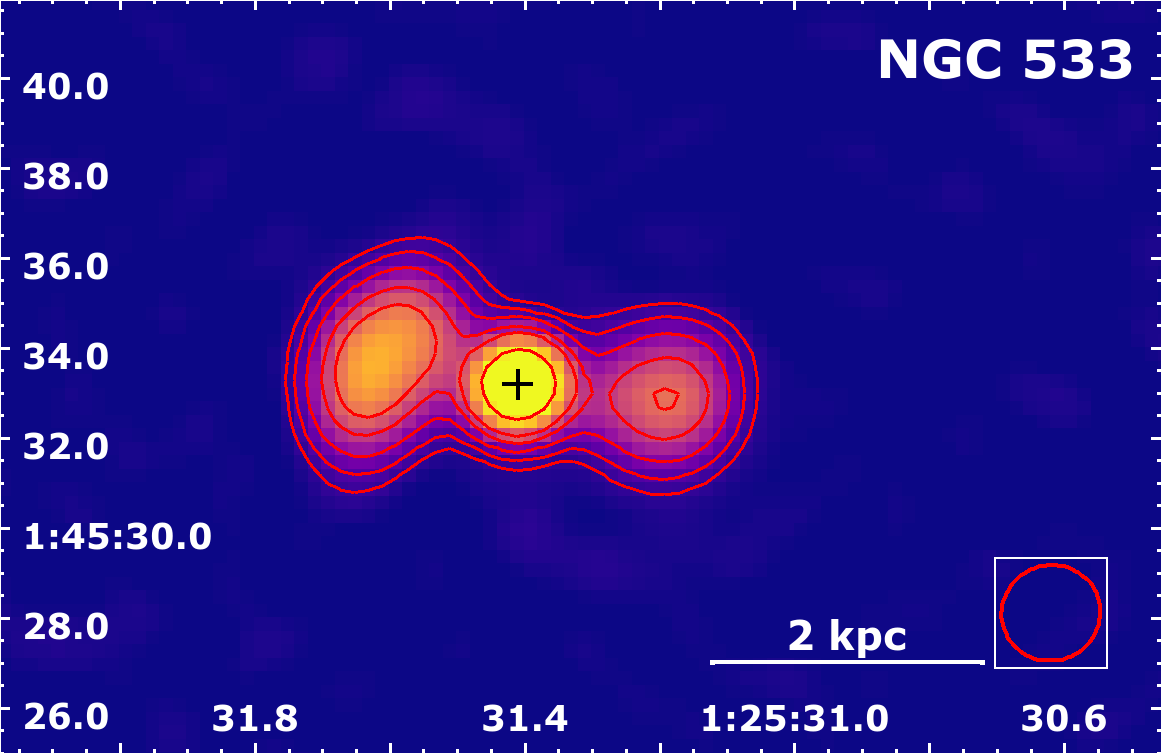}\\
\vspace{10pt}
\includegraphics[width=197pt]{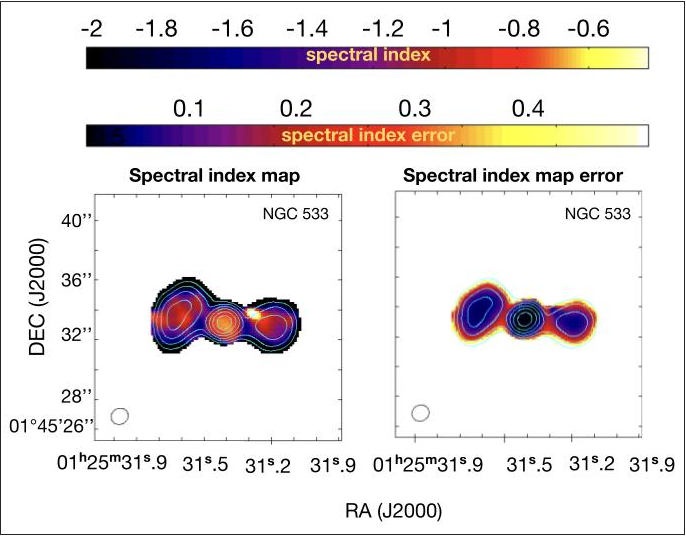}
\includegraphics[width=190pt,height=150pt]{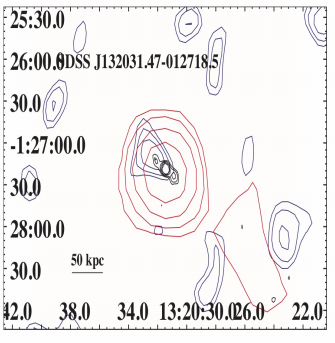}
    \caption{{\bf Top: }The total intensity image of NGC\,533 in VLA A configuration at 1--2\,GHz and ({\bf bottom left:)} its corresponding in-band spectral index map with measurement uncertainties produced by the multi-term `cleaning' algorithm in {\sc casa}. {\bf Bottom right: }The morphological structure of NGC\,533 is similar to the structure in the COMP2CAT source J132031.47-012718.5.}
    \label{fig:frii_ngc533}
\end{figure*}

The radio morphology of NGC\,533 (Figure\,\ref{fig:frii_ngc533}; top) was observed within our 15A-305 project in the VLA A configuration and has a resolved central region and two inflated almost symmetric small-scale spherical radio jets/lobes with a possible hot spot in the western lobe. 
We suggest that this source resembles the radio morphology of FR\,II radio galaxies based on the visual comparison with sources in the FR\,II FIRST Catalog (FRIICAT) \citep{capetti2017b}. Further analysis of the in-band spectral index map revealed a flattening of the spectral index at the position of the radio lobes especially in the region west of the radio core visible in Figure \ref{fig:frii_ngc533} (bottom left). Potentially, due to its small physical size (less than 30\,kpc), it could fall into the category of Compact Double Radio galaxies (COMP2CAT) \cite{jimenez-gallardo2019}. NGC\,533 could resemble the compact radio source J132031.47-012718.5 (Figure\,\ref{fig:frii_ngc533}; bottom right).

\begin{figure*}
\centering
\includegraphics[width=210pt]{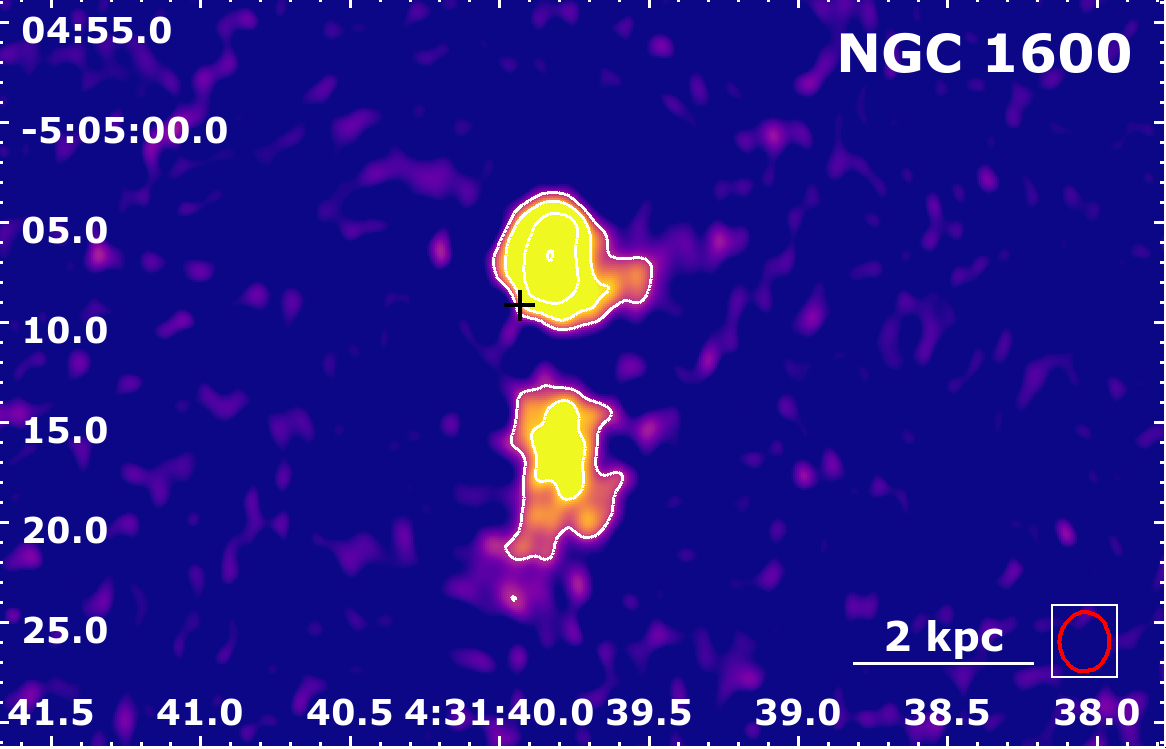}
\includegraphics[width=135pt]{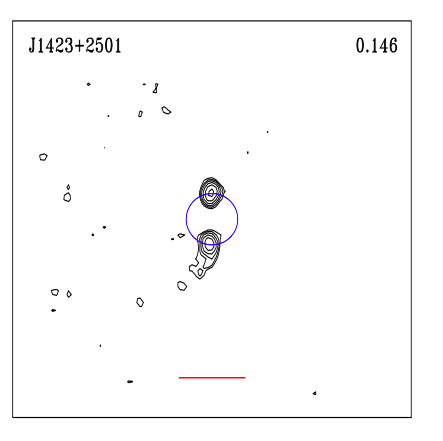}
\caption{{\bf Left: }The total intensity image of NGC\,1600 in VLA A configuration at 1--2\,GHz has a morphology resembling the FR\,II radio source J1423+2501 from the FRIICAT catalog \citep{capetti2017b} ({\bf right}).}
\label{fig:frii_ngc1600}
\end{figure*}

\subsubsection*{NGC\,1600}
\label{sec:ngc1600_frii}
The radio emission of the giant elliptical galaxy NGC\,1600 was previously published by \cite{birkinshaw1985}, where they used VLA C configuration data at 4.885\,GHz. They defined the source as a double radio source (in other words FR\,II)  and stated that the galaxy's core is located close to the northern radio lobe-like feature. The higher resolution VLA A configuration data at 1.4\,GHz shows more clearly the jets/lobes morphology and potential hotspots.

Additionally, we suggest that the radio morphology in NGC\,1600 at both frequencies resembles the radio morphology of the source J1423+2501, identified as an FR\,II radio source in the FRIICAT catalog \citep{capetti2017b} with an unidentified radio core (Figure\,\ref{fig:frii_ngc1600}).\ 

\clearpage
\newpage
\section{Summary and conclusions}
\label{sec:summary}
Here, we summarize the results of our statistical study of a large sample of optically and X-ray brightest nearby early-type galaxies. Our main goal was to study the radio properties of the AGN residing inside these galaxies. The final results are:

\begin{itemize}

\item A large, volume-, optical and X-ray luminosity-limited sample of 42 early-type galaxies observed with multiple configurations of the VLA reveals the presence of a radio source in almost all galaxies. Our radio source detection rate is as high as 41/42 (98\%), with the remaining source detected in archival single dish observations \citep{dressel1978,condon1988}, reported in the NVSS survey \citep{brown2011}, and at low frequencies using LOFAR \citep{birzan2020}.
The radio emission has an extended morphology in at least 67$\%$ (28 out of 42) of sample galaxies. 34 sources out of 42 (81$\%$) display a decrease in the X-ray surface brightness due to the hot gas being displaced out by jets creating X-ray cavities in the hot atmosphere.
\item Significant correlations are found between the radio power and the luminosity of the central X-ray point source, and between the radio flux density and the largest linear size of the emission. On the other hand, no correlation was found between the radio power and the luminosity of the X-ray atmospheres, the mass of the supermassive black hole, the cooling time or the entropy of the hot gas, or the luminosity of the H$\alpha +$[N II] line emission.
\item The central spectral indices show a large variety, with approximately 1/3 of the sources having relatively flat spectra, indicating the presence of fresh relativistic particles injected by the AGN, which could mean that the AGN has been switched on recently. The majority of our sources have steep spectra, indicating that the relativistic plasma has undergone significant aging in the nuclear regions of the host galaxies since the last major cycle of AGN activity.
\item On the one hand, the analysis of FUV fluxes suggests that the observed radio emission at 1.5\,GHz for 5 of 14 low-power early-type galaxies with point-like radio morphologies could be dominated by star formation. On the other hand, we also observe ghost cavities in 3 of those 5 galaxies, suggesting the interaction of radio lobes inflated by the AGN  with the hot X-ray atmospheres and thus, the presence of radio-mechanical AGN feedback.
\item Importantly, for four (potentially five) sources, we found an offset between the radio emission and the optical centre of the galaxy, which could be due to a merger.
\end{itemize}

\section*{Acknowledgements}

RG, NW, and TP are supported by the GACR grant 21-13491X.

This work is supported by the ``Departments of Excellence 2018 - 2022’’
Grant awarded by the Italian Ministry of Education, University and
Research (MIUR) (L. 232/2016).

This research has used resources provided by the Ministry of
Education, Universities and Research for the grant MASF\_FFABR\_17\_01.

This investigation is supported by the National Aeronautics and Space
Administration (NASA) grants GO9-20083X and GO0-21110X.

R.G. thanks the Smithsonian Astrophysical Observatory for support and hospitality.

R.G. would like to thank the NRAO help-desk staff for their assistance, discussions, and help with the VLA data reduction and analysis. Special thanks go to Dr. Amy E. Kimball.

The National Radio Astronomy Observatory is a facility of the National Science Foundation operated under cooperative agreement by Associated Universities, Inc. 
The scientific results reported in this article are based on observations made by the {\it Chandra X-ray Observatory} and published previously in the cited articles. 

Based on observations obtained at the Southern Astrophysical Research (SOAR) telescope, which is a joint project of the Minist\'{e}rio da Ci\^{e}ncia, Tecnologia, Inova\c{c}\~{o}es e Comunica\c{c}\~{o}es (MCTIC) do Brasil, the U.S. National Optical Astronomy Observatory (NOAO), the University of North Carolina at Chapel Hill (UNC), and Michigan State University (MSU).

This research has made use of resources provided by the Compagnia di San Paolo for the grant awarded on the BLENV project (S1618\_L1\_MASF\_01) and by the Ministry of Education, Universities and Research for the grant MASF\_FFABR\_17\_01.

This investigation is supported by the National Aeronautics and Space Administration (NASA) grants GO4-15096X, AR6-17012X and GO617081X.

This work is supported by the ``Departments of Excellence 2018 - 2022’’ Grant awarded by the Italian Ministry of Education, University and Research (MIUR) (L. 232/2016). 

F.Massaro acknowledges fi-cial contribution from the agreement ASI-INAF n.2017-14-H.0. A.C.F. acknowledges support from ERC Advanced Grant 340442. K.R. acknowledges financial support from the ERC Starting Grant ``MAGCOW'', no. 714196 and M. S. acknowledge the support from the NSF grant 1714764.

This research has made use of the SIMBAD database and VizieR catalog, operated at CDS, Strasbourg, France.

KG wishes to thank the Hungarian National Research, Development and Innovation Office (OTKA K134213) for support.

\software{CASA \citep[v4.7.2.;][]{mcmullin2007}, CIAO \citep[v4.13;][]{Fruscione2006}, SHERPA \citep[v4.13;][]{Freeman2001}}.

\bibliography{clusters}{}
\bibliographystyle{aasjournal}
\clearpage
\newpage
\appendix
\section{Appendix: Individual sources}
\label{app:individual}
A detailed description of the radio emission morphologies of each source in our sample and additional relevant information on the multifrequency emission are given in the following sections. An asterisk (` * ') after the name of the source indicates that it was observed as part of our VLA A-configuration project (ID: \,15A-305, PI: Werner) from 2015.

\subsection{3C\,449}
\label{sec:3c449}
The low-power FR\,I radio source 3C\,449 is the CDG of the 1.5\,keV group/cluster Zwicky 2231.2+3732 at a luminosity distance of $\sim$\,76\,Mpc. The radio emission extends out to almost 500\,kpc \citep{perley1979,andernach1992,feretti1999} in the form of relativistic (up to $\sim$\,3.3\,kpc) and well-collimated (up to $\sim$\,16.7\,kpc) radio jets terminating in large diffuse radio lobes with a slight bend to the west (Figure\,\ref{fig:chandra_vla_a}a).\ 

Associated X-ray emission from the hot gas and its interaction with the radio source has been studied by \cite{hardcastle1998}, \cite{croston2003}, and \cite{lal2013}. The X-ray cavities were observed coincident with the inner southern radio jet and the {\it Chandra} X-ray Observatory detected cold fronts, which likely originate from sloshing of the hot gas that could also contribute to the bending of the radio jets.\

HST observations revealed a~600\,pc dusty disk, the orientation of which is not perpendicular to the orientation of the radio jet axis \citep{feretti1999,tremblay2006}. The CO(1-0) transition line, as a tracer of molecular gas was detected by \cite{flaquer2010}. The study of SOAR data revealed H$\alpha +$[N II] line emission within the innermost 2\,kpc from the core \citep{lakhchaura2018}.

\subsection{IC\,1860}
\label{sec:ic1860}
IC\,1860 is CDG of the 1.3\,keV group IC\,1860 \citep[Abell\,S301;][]{abell1989}.

The observed radio emission in the VLA A configuration has a form of point-like\footnote{Point-like morphology was found by \cite{dunn2010} using observations at 1.4\,GHz from 2007.} radio source slightly elongated in the north-eastern and south-western direction (Figure\,\ref{fig:chandra_vla_ps_a}a).\

The {\it XMM-Newton} and {\it Chandra} data were analysed by \cite{gastaldello2013} who found sharp surface brightness discontinuities (cold fronts) which they interpreted as evidence for sloshing, which could be a result of a minor merger with the spiral galaxy IC\,1859 \citep[this is also consistent with numerical simulations for relaxed clusters;][]{ascasibar2006}. Moreover, their radio observations at lower frequencies with Giant Metre-wave Radio Telescope (GMRT) revealed a steep-spectrum diffuse radio emission enclosed by these cold fronts.

\subsection{IC\,4296*}
\label{sec:ic4296}
IC\,4296 is a giant elliptical galaxy and the CDG of Abell\,3565. 

The radio emission identified with the radio source PKS\,1333-33, extends well beyond the host galaxy IC\,4296. Our VLA A and D configuration data (Figure\,\ref{fig:chandra_vla_a}b) show a unresolved central radio source and extended radio emission in the form of two well-collimated (within the first $\sim$\,10\,kpc from the core) radio jets expanding up to $\sim$\,150\,kpc on both sides with many knots and wiggles and terminating in large radio lobes with a radius of $\sim$\,60-80\,kpc. IC\,4296 and its radio source (in VLA C configuration) was extensively studied by Kileen et. al \citep{killeen1986I,killeen1986II,killeen1988}.\

The X-ray emission from this giant elliptical galaxy was described by \cite{pellegrini2003}.
Moreover, our analysis of the archival {\it XMM-Newton} data revealed an X-ray cavity associated with the north-western radio lobe \citep{grossova2019}. The more in-depth X-ray spectral analysis of thermodynamical profiles show an unusually low central entropy and cooling time. 

The presence of cooling gas in the form of warm H$\alpha +$[N II] (at kpc scale) \citep{lakhchaura2018}, molecular CO(2-1) emission (at 100\,pc scale) \citep{boizelle2017,ruffa2019} and a warped dust disk (at $<$\,pc scale) \citep{schmitt2002} has been revealed in the central region. \

More details of the potential unbalanced cooling in the nucleus and the relationship between the radio, X-ray and optical emission for IC\,4296 is summarized in our paper \citep{grossova2019}.\\

\subsection{NGC\,57*}
\label{sec:ngc57}
NGC\,57 is another giant elliptical galaxy from our proposal VLA A configuration data and identified as an isolated elliptical galaxy \citep{smith2004, o'sullivan2007}.\ 

Our high resolution VLA A configuration total intensity image (Figure\,\ref{fig:chandra_vla_ps_a}b) reveals, for the first time, a point-like radio morphology previously unseen in the TIFR GMRT sky survey (TGSS), the NRAO VLA sky survey (NVSS) or in the study by \cite{o'sullivan2007}.
No dusty features \citep{goullaud2018} or warm gas \citep{lakhchaura2018} were found in this galaxy.

\subsection{NGC\,315}
\label{sec:ngc315}
NGC\,315 is well-known giant elliptical galaxy classified as a FR\,I radio source \citep{fanaroff1974}, which resides in the central region of the NGC\,315 (WBL\,22) group. 

The radio emission has been thoroughly studied at various radio frequencies by many research groups \citep[; and references therein]{bridle1979,willis1981,klein1994,mack1997, laing2006,worrall2007,giacintucci2011} and has the form of relatively symmetric radio jets with several wiggles and knots, which are inflating wiggles and knots, which extend almost 200 kpc. The SE jet is inflating a radio lobe, while the NW jet undergoes an apparent 180 degree bend and forms an extended plume (Figure\,\ref{fig:chandra_vla_a}c).\  

\cite{chen2012} investigated the kinematic and spatial properties of the close galaxy members of NGC\,315 and found a low density ambient X-ray gas environment, which could be the reason why the radio jet can expand to such a large distance from the core. \

The X-ray analysis revealed a strong X-ray jet, which traces the radio synchrotron emission and extends up to about $\sim\,$10\,kpc \citep[Figure\,\ref{fig:chandra_vla_a}c]{worrall2003,worrall2007,donato2004}.

The optical emission is represented by a dusty disk perpendicular to the direction of the jets and a few dust patches \citep{capetti2000}. Moreover, at the position of the HST dust patches, \cite{morganti2009} found two HI absorption components. The cold molecular gas, traced by CO(1-0) line emission, was found with the IRAM telescope \citep{flaquer2010}. Moreover, \cite{boizelle2021} detected a sub-kpc CO disk. The warmer nebulae, traced by H$\alpha +$[N II] line, was observed by \cite{ho1997}, but not confirmed in the more recent study as presented in \cite{lakhchaura2018} combing narrow-band imaging with long slit spectroscopic observations with SOAR.\

\subsection{NGC\,410*}
\label{sec:ngc410}
Another galaxy from our VLA A configuration project is a giant elliptical NGC\,410 (Figure\,\ref{fig:chandra_vla_ps_a}c), which is the CDG of the LGG\,18 group \citep{kolokythas2018}.\ 

The radio emission with a point source-like morphology was previously observed with the VLA at various frequencies \citep{condon1998,filho2002}.
The more recent study by \cite{kolokythas2018} at lower frequencies (610~MHz) with the GMRT found a small extension of the radio emission to north-east with a total extent $\sim\,$10\,kpc.\

\cite{gonzalez-martin2009} claimed no point source detection in the X-ray hard band (4.5-8.0\,keV) with {\it XMM-Newton} (together with other evidence from the optical, UV and radio bands) and therefore classified NGC\,410 as a galaxy without an AGN. The X-ray halo emission was detected in 0.5-7\,keV band with {\it Chandra} \citep{lakhchaura2018}. 

A recent study with SOAR data, as summarized by \cite{lakhchaura2018}, revealed H$\alpha +$[N II] line emission within the innermost 2\,kpc from the core.

\subsection{NGC\,499}
\label{sec:ngc499}
NGC\,499 is the only one galaxy in our sample of 42 early-type sources without radio emission  (Figure\,\ref{fig:chandra_vla_ps_a}d) observed in the nucleus or in the extended diffuse region.\

A faint radio source was observed within the 1.4\,GHz NVSS survey \citep{condon1998} with a flux density of 0.7\,mJy at $2\,\sigma_{\rm RMS}$ \citep{brown2011}, which is well above the sensitivity limit of our VLA observations (for more details see Section\,\ref{sec:radio_central}).

NGC\,499 is a CDG of the NGC\,499 group, which is a part of the Pisces cluster \citep{kim1995}. \cite{paolillo2003} have shown that NGC\,499 group is in the process of merging with the NGC\,507 group (Section\,\ref{sec:ngc507}).\

The luminosity of the X-ray halo is relatively high \citep[$\sim\,2.3\cdot 10^{42}$\,erg/s;][]{dunn2010,lakhchaura2018}. Moreover, the {\it Chandra} image shows possible ghost cavities at about $\sim$\,10-15\,kpc to the north and south of the nucleus, a sign of a previous AGN activity \citep{panagoulia2014b,kim2019}. 

\cite{lakhchaura2018} present a detection of a centrally located warm nebular emission.

\subsection{NGC\,507*}
\label{sec:ngc507}
NGC\,507 is the CDG of the NGC\,507 group, which belongs to the Pisces cluster. As mentioned in the previous section, the NGC\,507 group is likely to be merging with the NGC\,499 group (Section\,\ref{sec:ngc499}).\

The large-scale diffuse radio emission from a low-power FR\,I radio source B2\,0120+33 \citep{parma1986} has a shape resembling a letter `$\Gamma$' formed by two radio lobes with diameters of $\sim$\,40 and $\sim$\,20\,kpc \citep{dunn2010, murgia2011}. 
Interestingly, this emission has a very steep radio spectrum, possibly due to the `fossil' radio material from previous AGN activity \citep{murgia2011}.\

The central part of the brighter western lobe was resolved also by our high resolution A configuration data (Figure\,\ref{fig:chandra_vla_a}d) and coincides with a {\it Chandra} X-ray cavity at $\sim$\,7\,kpc to the south-west  of the galaxy core \citep{dong2010}.\\
On the other hand, the eastern radio lobe is bending to the south with its edge tracing a {\it Chandra} X-ray cold front \citep{fabbiano2002,kraft2004}, a result of sloshing. The disturbed morphology could be due to the ongoing merger with NGC\,499 as studied and simulated by \cite{ascasibar2006}. A different scenario is proposed by \cite{kraft2004}: the sharp-edged X-ray surface brightness profile could be entrained ICM material from the expanding eastern radio lobe.\\
{\it XMM-Newton} data also confirms the observed discontinuity \citep{fabbiano2002}. 

No warm \citep{lakhchaura2018} and cold gas was found inside of this giant elliptical galaxy.

\subsection{NGC\,533*}
\label{sec:ngc533}
NGC\,533 is the CDG of the NGC\,533 group of galaxies.

The new high resolution A configuration data observed within our project 15A-305 revealed a central bow-tie-shaped radio emission with a total extent of $\sim$\,5\,kpc (Figure\,\ref{fig:chandra_vla_a}e), which resembles the radio morphology of FR\,II radio sources (see Discussion \ref{sec:frII}).\ 

NGC\,533 is one of the two galaxies in our sample (besides NGC\,708; Section\,\ref{sec:ngc708}) which shows signatures of being a young radio galaxy \citep{o'dea1998}: a compact morphology of the radio emission within the host galaxy without extended features, cavities in X-rays as a sign of small-scale interaction of the radio and X-ray plasma, a double morphology of the radio emission.

In comparison with the central X-ray emission, previously studied using {\it Chandra} X-ray images by \cite{dunn2010} and \cite{shin2016}, we suggest that the radio lobes visible in the red contours in Figure\,\ref{fig:chandra_vla_a}e could correspond to the presence of the north-eastern and especially south-western X-ray cavity. 

The optical analysis presented in \cite{ferrari1999}, \cite{temi2007}, and \cite{lakhchaura2018} showed co-spatial H$\alpha +$[N II] emission coincident with a filamentary dust distribution.

\subsection{NGC\,708}
\label{sec:ngc708}
The giant elliptical galaxy, NGC\,708 is a CDG of the cooling flow cluster Abell\,262. \cite{parma1986} and \cite{blanton2004} defined this galaxy as a low-power FR\,I radio source.\

VLA archival data in the A and C configuration (Figure\,\ref{fig:chandra_vla_a}f), also studied by \cite{parma1986}, \cite{blanton2004}, and \cite{clarke2009}, reveal a diffuse two-sided radio lobe-like emission with a total extent of $\sim$\,40\,kpc from the central radio source B2\,0149+35.

The radio emission from NGC\,708, like NGC\,533 (see Section\,\ref{sec:ngc533}), resembles the emission of a young radio galaxy \citep{o'dea1998} due to its confinement within 10\,kpc of the host galaxy without extended emission, interaction with the inner X-ray gas in form of cavities \citep{blanton2004, clarke2009,panagoulia2014b}, double radio morphology, and presence of the optical emission line [OIII] \citep{liuzzo2013}.\

The X-ray emission shows surface brightness dips, which could be signatures of cavities on both sides (especially on the eastern side of the core) consistent with the expanding diffuse radio emission  \citep{blanton2004, clarke2009}. The X-ray analysis shows that the enthalpy of the radio lobes is not sufficient to balance radiative losses from the ICM, suggesting that NGC 708 may have shut down or be in a low-power phase of activity. This is also supported by the non-detection of radio emission on smaller scales with VLBI \citep{liuzzo2010}.\

A search for multiphase gas in NGC\,708 was performed by \cite{sahu2016}. A molecular CO gas disk, with a total extent of $\sim$\,3\,kpc, was discovered with ALMA by \cite{braine1994} and in more detail studied also by \cite{russell2019}, \cite{olivares2019}, and \cite{north2021}.
The peak of the molecular gas emission coincides with the AGN core and is perpendicular to the projected orientation of the radio jets. The cool gas is also surrounded by dust lanes in the nuclear region \citep{sahu2016, kulkarni2014}.\\
The extended ($\sim$\,13\,kpc) warm nebula traced by H$\alpha +$[N II] emission \citep{lakhchaura2018} is also consistent with the {\it Chandra} X-ray surface brightness peaks, which suggests that at least part of the hot X-ray gas is cooling at even larger distances from the core \citep{blanton2004}.

\subsection{NGC\,741}
\label{sec:ngc741}
NGC\,741 is the CDG in the group holding the same name.

The radio structure, first studied by \cite{venkatesan1994} and \cite{birkinshaw1985}, was probably formed by the close passage of the second brightest galaxy in the group, the head-tail radio galaxy NGC\,742 with the same redshift as NGC\,741 \citep{schellenberger2017}.

Our analysis, using a new high resolution A configuration total intensity image, confirms the radio emission in the form of a bridge connecting NGC\,471 and NGC\,742 and a prominent bent radio tail (to the southwest) in the C configuration total intensity image (Figure\,\ref{fig:chandra_vla_a}g).\

X-ray analysis by \cite{jetha2008} and \cite{schellenberger2017} has shown several generations of X-ray cavities.

The warm gas was found in the nuclear regions of NGC\,741 by \cite{macchetto1996}, but more recent narrow-band and long slit spectroscopic SOAR observations are disfavoring their findings \citep{lakhchaura2018}. Dust is missing in this giant elliptical galaxy \citep{verdoes2005}.

\subsection{NGC\,777*}
\label{sec:ngc777}
Compact radio emission is detected in our new VLA A configuration total intensity image with a resolution of $\sim$\,1.5\,arcsec for NGC\,777 (Figure\,\ref{fig:chandra_vla_a}h), the giant elliptical galaxy and the CDG of LGG\,42 group. This compact morphology was also detected by \cite{kolokythas2018} with GMRT.\ 

Additionally, we also imaged the archival VLA C configuration data, where no radio emission was detected at the angular resolution of $\sim$\,3.2\,arcsec, possibly due to the flux density being very close to the sensitivity limit of the VLA observations (see Section\,\ref{sec:radio_central}).

The X-ray emitting halo has an almost undisturbed morphology ($\sim$\,40\,kpc across) with a quite high luminosity of $\sim\,4\cdot10^{42}$\,erg/s \citep[when compared to the rest of the galaxies in the sample][]{lakhchaura2018}. The X-ray surface brightness depressions were detected in the hot atmosphere of NGC\,777 by \cite{panagoulia2014b}.

\citet{lakhchaura2018} reports a non-detection of any trace of warm ionized nebulae.

\subsection{NGC\,1132}
\label{sec:ngc1132}
NGC\,1132 is a giant elliptical, which resides inside of the well-known fossil group NGC\,1132.\

Radio emission from the archival VLA C configuration data (Figure\,\ref{fig:chandra_vla_ps_a}e) shows a point source morphology from the unresolved central source.\

The outer contour level ($5\times\sigma_{\rm RMS}$; see Table\,\ref{tab:results}) coincides with the possible shock front in the X-rays noticed by \cite{kim2018} and could be connected to the merging history of NGC\,1132 as suggested by \cite{mulchaey1999} \citep[and predicted by the simulations of][]{barnes1989,governato1991}. Moreover, \cite{dong2010} found an X-ray cavity at a distance of 4\,kpc south of the host galaxy's center, which is consistent with the potential set of cavities detected in the southern regions by Plšek et al. (in prep).\

No emission from warm gas emission was detected for this source \citep{lakhchaura2018}.

\subsection{NGC\,1316}
\label{sec:ngc1316}
The giant elliptical galaxy, NGC\,1316, hosts a well-known strong radio source Fornax\,A located at the outskirts of the Fornax cluster.

Radio emission in the higher resolution BA configuration image extends from the northwest to southeast in the form of an oblong letter 'S' and represents the most recent of its AGN outbursts with the south-eastern extension also visible in the C configuration image (Figure\,\ref{fig:chandra_vla_b}a), \citep[both previously published; ][]{fomalont1989, maccagni2020a,maccagni2021b}.\ 

Several generations of X-ray cavities are visible in the {\it XMM-Newton} and {\it Chandra} images \citep{lanz2010}. \ 

The S-shaped jet morphology together with the disturbed morphology of the multi-phase gas \citep{morokuma-matsui2019,lakhchaura2018} and dust \citep{duah2014} supports the fact that NGC\,1316 is in the process of merging into the Fornax cluster (Section\,\ref{sec:ngc1399}).

\subsection{NGC\,1399*}
\label{sec:ngc1399}
The radio emission of NGC\,1399, the CDG of the Fornax cluster, has a two-sided extended radio morphology in the form of two well-collimated radio jets terminating in two mildly diffuse radio lobes (Figure\,\ref{fig:chandra_vla_b}b). Similar morphological features were previously discussed in \cite{shurkin2008} and 
\cite{dunn2010}.\ 

Moreover, \cite{paolillo2002}, \cite{shurkin2008}, and \cite{panagoulia2014b} found signatures of the interaction between the AGN and the hot gas, in the form of X-ray cavities, with bright rims coincident with the outer edges of both radio lobes.\

Even though the presence of CO(2-1) line emission was confirmed by \cite{prandoni2010}, no dust features were found in the core of NGC\,1399. Moreover, \cite{werner2014} showed that the central regions of this galaxy lack [C II] line emission.

\subsection{NGC\,1404*}
\label{sec:ngc1404}
The giant elliptical NGC\,1404, also part of our new A configuration VLA observation, is another member of the Fornax cluster. \cite{machacek2005} suggested that NGC\,1404 is falling towards the center of the Fornax cluster, where NGC\,1399 (Section\,\ref{sec:ngc1399}) resides.\ 

Our new VLA A configuration data (Figure\,\ref{fig:chandra_vla_ps_a}f) at high sensitivity with the RMS noise of the total intensity image reaching $25\,\mu$Jy per beam revealed a faint (flux density of $210\,\mu$Jy per beam) central radio source and thus supporting the previous findings of \cite{dunn2010}, where the detection was very close to the sensitivity limit. Moreover, the central radio source was also detected in VLBI observations, which can be found in the archive of Leonid Petrov\footnote{\href{http://astrogeo.org/cgi-bin/imdb_get_source.csh?source=J0338-3527}{http://astrogeo.org/cgi-bin/imdb$\_$get$\_$source.csh?source=J0338-3527}})
On the other hand, a more compact archival VLA CD configuration observation did not detect radio emission, maybe due to a lower flux density than the sensitivity limit (see Section\,\ref{sec:radio_central}). \ 

There are no signatures of ongoing AGN-gas interactions in this system, as supported by a quite smooth circumnuclear morphology of the hot X-ray halo \citep{machacek2005,dunn2010}. Cold \citep{werner2014} and warm gas \citep{lakhchaura2018} is missing in this galaxy.

\subsection{NGC\,1407*}
\label{sec:ngc1407}
Another object within our sample with a newly observed VLA A configuration data is NGC\,1407. This giant elliptical galaxy is located inside in the dynamically relaxed NGC\,1407 (Eridanus\,A) group, part of the Eridanus supergroup. 

The radio emission from the new A configuration data does not show a very distinctive radio point source in the nucleus (Figure\,\ref{fig:chandra_vla_b}c). More prominent is the diffuse morphology with asymmetric radio jet-like structures, extending especially to the east as seen also in the archival VLA B and C configuration data \citep{giacintucci2012}.  \ 

From the multiband and multifrequency study \citep{giacintucci2012} reoccurring activity of the AGN has been suggested, similar to that found for another fossil giant elliptical galaxy in our sample, NGC\,5044 (Section\,\ref{sec:ngc5044}).\\
Only an upper limit on the flux density of CO(1-0) emission is found for NGC\,1407 \citep{babyk2019}.

\subsection{NGC\,1550*}
\label{sec:ngc1550}
NGC\,1550 is the CDG of the group of galaxies having the same name. 

The radio emission observed within our new high resolution A configuration observation revealed asymmetric small-scale jets with a physical extent of $\sim$\,4\,kpc. At the lowest contour level created at $5\times\,\sigma_{\rm RMS}$, we find a small hotspot-like feature at the distance of 6\,kpc. This feature seems not to be the extension of the jet and is located within the diffuse lobe-like emission visible in the compact archival C configuration total intensity image (Figure\,\ref{fig:chandra_vla_b}d), previously published by \cite{dunn2010}. \\
The lower frequency GMRT study supports the presence of the diffuse lobe-like morphology \citep{kolokythas2018}. They also found that the inward-bended eastern jet seems to be aligned with the potential hotspot feature seen in the VLA A configuration contours (Figure\,\ref{fig:chandra_vla_b}d).\

\cite{kolokythas2020}, in their recent radio and X-ray study found a sign of sloshing of the ambient IGM around NGC\,1550, previously known as relaxed. They present arc-shaped cold fronts together with a potential X-ray cavity \citep[supported by the findings of ][]{panagoulia2014b} coincident with the position of the radio lobe.\ 
 
Neither molecular gas nor cold dusty features have not yet been observed and \cite{lakhchaura2018} presents non-detection of warm ionized nebulae for this giant elliptical galaxy.

\subsection{NGC\,1600}
\label{sec:ngc1600}
NGC\,1600 is a relatively isolated CDG of the NGC\,1600 group of galaxies \citep{smith2008}. 

The radio emission was previously published by \cite{birkinshaw1985} at 4.85\,GHz with VLA in C configuration and shows a potential double radio source without a clearly defined radio core. The higher resolution A configuration data at 1.4\,GHz (Figure\,\ref{fig:chandra_vla_b}e) shows in more detail the radio jets/lobes features and potential hotspots, which could resemble the morphology of FR\,II radio source (see Discussion\,\ref{sec:ngc1600_frii}).

The X-ray emission from {\it Chandra} X-ray image shows a diffuse structure, which seems to show two surface brightness depressions at the positions of the two radio lobe structures, previously studied by \cite{sivakoff2004}.\

Moreover, they found excess emission in the X-rays, which more or less coincides with the H$\alpha +$[N II] emission \citep{singh1995,lakhchaura2018} of the ionized gas (extending up to $\sim$\,30\,arcsec from the core) and dust \citep[extending up to $\sim$\,10\,arcsec from the core;][]{ferrari1999}.\

The signatures of properties similar to the `fossil group' were for NGC\,1600 identified by \cite{santos2007} and \cite{smith2008}.

\subsection{NGC\,2300*}
\label{sec:ngc2300}
The radio emission from our new VLA A configuration data for NGC\,2300, the central dominant galaxy of the NGC\,2300 group, shows a point-like morphology (Figure\,\ref{fig:chandra_vla_ps_a}g). The archival VLA D configuration data reveals radio point source emission, as well (Figure\,\ref{fig:chandra_vla_ps_a}g) .\ 

NGC2300 is known to undergo tidal interactions with a late-type active star-forming galaxy NGC\,2276 \citep{wolter2015}. Plšek et al. in prep. detected potential `ghost' X-ray cavities in NGC\,2300. \

The optical study of the HST  and SOAR data, done by \cite{xilouris2004}, showed that there is no detection of dust or warm ionized nebulae \citep{lakhchaura2018} inside NGC\,2300.

\subsection{NGC\,3091*}
\label{sec:ngc3091}
NGC\,3091 is a part of our VLA high resolution project from 2015 and is the CDG of a poor Hickson Compact Group, HCG\,42 \citep{hickson1982} \citep{colbert2001}. 

The radio morphology, seen in the highest resolution $\sim$\,2\,arcsec VLA A configuration data (Figure\,\ref{fig:chandra_vla_ps_a}h), is represented by a faint point source-like radio emission in the center of the galaxy. The AB configuration data with $\sim$\,14\,arcsec resolution at the same frequency (1.4\,GHz) reveals no detection, maybe because of a lower flux density limit of our VLA observation  (see Section\,\ref{sec:radio_central}).\

Even though radio emission has a point-like morphology, Plšek et al. (in prep.) detected potential `ghost' cavities in the X-ray hot atmosphere.

In the analysis of the optical and NIR morphology, no H\,I, dust features \citep{colbert2001} or warm gas \citep{lakhchaura2018} have been detected in the central regions of NGC\,3091.

\subsection{NGC\,3923*}
\label{sec:ngc3923}
The biggest early-type galaxy, NGC\,3923, located in the optical group (LGG 255) with only galaxy-scale X-ray halo has an oval-shaped radio morphology in our new VLA A configuration data (Figure\,\ref{fig:chandra_vla_ps_b}a), while the archival C and CD configuration data show no radio emission. This source was observed in the 5\,GHz survey done by \cite{disney1977}.

{\it Chandra} data have shown potential X-ray surface brightness decrements in the form of `ghost' cavities without corresponding radio emission from radio jets and lobes in this giant elliptical galaxy (Plšek et al. in prep.). 

NGC\,3923 is a famous shelled\footnote{NGC\,3923 has between 22 and 42 shells, the largest number of shells observed in the elliptical galaxy.} giant elliptical with dusty filamentary features and H${\alpha}$ emission in the central region \citep{bilek2016,miller2017}. On the other hand, \cite{lakhchaura2018} reported non-detection of warm gas in the core.

\subsection{NGC\,4073*}
\label{sec:ngc4073}
The giant elliptical NGC\,4073 is dominant member of the poor cluster MKW\,4 and shows a point source-like radio morphology in our new high resolution VLA A configuration observation (Figure\,\ref{fig:chandra_vla_ps_b}b). A weak detection of the central radio source has been reported by \cite{hogan2014} with VLA in the C-band (4--8\,GHz).\ 

The X-ray emission was studied by \cite{o'sullivan2003} who found a 1.7\,keV halo within the relaxed system. Moreover, Plšek et al. (in prep.) found potential `ghost' cavities in the X-ray atmosphere. Warm emission line nebula was not detected in this source \citep{lakhchaura2018}. 

\subsection{NGC\,4125}
\label{sec:ngc4125}

NGC\,4125 is a luminous giant elliptical galaxy residing in the NGC\,4125 group (alternatively called LGG\,266).

The radio emission (Figure\,\ref{fig:chandra_vla_ps_b}c), detected in the archival D configuration data, has a diffuse morphology with an extension to the northeast, corresponding to a point-like radio source visible in the archival 1.4\,GHz Westerbork Synthesis Radio Telescope (WSRT) observations. More prominent structure has a form of inflated radio lobes, which are offset by 4\,kpc (46\,arcsec) in projection with respect to the radio, optical, and X-ray core of NGC\,4125. This offset was also noticed in the NVSS data at 1.4\,GHz by M. Rupen et al. in their online notes\footnote{ https://www.cv.nrao.edu/$\sim$\,jhibbard/HIinEs/HIinE.html} about VLA H\,I data of early-type galaxies and by \cite{krajnovic2002} in the VLA C configuration observations at 8.3\,GHz. In both studies, only the central regions of the offset radio lobes were detected. The radio core corresponding to the X-ray and optical center of NGC\,4125 was missing and is only clearly visible in the WSRT images. Moreover, due to the missing optical counterpart of the extended radio-lobe like emission, \cite{krajnovic2002} defined NGC\,4125 as a background source (see Discussion \ref{disc:offset_ngc4125} for a possible explanation of this unusual emission).

\cite{gonzalez-martin2006} studied NGC\,4125 as a part of the X-ray study of LINER galaxies and identified a centrally located hard X-ray point source-like emission, which is a sign of ongoing AGN activity. The extended cold emission from the large amount of dust \citep{kulkarni2014}, [C II] and [N II] has been observed, but no CO\footnote{\cite{wiklind1995} claimed detection of CO(1-0) line, which was not confirmed by \cite{welch2010}.}, H I \citep{welch2010,wilson2013} or H$\alpha +$[N II] emission \citep{lakhchaura2018}. This could be explained by a merger-triggered star formation outburst providing heated gas and dust to the surrounding medium of the galaxy.

\subsection{NGC\,4261}
\label{sec:ngc4261}
NGC\,4261 is the optical counterpart to the radio source 3C\,270 \citep{birkinshaw1985}, which is a very well known FR\,I radio source inside the Virgo cluster \citep[W cloud;][]{garcia1993} with well-defined radio jets and large radio lobes piercing through the atmosphere of the host galaxy and extending beyond from the core (Figure\,\ref{fig:chandra_vla_b}f). 

A detailed study, combining data from the GMRT and VLA as well as comparing the radio emission with the X-ray emission of the hot gas, was published by \cite{kolokythas2015} and \cite{o'sullivan2011}.\

From the X-ray point of view, NGC\,4261 has an extended X-ray halo \citep{davis1995} with features showing the interaction with radio source in the form of X-ray cavities \citep{croston2008,o'sullivan2011}. On small scales, X-ray jets are present corresponding to the innermost collimated radio jets \citep{gliozzi2003,zezas2005,worrall2010}.\

On the parsec scale, \cite{jaffe1994} and \cite{ferrarese1996} found a warped dusty disk with the rotational axis perpendicular to the direction of the radio jets streaming out of the nuclear region and extending up to $\sim$\,100\,pc from the core. The presence of atomic and molecular gas from H\,I and CO (2--1) was detected too. Moreover, a sub-kpc CO disk was recently detected by \cite{boizelle2021}.\

Very similar morphology and multiband features are observed for the giant elliptical galaxy IC\,4296 (Section\,\ref{sec:ic4296}), which has, however, $\sim$\,6$\times$larger radio lobes, what leads to $\sim$\,30$\times$larger total energy output supplies for the inflation of those radio lobes \citep{frisbie2020,grossova2019}. 

\subsection{NGC\,4374}
\label{sec:ngc4374}
Messier\,84 or NGC\,4374 is a giant elliptical galaxy in the Virgo A group, which is a part of the Virgo cluster together with NGC\,4406 (Section\,\ref{sec:ngc4406}), NGC\,4486 (Section\,\ref{sec:ngc4486}), and NGC\,4552 (Section\,\ref{sec:ngc4552}). \ 

The radio emission in the archival B configuration data (Figure\,\ref{fig:chandra_vla_b}g) shows well-defined and in the central regions well-collimated jets terminating in diffuse plum-like lobes \citep[first published by][]{laing1987} with the northern lobe brighter than the southern and bent to the east. VLA D configuration data reveal pear-like diffuse emission extending beyond the B configuration data.\
Moreover, the VLBA data analysed by \cite{ly2004} revealed a jet-like extension in the direction consistent with the northern radio jet from the VLA.

The extended radio lobe emission is filling the region of the drops in the surface X-ray brightness, the X-ray cavities \citep{finoguenov2001}, and extending far beyond the host galaxy \citep{devereux2010}.\ 

The nuclear region in NGC\,4374 revealed a small ionized gas disk \citep{bower1997}, faint circumnuclear CO(2-1) absorption and emission together with an asymmetric dust lane \citep{verdoes1999,boizelle2017}. All features are perpendicular to the innermost radio jets.

\subsection{NGC\,4406*}
\label{sec:ngc4406}
NGC\,4406, know as M\,86, is located in the Virgo cluster and flying in the direction towards us. Our new VLA A configuration data at 1.5\,GHz  (Figure\,\ref{fig:chandra_vla_ps_b}d) reveal a faint, point-like radio emission in the center of this giant elliptical. Similarly, \cite{dunn2010} found point-like radio emission in the nuclear region at higher frequencies at around 4.9\,GHz in the C configuration data. On the other hand, complementary archival data in VLA D configuration at 1.4\,GHz have not detected any radio emission from the core, possibly due to the high noise level of the observation (for more details see Table\,\ref{tab:results}).\ 

NGC\,4406 is the second X-ray brightest giant elliptical in Virgo cluster (after M\,87) and its negative radial velocity suggests a supersonic movement through the ICM. This could create the observed X-ray features (plume and tail) by ram pressure stripping \citep{forman1979, randall2008, ehlert2013, kim2019}.\ 

Signs of possible galaxy-galaxy interactions are observed in the distribution of the atomic gas and dust \citep{smith2012} as well as the H$\alpha$ features connected to the neighboring spiral galaxy NGC\,4438 \citep{kenney2008,lakhchaura2018}.

\subsection{NGC\,4472*}
\label{sec:ngc4472}
NGC\,4472, or M\,49, is an early-type elliptical and CDG of the M\,49 group falling from the south into the Virgo cluster and it is possibly in the second turn around the cluster center \citep{su2019}.\ 

NGC\,4472 was observed in our 2015 project in VLA A configuration (Figure\,\ref{fig:chandra_vla_b}h). Additionally, we also reduced archival VLA C configuration data \citep{condon1988}.
The high angular resolution of the A configuration total intensity image (Figure\,\ref{fig:chandra_vla_b}h contours) showed the radio emission from the innermost central region in the form of an elongated amorphous feature with the extended tail to the west and a small sign of radio emission to the north-east.\\
On the other hand, using the more compact C configuration array, the extended radio emission with a total extent of $\sim$\,10\,kpc is seen  \citep{ekers1978}. The central regions in the C configuration image are consistent with the radio features in the A configuration. Additional extended lobe-like structures are observed to the east and west of the nucleus. 

The eastern lobe is consistent with a decrement in the X-ray surface brightness, one of the X-ray cavities observed in this source \citep[for more details see, e.g.][]{biller2004,kraft2011,panagoulia2014b,su2019}. Warm gas is missing in the central or extended regions of this giant elliptical galaxy \citep{lakhchaura2018}.\

\subsection{NGC\,4486}
\label{sec:ngc4486}
NGC\,4486 or Messier\,87, is the central dominant galaxy of the Virgo cluster. Here, we summarize the results obtained from the archival A, B, and C configuration data ({Figure\,\ref{fig:chandra_vla_c}a}).\ 

The bright radio core, 3C\,274, was detected along with its innermost radio jet, which is coincident with the X-ray emission of the jet \citep{marshall2002}.
On the opposite side of the core, an elongated lobe-like emission is visible and coincides with an X-ray cavity \citep{young2002,forman2005,forman2007}. The archival C configuration VLA map, which is more sensitive to extended structures, presents a western tail extending up to $\sim$\,10\,kpc from the nucleus. However, we note that at 300\,MHz the entire structure of NGC\,4486 is far more extended \citep[up to 80\,kpc, e.g.:][]{owen2000}.  

The presence of multiphase gas in the form of cold molecular CO clouds \citep[e.g.:][]{simionescu2018}, cool [C II] \citep{werner2014}, nuclear ionized gas \citep{arp1967,macchetto1996}, as well as extended warm ionized H$\alpha +$[N II] nebulae \citep{lakhchaura2018} has been reported.

\subsection{NGC\,4552}
\label{sec:ngc4552}
The radio emission from the giant elliptical NGC\,4552 or M\,89, which resides inside of the subgroup A of the Virgo cluster, is presented in Figure\,\ref{fig:chandra_vla_c}b.\ 

The red contours are showing the innermost radio point-like emission from the archival high resolution VLA A configuration. More recent (Project ID: 16A-275) previously unpublished C configuration data revealed a more diffuse butterfly-like emission extending on both sides to about 0.8\,kpc. \

The relativistic plasma from the jet-like emission is clearly interacting with the hot X-ray emitting gas and forming on both sides X-ray cavities. \cite{machacek2006} and \cite{allen2006} analysed the {\it Chandra} X-ray data and found two bright rings surrounding the two cavities to the north and south, which are perpendicular to the radio structures seen in the VLA images. Moreover, \cite{machacek2006} also found signatures of ram pressure stripping of hot gas as the galaxy moves through the ICM of the Virgo cluster \cite[see also:][]{kraft2017}.\

\cite{ferrari1999} has shown that the emission from the ionized gas extends up to 10\,arcseconds, whereas the dust absorption is very weak. The warm ionized gas, traced by H$\alpha +$[N II] has not been confirmed in a more recent study described in \cite{lakhchaura2018}.

\subsection{NGC\,4636}
\label{sec:ngc4636}
NGC\,4636 is a central early-type galaxy within a poor group in the far outskirts of the Virgo Cluster.\ 
VLA A configuration images, as presented in Figure\,\ref{fig:chandra_vla_c}c, have been previously published and discussed by \cite{dunn2010}. The additional archival C configuration VLA data from 2017 revealed a more diffuse, but still quite compact and weak radio-jet-like emission with a total extent of $\sim$\,3.5\,kpc. Although the majority of detected radio emission in VLA is consistent with the radio emission at lower frequencies (235 and 610\,MHz) with GMRT observed by \cite{giacintucci2011}, they also found a more extended north-eastern radio lobe, which is coincident with the observed X-ray (NE) cavity\citep{stanger1986,jones2002, allen2006, baldi2009,panagoulia2014b}. The SW radio lobes are filled only partially with relativistic plasma.\

There have been reports of very weak CO(2-1) emission  \citep[indicating a cloud of cold molecular gas; ][]{temi2018}, [CII] emission and warm ionized H$\alpha$+[NII] nebulae \citep{werner2014}, as well as dusty features \citep{temi2003}.

\subsection{NGC\,4649}
\label{sec:ngc4649}
The giant elliptical galaxy NGC\,4649, or M\,60, is the third most luminous galaxy in the Virgo Cluster. Archival VLA data obtained in the A configuration reveal point-like emission from a weak central radio source in NGC\,4649.\

Using D-configuration data, \cite{shurkin2008} and \cite{dunn2010} found radio lobes filling the innermost X-ray cavities.\ 

Sharp edges seen in the X-ray images are spatially coincident with the radio emission detected in the D configuration (Figure\,\ref{fig:chandra_vla_c}d).\
The X-ray properties of the hot gas are discussed in more depth by \cite{shurkin2008,dunn2010, paggi2014, wood2017} and \cite{kim2020}.\

Molecular gas, traced by CO (1-0) emission, was detected by \cite{sage1989}. However, \cite{young2002} failed to confirm their findings with the observations from IRAM and provide only upper limits on CO flux.
The galaxy does not show the presence of warmer gas traced by H$\alpha$+[NII] emission. 

\subsection{NGC\,4696}
\label{sec:ngc4696}
Figure\,\ref{fig:chandra_vla_c}e reveals a radio emission in NGC\,4696, the CDG of Abell\,3526 (Centaurus Cluster) from archival VLA observations obtained in the A and BC configurations.\

A thorough study of the radio properties was performed by \cite{taylor2002} and our total intensity image of VLA A configuration data is consistent with their results, where the radio emission shows a bright nuclear source, radio jets and lobes with the western lobe bending towards east.\ 

The interaction with the hot X-ray atmosphere is clearly present, as the radio plasma pushed out the hot gas and created cavities \citep{taylor2006,panagoulia2014b,sanders2016}. Interestingly, \cite{gonzalez-martin2006} found only a diffuse morphology without a clear point-like emission in the hard X-ray bands as a sign of a central star-bursting activity. They also noted that the X-rays could be just obscured by the dust and gas. The ongoing AGN activity is supported by our VLA A configuration data together with the VLBA observation of a small-scale one-sided jet detected by \cite{taylor2006}.\

The cold extended molecular gas and warm ionized gas and dust were discussed in the recent publication by \cite{olivares2019} and references therein. Moreover, \cite{mittal2011} observed a cooling [C II] line in the central region.

\subsection{NGC\,4778}
\label{sec:ngc4778}
NGC\,4778 or NGC\,4761 is the only one lenticular (S0) galaxy in our sample and the CDG of a bright compact group, HCG\,62 \citep{hickson1982}. It is in the process of merging with its companion NGC\,4776 \citep[or NGC\,4759][]{spavone2006}. A small point source-like central radio emission is observed in both archival VLA A \citep{vrtilek2002} and D configuration total intensity images  (Figure\,\ref{fig:chandra_vla_ps_b}e).

Moreover, looking at the lower radio frequencies with GMRT, \cite{gitti2010} found the expected radio lobes, coincident with the observed X-ray cavities \citep{morita2006,panagoulia2014b} and \cite{giacintucci2011} detected a possible second outer set of radio lobes.\

The H$\alpha$ emission in NGC\,4778 was detected and studied by \cite{valluri1996}.

\subsection{NGC\,4782}
\label{sec:ngc4782}
The radio source known as 3C\,278 is hosted by the giant elliptical galaxy NGC\,4782 in the center of the group LGG\,316. 

Radio emission, shown in the archival high resolution A configuration image (Figure\,\ref{fig:chandra_vla_c}f) is rather peculiar as already noted by \cite{borne1996} and \cite{machacek2006}. The radio jets emanating from the core are, at first invisible due to the high relativistic velocities. After $\sim$\,2\,kpc the eastern side of the jet is tilted to the north, whereas the western side continues in a straight line up to 15\,kpc, where it bents to the north as well. When using a more compact B configuration, a more diffuse radio emission is observed, which not only traces the peculiar bend of the jets, but also fills the space between and around the jet structure.\

The X-ray images reveal an interaction with the neighbor NGC\,4783, which could be responsible for the observed bend of the jet. Moreover, X-ray cavities have been observed at the position of both radio lobes  \citep{borne1996,machacek2006}.\ 

A nuclear (within 2\,kpc from the core) warm ionized nebula was detected with SOAR \citep{lakhchaura2018}, otherwise no cool gas or dust was found in this giant elliptical galaxy.

\subsection{NGC\,4936*}
\label{sec:ngc4936}
The radio emission from the giant elliptical galaxy NGC\,4936 in our new VLA A configuration total intensity image (Figure\,\ref{fig:chandra_vla_ps_b}f), shows a point-like radio emission from the nucleus.\

\cite{macchetto1996} studied the warm ionized gas and found H${\alpha} +$[N II] emission concentrated in a small disk similar to NGC\,4872 (Section\,\ref{sec:ngc4782}).
The neutral gas, in the form of a double peaked H\,I line profile, was observed for this giant elliptical by \cite{reid1994} (and references therein), together with strong emission from N II, S II and O I in the core of NGC\,4936.

\subsection{NGC\,5044}
\label{sec:ngc5044}
NGC\,5044 is the central dominant galaxy of the X-ray bright group NGC\,5044, which is well known for its large cold gas reservoir. 

Archival radio observations in the A, BnA and D configurations of the VLA presented in Figure\,\ref{fig:chandra_vla_c}g reveal emission from the central radio source with small-scale jets pointing to north-east and south-west \citep[partly visible in the images analysed by ][]{dunn2010}. The previously unobserved indications for diffuse jet-like emission extending to about 6\,kpc from the core in the east-west direction are seen in VLA B configuration contours. More diffuse oval-shaped radio emission extending up to $\sim$\,20\,kpc is seen in the D configuration.\ 

NGC\,5044 has undergone (at least) three AGN outbursts. \cite{gastaldello2009}, \citet{david2011}, \citet{giacintucci2011}, \citet{david2017}, and \citet{schellenberger2020a} presented GMRT and {\it Chandra} data, which revealed that the radio emission is consistent with the presence of the innermost southern X-ray cavity. They also found a second pair of more extended (up to $\sim$\,10\,kpc) ghost X-ray cavities, which suggest another episode of an AGN outburst in NGC\,5044. 
The last, the third, and most recent activity, is traced in the most recent study \citep{schellenberger2020b}, where they also show the presence of small-scale radio jets seen in VLBI data.\ 

Sloshing features are also observed at further distances from the center caused by the accretion of a less massive group \citep{gastaldello2009}.\

The filamentary morphology of the H$\alpha +$[N II] ionized nebulae extends up to 6\,kpc \citep{ferrari1999}, whereas the corresponding dust emission extends to only 1.3\,kpc \citep{temi2007,temi2007b}, similar to NGC\,4472 (Section\,\ref{sec:ngc4472}).\ 

Studies done by \cite{david2014,temi2018}, and \cite{schellenberger2020a} showed that CO emitting molecular gas clouds within $\sim$\,10\,kpc from the core coincide well with the presence of the strong emission from the warm ionized gas (H$\alpha$ filaments), the hot X-ray emitting gas, the innermost X-ray cavities \citep{schellenberger2020b} and the cold dust features in the central regions of NGC\,5044  as well as [C II] line emission \citep{werner2014}. Moreover, two CO absorption features were also detected within 5\,pc from the core \citep{schellenberger2020b}.

\subsection{NGC\,5129*}
\label{sec:ngc5129}
NGC\,5129 is a radio-quiet giant elliptical and the dominant galaxy of a small group. 

Our new A configuration data (Figure\,\ref{fig:chandra_vla_c}h) revealed a bright radio core with small-scale radio jets extending to about 2\,kpc to the north-west and south-east from the core. We categorized this source as a compact radio source with radio jets. We find that the radio emission is offset by $\sim$\,1.2\,arcsec from the X-ray and optical core of the galaxy. The point-like radio emission in NGC\,5129 was previously observed by the NVSS survey \citep{condon1998}. \ 
 
The hot  X-ray gas in the {\it Chandra} images has a disturbed morphology, which is not aligned with the direction of the radio jets. The X-ray gas properties were previously studied by \cite{bharadwaj2014}.\

Warm ionized gas extending up to 2\,kpc from the core was detected \citep{lakhchaura2018}.

\subsection{NGC\,5419*}
\label{sec:ngc5419}
NGC\,5419 is the CDG of the poor cluster A753, which hosts the radio source PKS\,B1400-33.\ 

Our new A configuration VLA observations revealed extended radio emission in the core of the galaxy. Archival VLA CD configuration data (Figure\,\ref{fig:chandra_vla_d}a) \citep{goss1987} show a more extended emission, which forms an L-shaped radio tail to the south extending to about 35\,kpc and a second prominent diffuse 120\,kpc-in-diameter steep spectrum radio relic. This is a unusual feature for a poor cluster \citep[previously observed by][]{subrahmanyan2003}. 

From the multifrequency data analysis, \cite{subrahmanyan2003} suggests that this diffuse extended emission could be the remnant plasma of a radio lobe (although no parent optical source has been identified) injected into the poor cluster.\

X-ray images showed that the hot gas coincides with the central radio emission of NGC\,5419 \citep{balmaverde2006}.

\subsection{NGC\,5813}
\label{sec:ngc5813}
NGC\,5813 resides in the NGC\,5846 group, where NGC\,5846 (Section\,\ref{sec:ngc5846}) is the brightest central galaxy.\ 

Archival data in VLA A, B, and C configuration at $\sim$\,\,1.4\,GHz \citep{randall2011} revealed signatures of AGN activity in radio images, when compared with the {\it Chandra} X-ray images \citep[][Figure\,\ref{fig:chandra_vla_d}b]{randall2015}. The position of the central radio core is consistent with the X-ray brightness peak and the radio lobe-like emission corresponds well with the innermost X-ray cavities, especially the northern part of the inner region \citep{randall2011,panagoulia2014b}.
Additionally, the compact C configuration data reveal more diffuse emission with small ripples to the north-east and south-west, but without a clear interaction with a second pair of more extended X-ray cavities \citep{randall2015}. The radio emission filling the second generation of cavities is found at lower frequencies with LOFAR \citep{birzan2020} and GMRT \citep{giacintucci2011}.

The ionized gas shows a filamentary morphology and is coincident with dust emission \citep{ferrari1999}. The ionised gas appears to be located in the wake of the rising AGN inflated bubbles \citep{randall2011}. The cooling lines of [C II], [N II], and O I  and warm nebulae were observed too \citep{werner2014,lakhchaura2018}.

\subsection{NGC\,5846}
\label{sec:ngc5846}
NGC\,5846 is a giant elliptical galaxy and a CDG of the massive group holding the same name.\ 

The innermost central region hosts a radio core surrounded by a more diffuse radio emission (Figure\,\ref{fig:chandra_vla_d}c), which also appears to fill the innermost X-ray cavities \citep{trinchieri2002,dunn2010, machacek2011,panagoulia2014b}. The extended and diffuse emission, with a total extent of $\sim$\,\,12\,kpc, is traced by the contours of the archival CnD configuration image. The results of the presented VLA data were previously published in \cite{machacek2011}. \

The X-ray image analysis revealed multiple signatures of sloshing in NGC\,5846 \citep{machacek2011}, although, without a distinctive X-ray core in the central regions \citep{trinchieri2002,satyapal2005,gonzalez-martin2006}. \ 

The ionized and molecular plasma coincides with the disturbed nuclear X-ray emission, [C II] line emission and cold, dusty features in the central regions of NGC\,5846. \citep{macchetto1996,caon2000,trinchieri2002,werner2014,temi2007b,temi2009,mathews2013,lakhchaura2018}. \\

\subsection{NGC\,7619*}
\label{sec:ngc7619}
The giant elliptical galaxy, NGC\,7619, is the CDG of the Pegasus\,I group.\ 

Our new VLA data in the A configuration have revealed a central radio source with a small extension to the east of the core (Figure\,\ref{fig:chandra_vla_ps_b}g), consistent with previously observed point-like morphology in the archival VLA C configuration data. This findings are consistent with the emission in NVSS \citep{condon1998} and with GMRT observations at lower frequencies (at 610\,MHz) by \cite{giacintucci2011}. 

The X-ray image analysis revealed a ram-pressure-stripped tail (to the south-west of the nuclear region) as a result of NGC\,7619 falling towards the center of the Pegasus cluster and interacting with the companion NGC\,7626 with radio jets/lobes morphology \citep{randal2009} located 415\,arcseconds (102\,kpc) from NGC\,7619.\

The study of ionized gas performed by \cite{macchetto1996} showed a circumnuclear emission with an additional extended tail from the north-east to south-west. Although, the subsequent long slit spectroscopy and narrow band imaging observations with SOAR did not confirm the previous findings of H$\alpha +$[N II] emission \citep{lakhchaura2018}. CO line emission was detected by \cite{temi2007} for this giant elliptical galaxy.

\onecolumngrid

\renewcommand{\thefigure}{B1.1}
\begin{figure}
\onecolumngrid
\section{Appendix: X-ray and radio images}
\label{app:xray_radio}
\subsection{X-rays and point-like radio emission}
\label{app:xray_radio_ps}
    \centering
    \includegraphics[width=230pt]{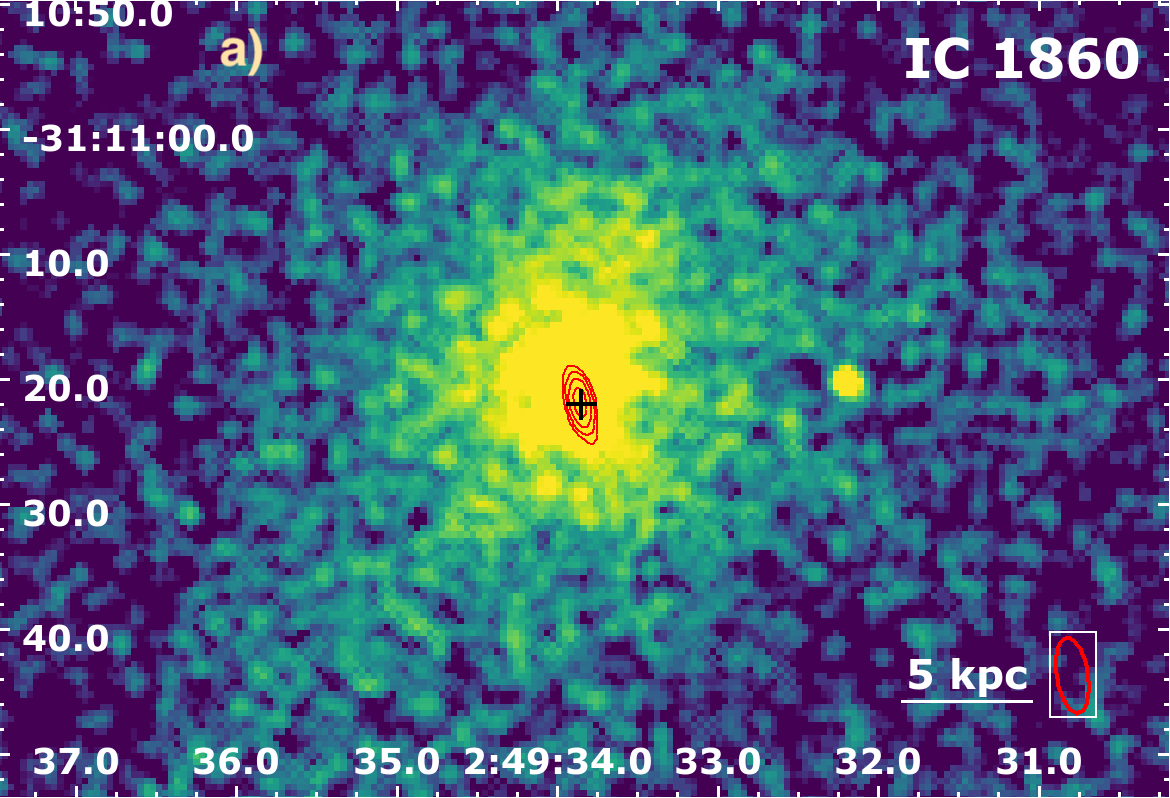}
    \includegraphics[width=230pt]{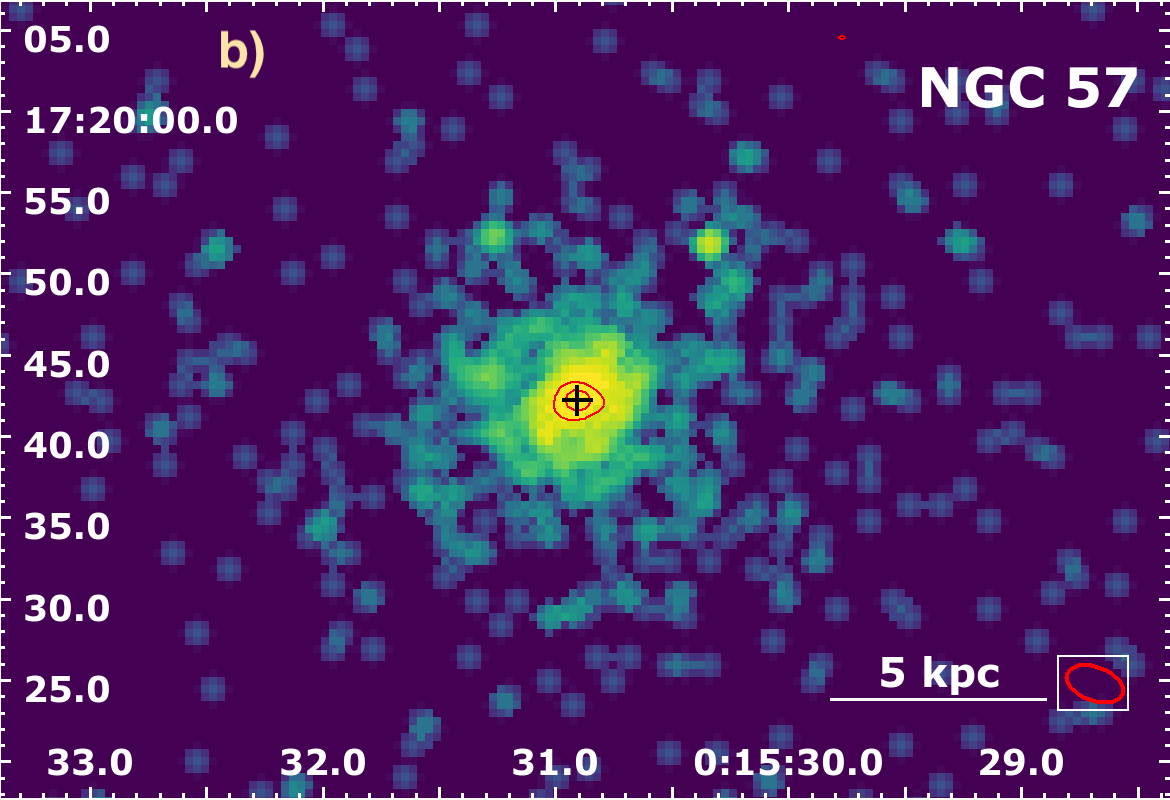}
    \includegraphics[width=230pt]{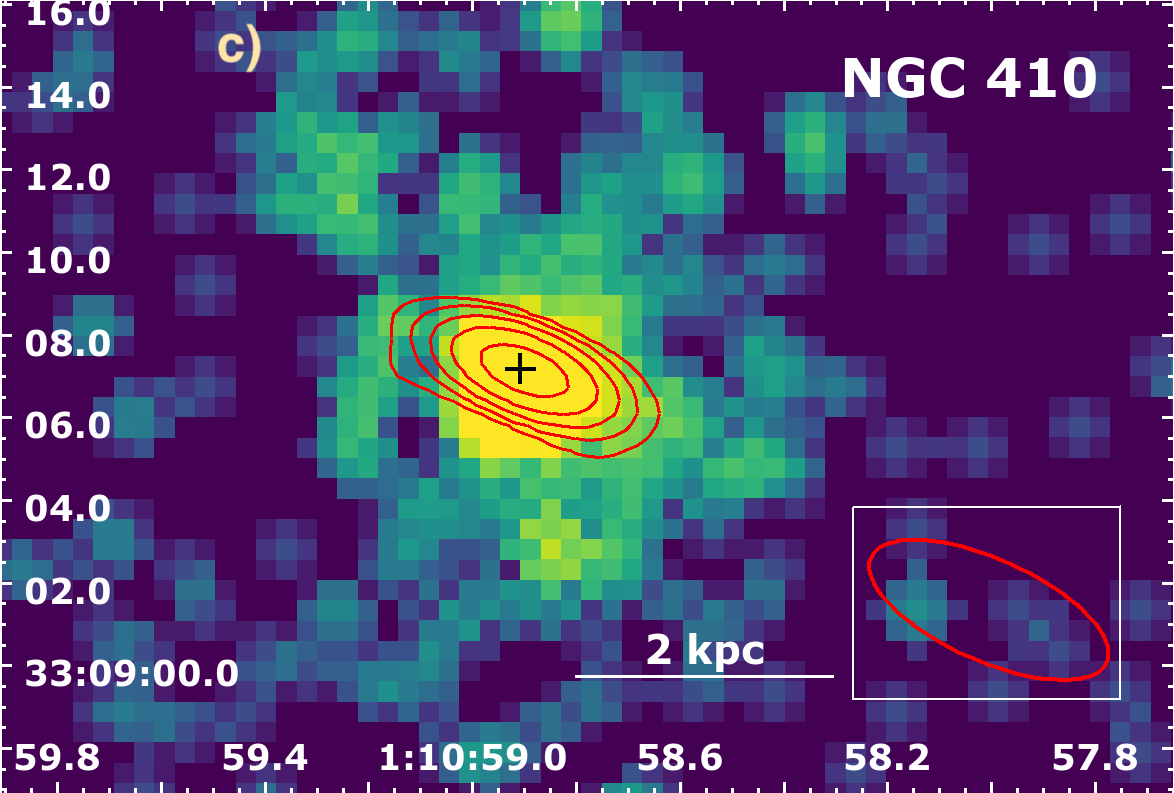}
    \includegraphics[width=230pt]{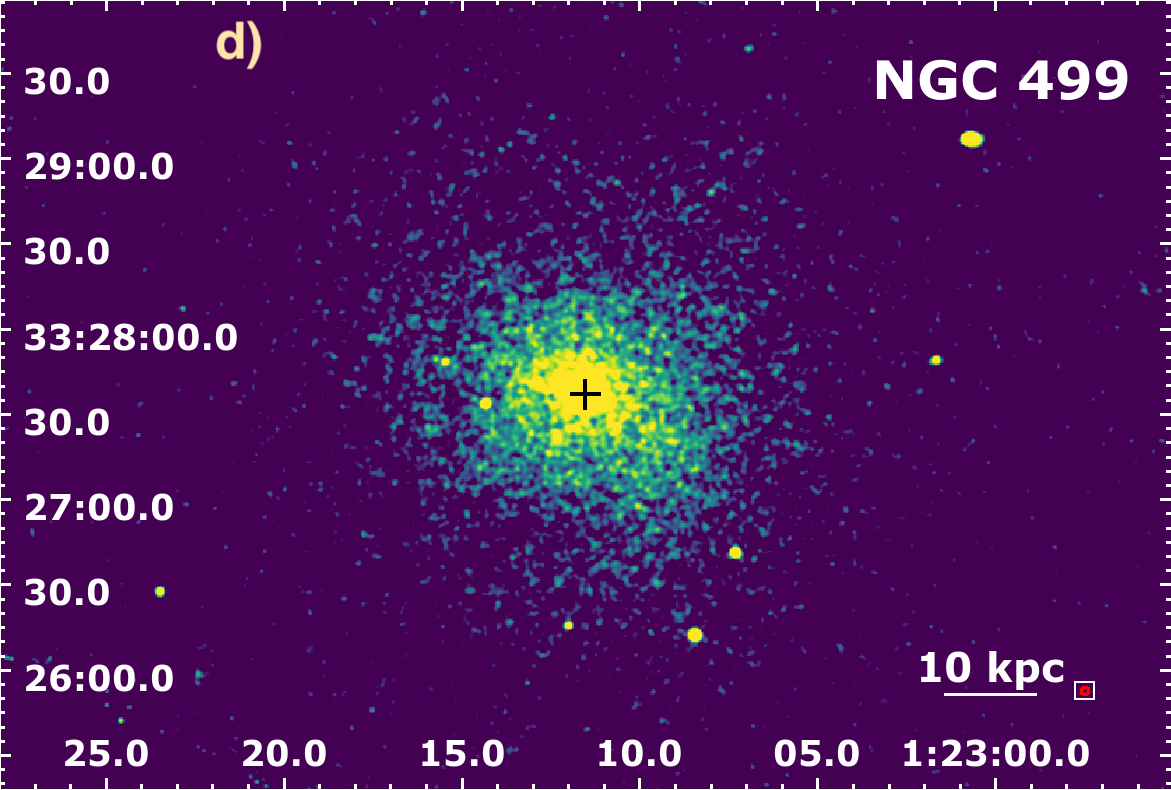}
    \includegraphics[width=230pt]{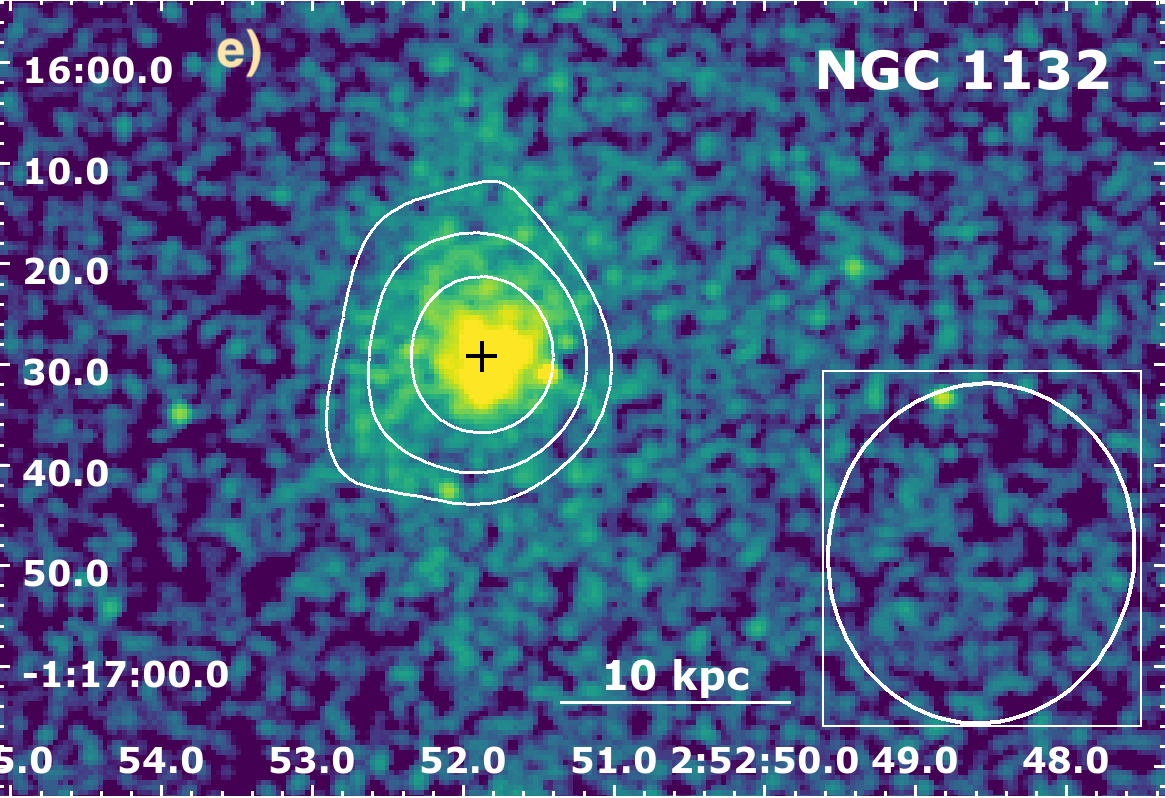}
    \includegraphics[width=230pt]{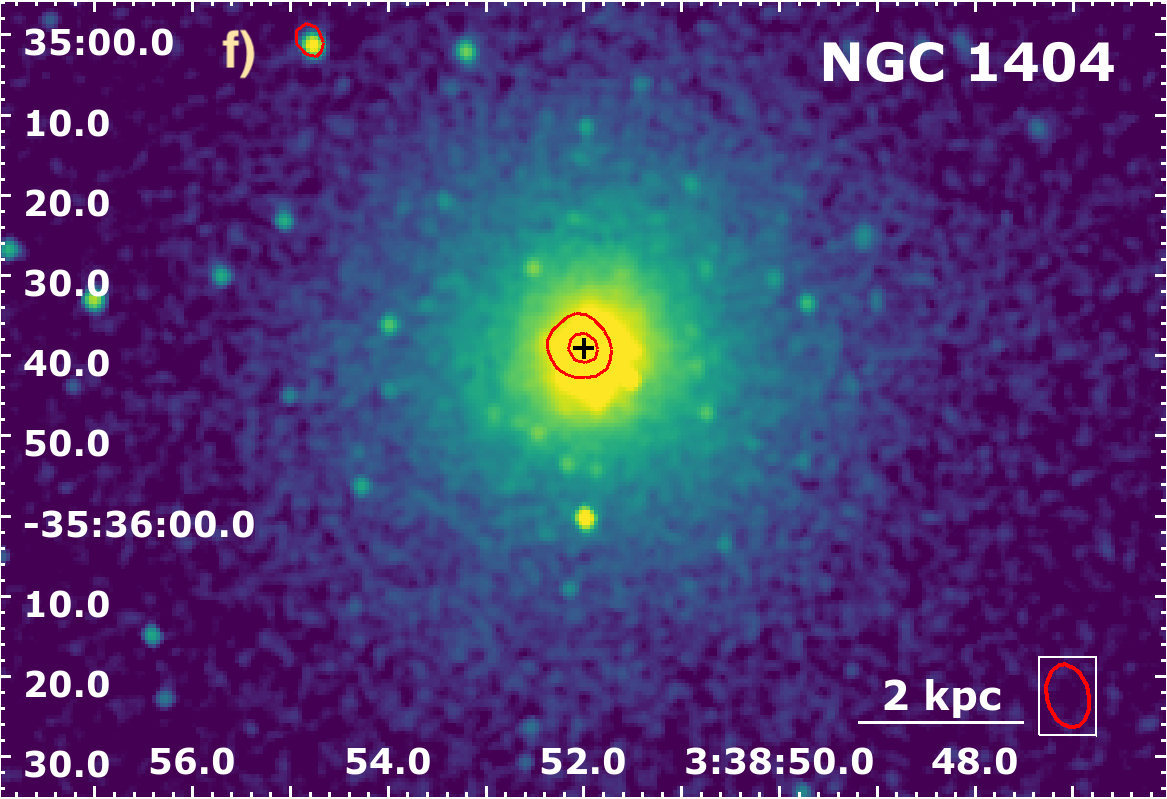}
    \includegraphics[width=230pt]{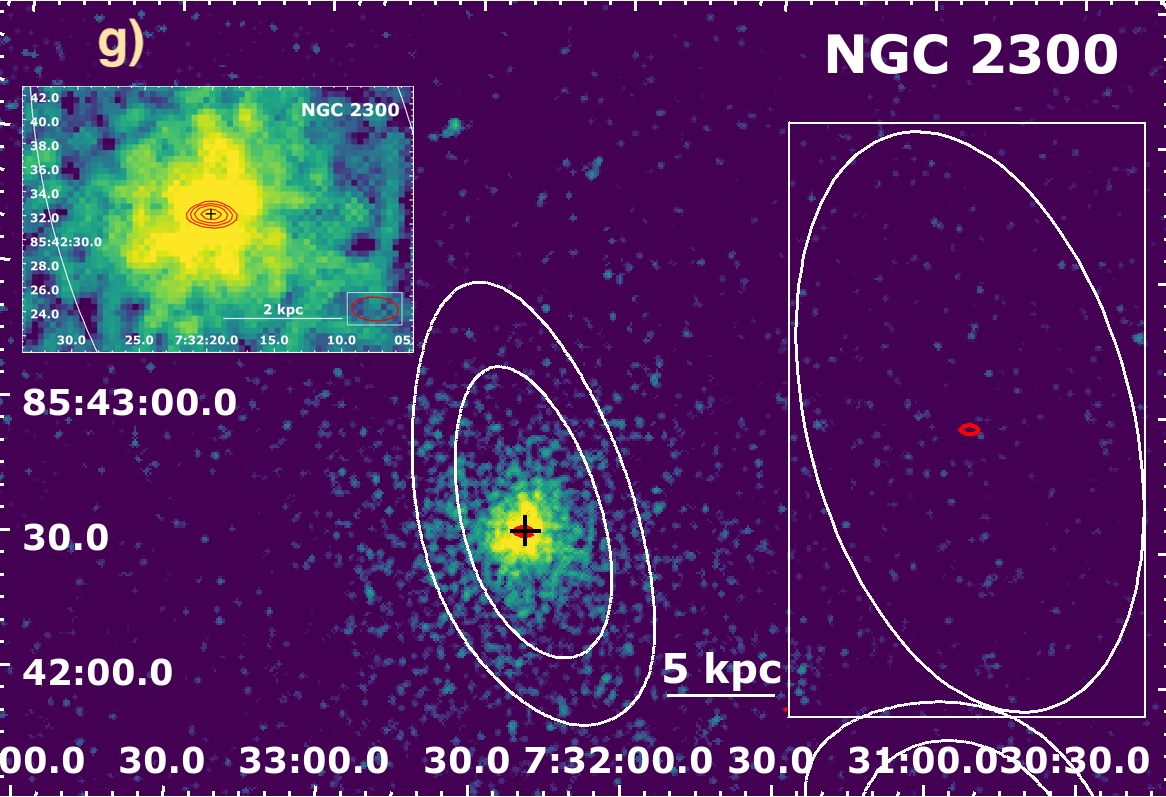}
    \includegraphics[width=230pt]{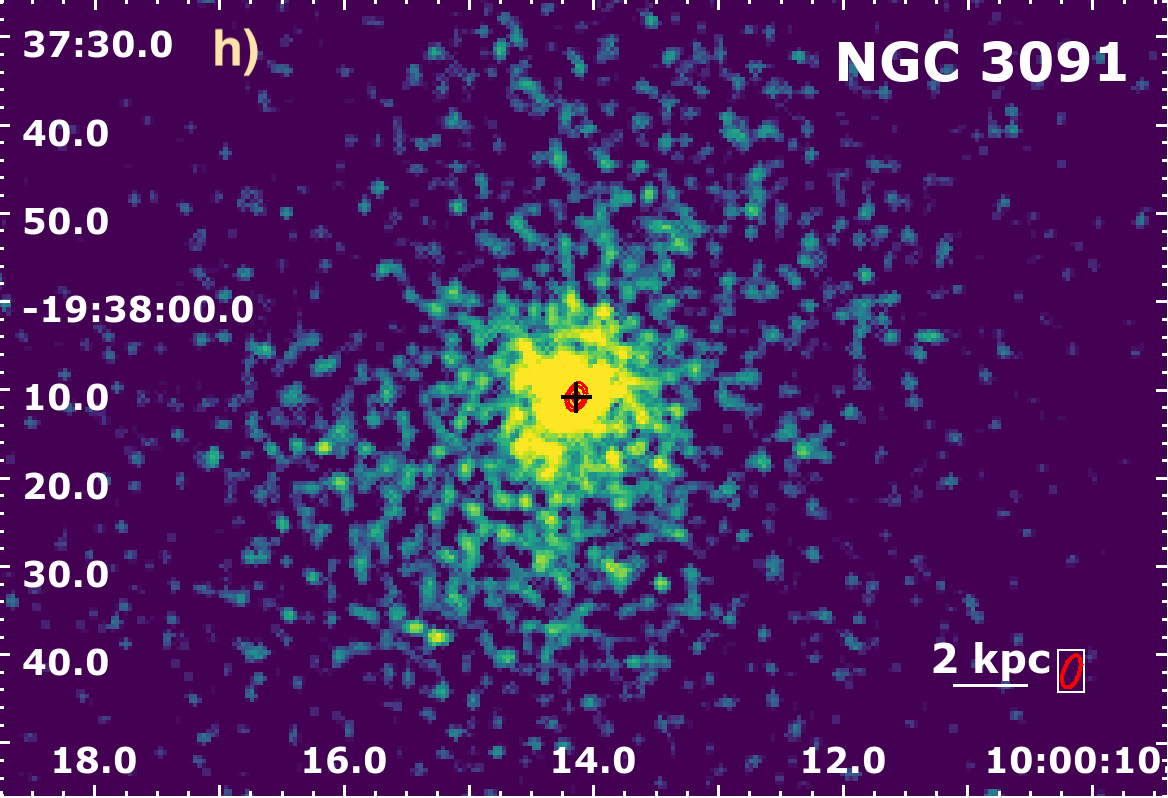}
    
     \caption{The {\it Chandra} smoothed (with a 2\,arcseconds Gaussian kernel) images overlaid by 1.5\,GHz VLA A, AB/B and C/D configuration contours in the red, cyan and white, respectively. In all cases, the contours are created at $5\,\times\,\sigma_{\rm RMS}$ and increase by the power of 2 up to the peak intensity. The center of the galaxy is represented by a black `+' sign. RMS noise ($\sigma_{\rm RMS}$) and peak intensity values are given in Table\,\ref{tab:results}.}
    \label{fig:chandra_vla_ps_a}
   \centering
\end{figure}

\renewcommand{\thefigure}{B1.2}
\begin{figure}
    \centering
    \includegraphics[width=230pt]{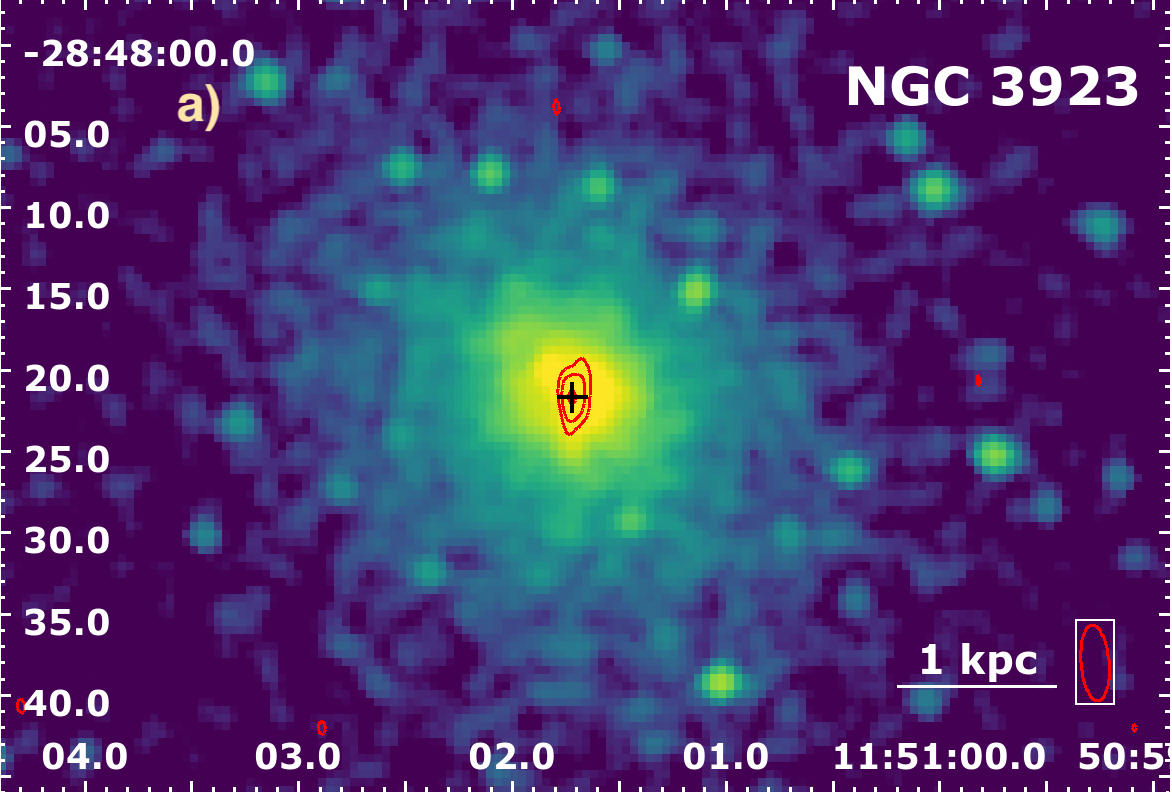}
    \includegraphics[width=230pt]{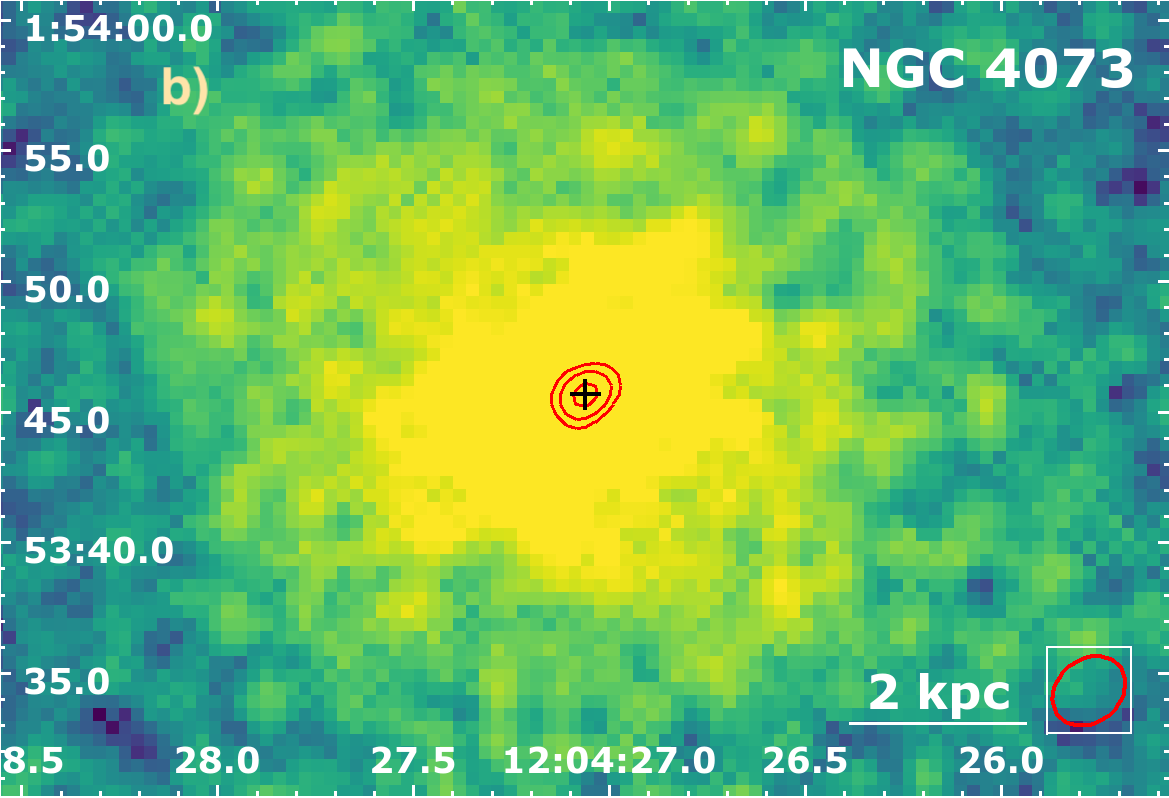}
    \includegraphics[width=230pt]{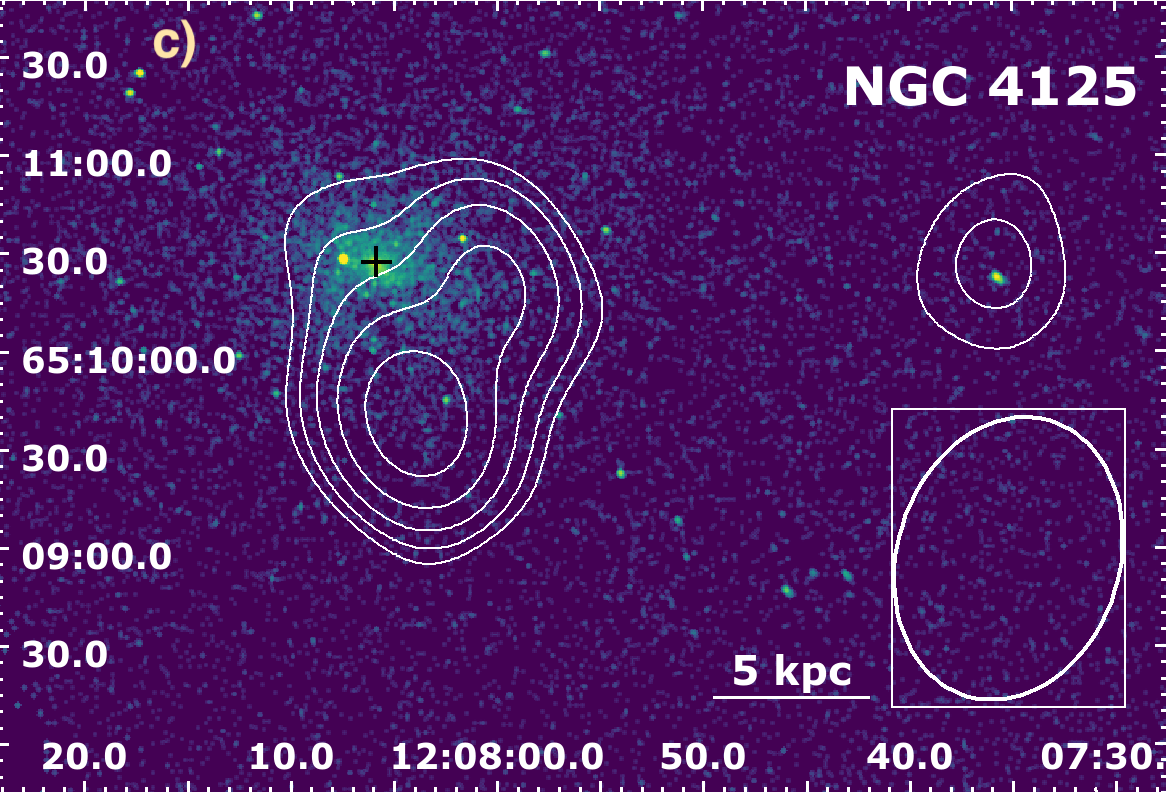}
    \includegraphics[width=230pt]{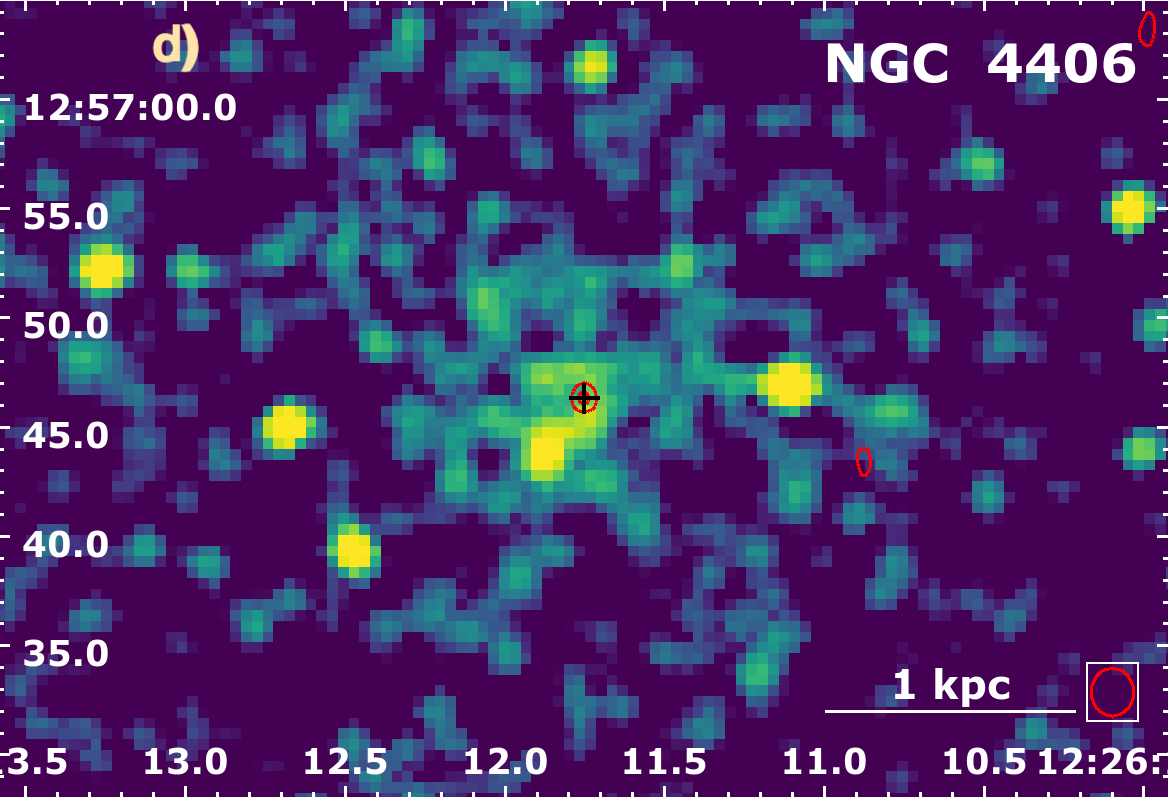}
    \includegraphics[width=230pt]{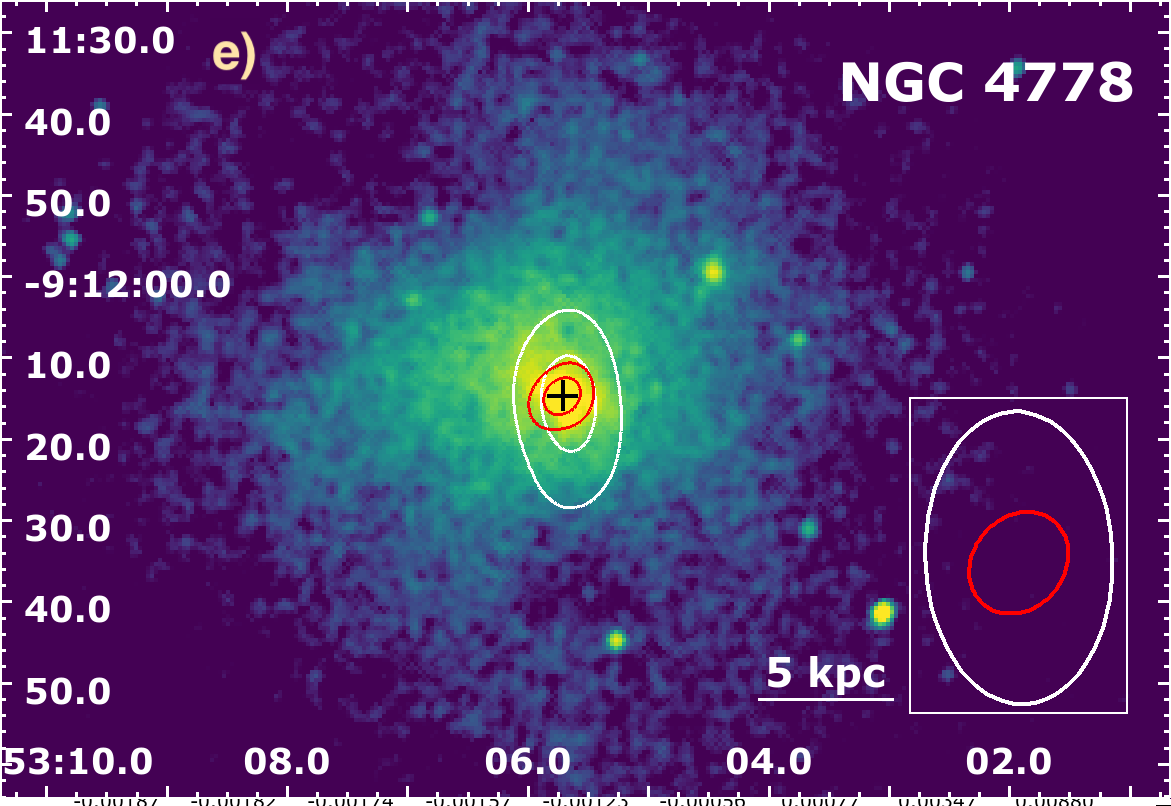}
    \includegraphics[width=230pt]{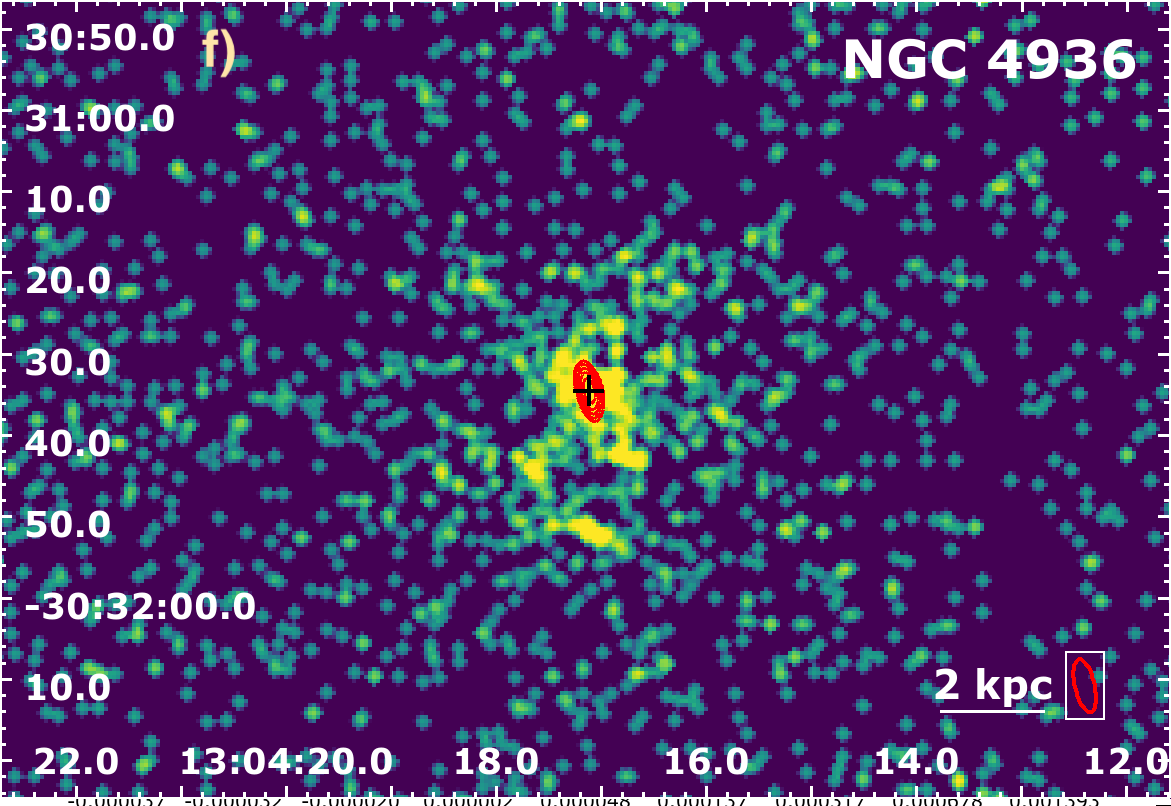}
    \includegraphics[width=230pt]{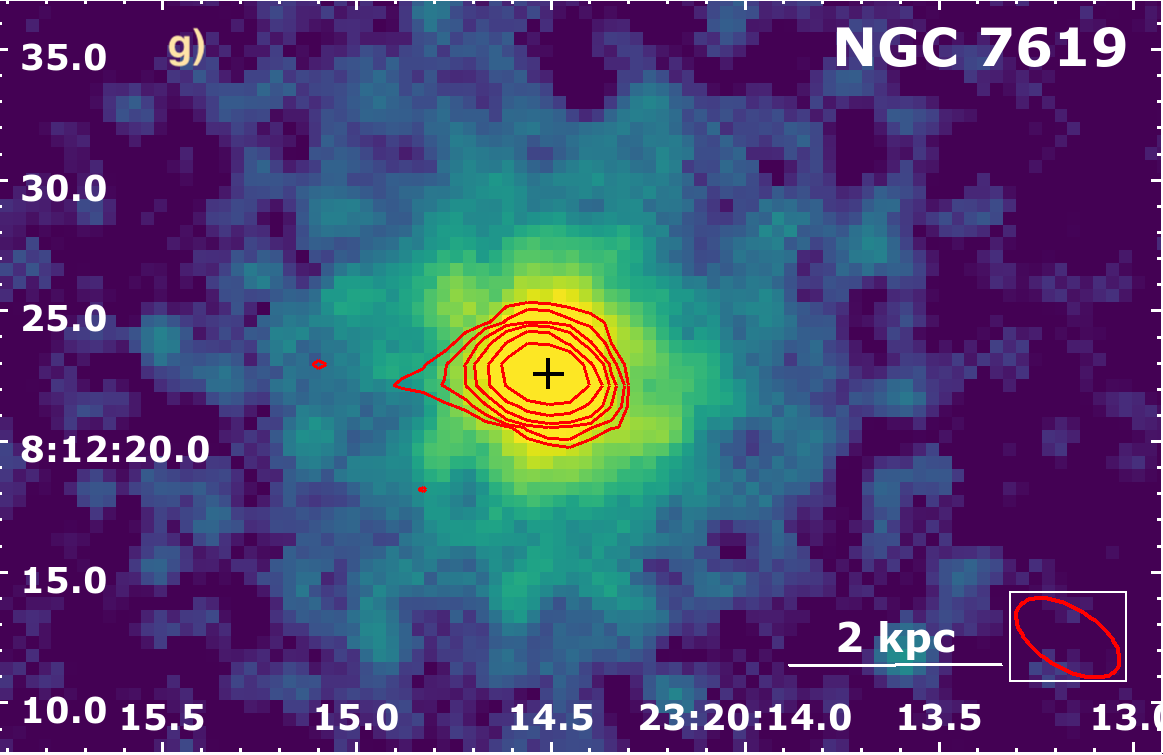}
     \caption{The X-ray {\it Chandra} smoothed (with a 2\,arcseconds Gaussian kernel) images overlaid by 1.5\,GHz VLA A, AB/B and C/D configuration contours in the red, cyan and white, respectively. In all cases, the contours are created at $5\,\times\,\sigma_{\rm RMS}$ and increase by the power of 2 up to the peak intensity. The center of the galaxy is represented by a black `+' sign. RMS noise ($\sigma_{\rm RMS}$) and peak intensity values are given in Table\,\ref{tab:results}.}
    \label{fig:chandra_vla_ps_b}
   \centering
\end{figure}

\renewcommand{\thefigure}{B2.1}
\begin{figure}
\subsection{ X-rays and extended radio sources}
\label{app:xray_radio_ext}
    \centering
    \includegraphics[width=251pt]{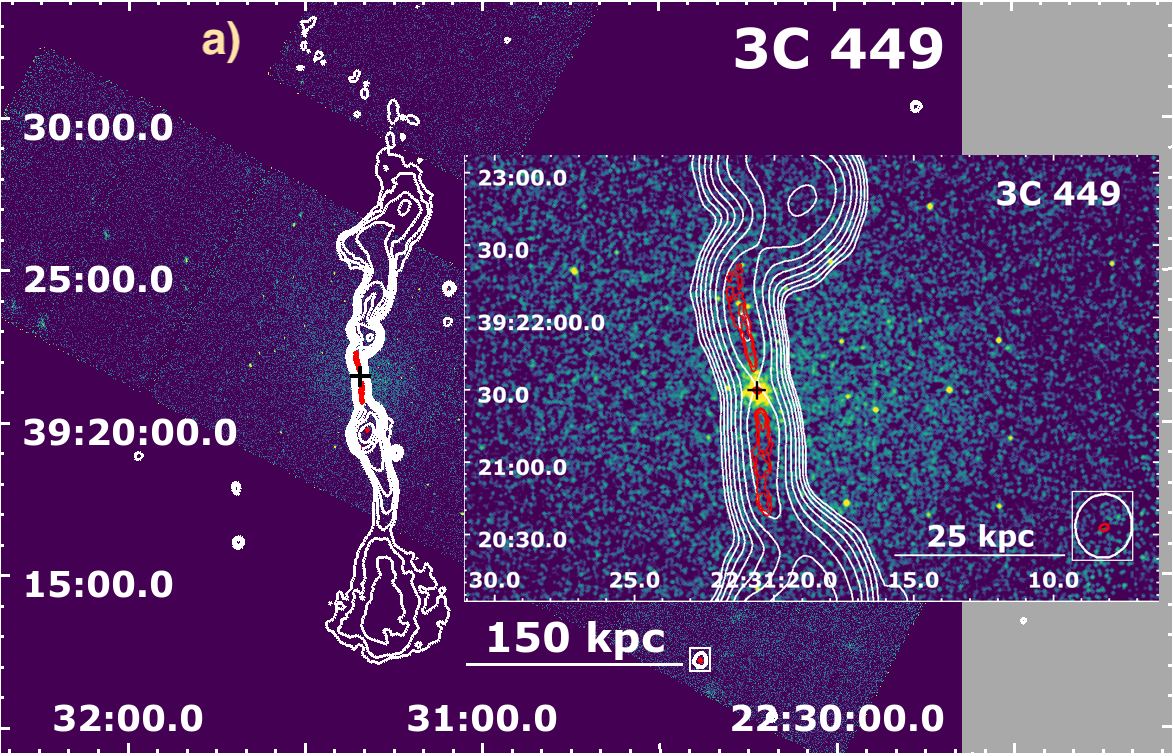}
    \includegraphics[width=250pt]{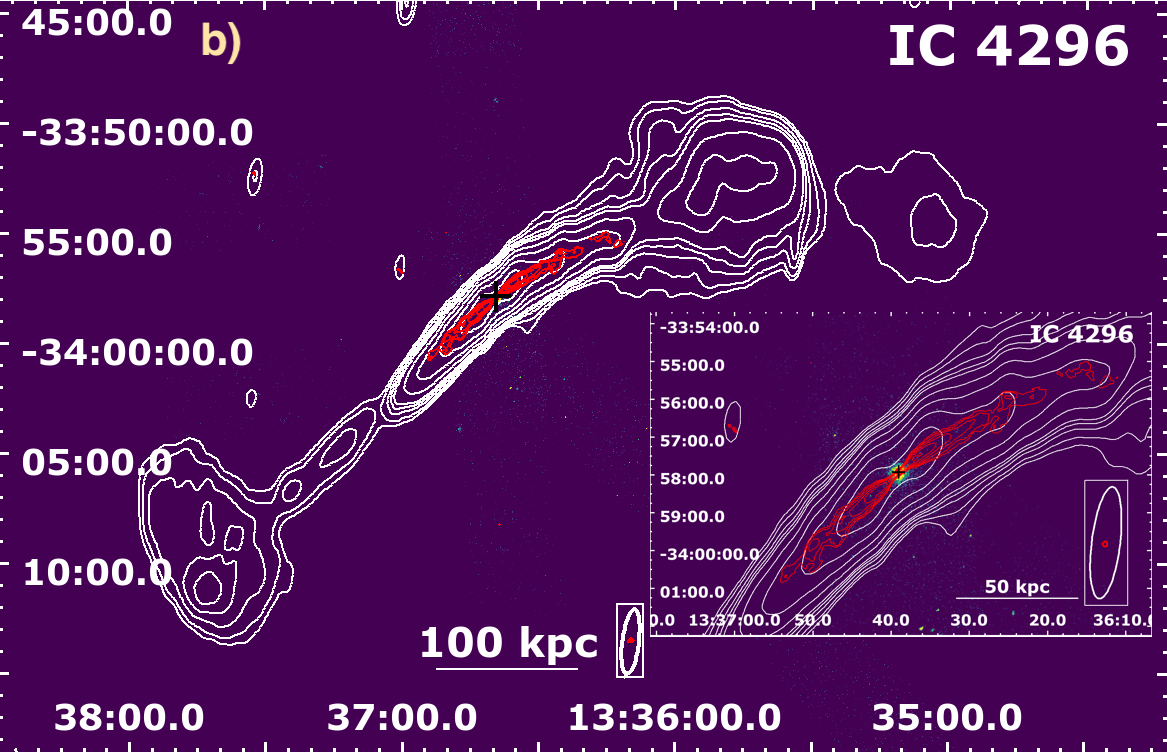}
    \includegraphics[width=252pt]{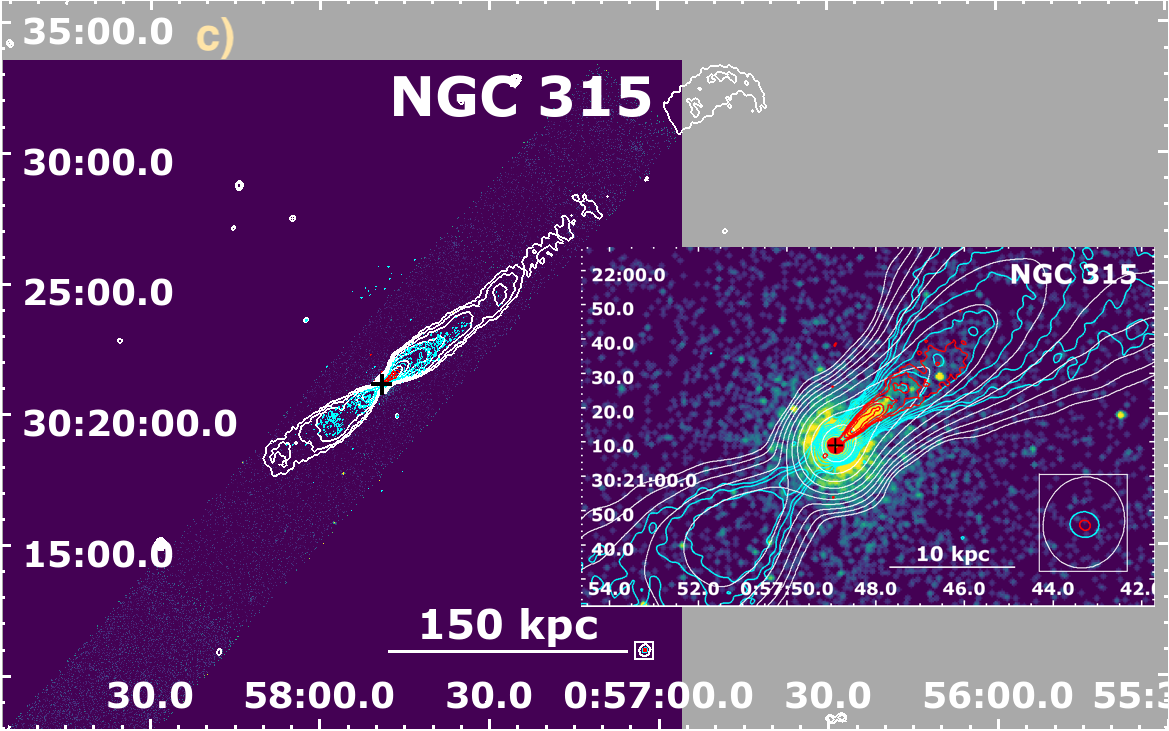}
    \includegraphics[width=250pt]{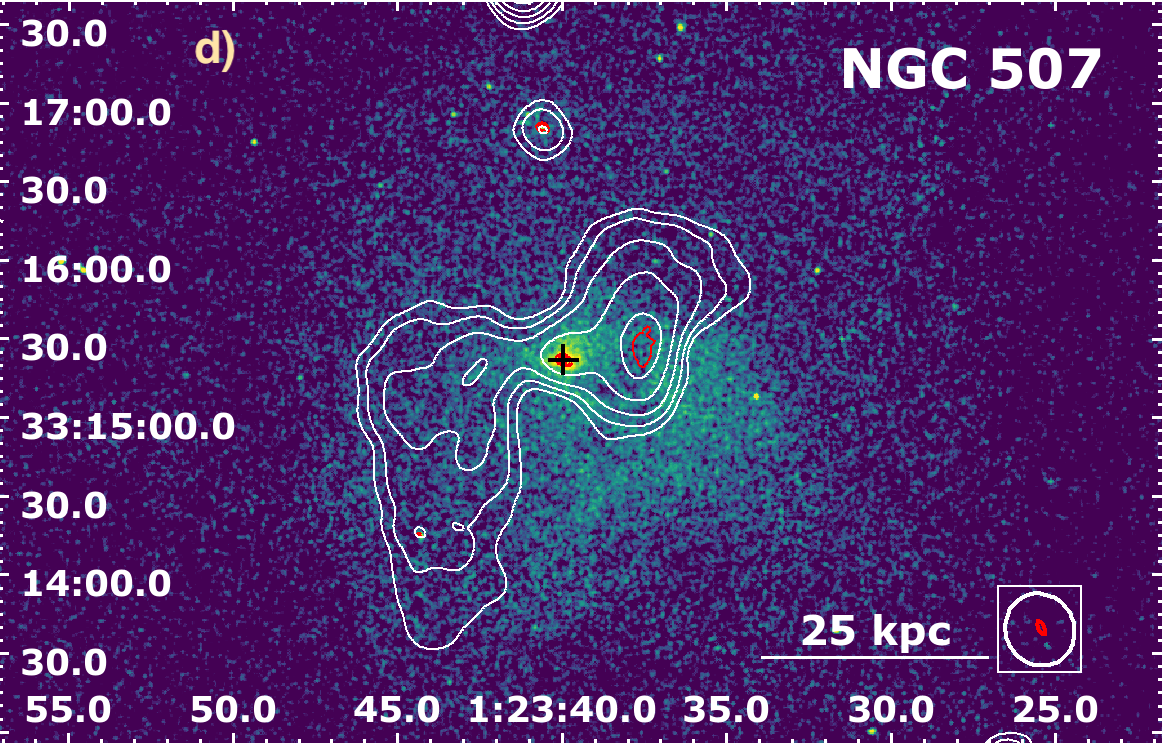}
    \includegraphics[width=250pt]{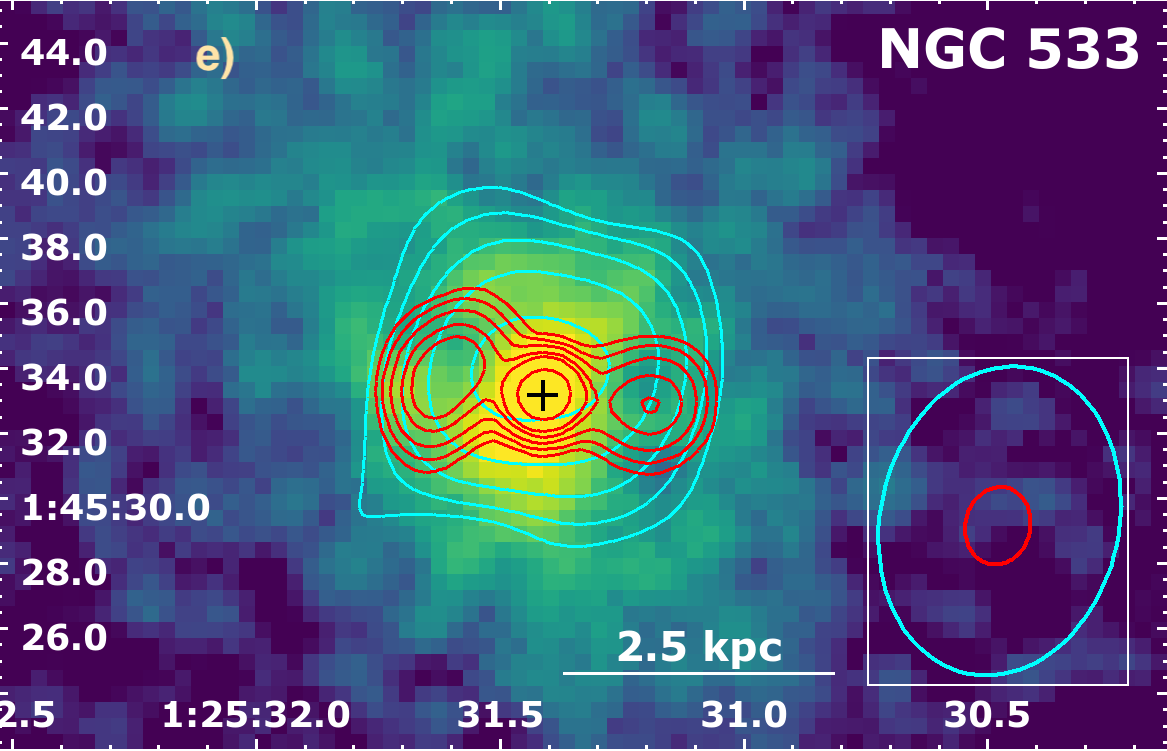}
    \includegraphics[width=250pt]{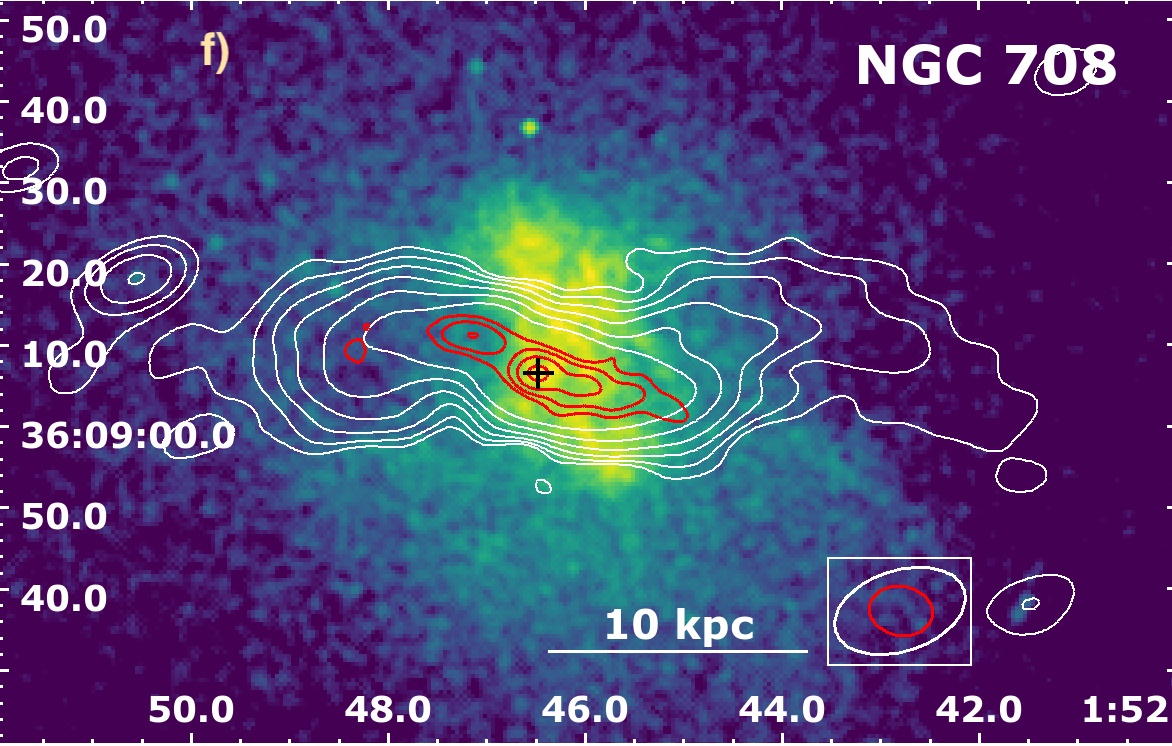}
    \includegraphics[width=250pt]{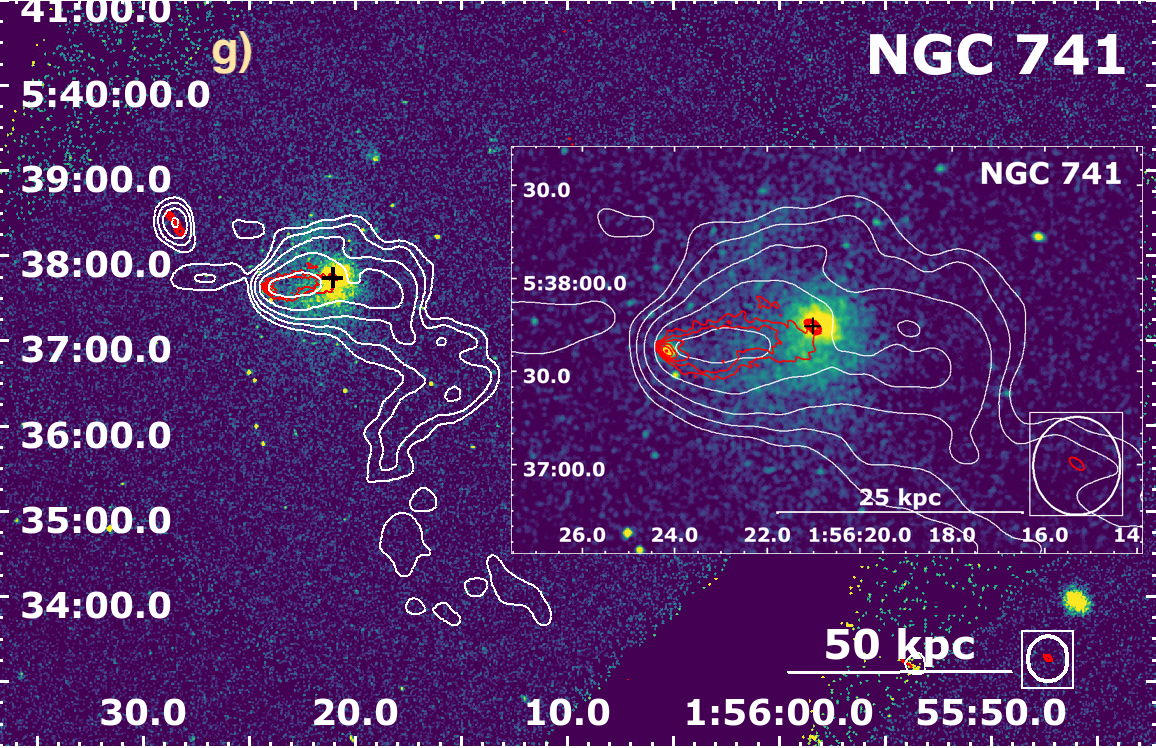}
    \includegraphics[width=250pt]{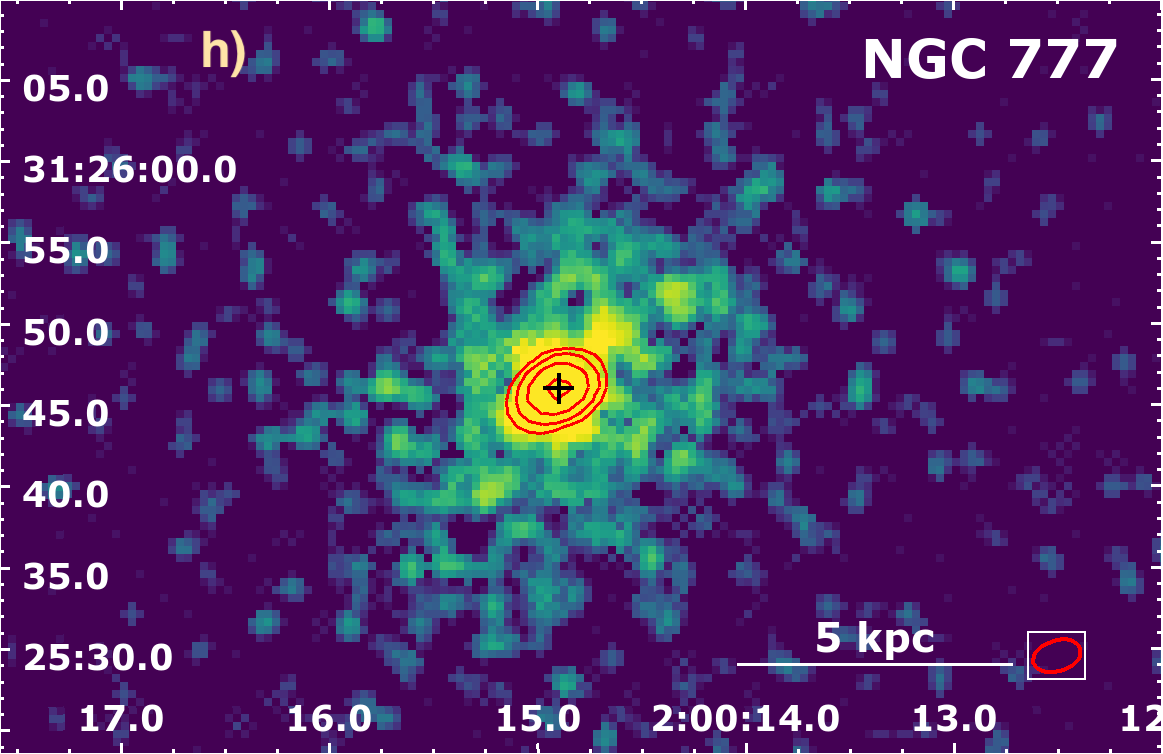}
     \caption{The X-ray {\it Chandra} smoothed (with a 2\,arcseconds Gaussian kernel)  images overlaid by 1.5\,GHz VLA A, AB/B and C/D configuration contours in the red, cyan and white, respectively. In all cases, the contours are created at $5\,\times\,\sigma_{\rm RMS}$ and increase by the power of 2 up to the peak intensity. The center of the galaxy is represented by a black `+' sign. RMS noise ($\sigma_{\rm RMS}$) and peak intensity values are given in Table\,\ref{tab:results}.}
    \label{fig:chandra_vla_a}
   \centering
\end{figure}

\renewcommand{\thefigure}{B2.2}
 \begin{figure}
    \centering
    
    \includegraphics[width=250pt]{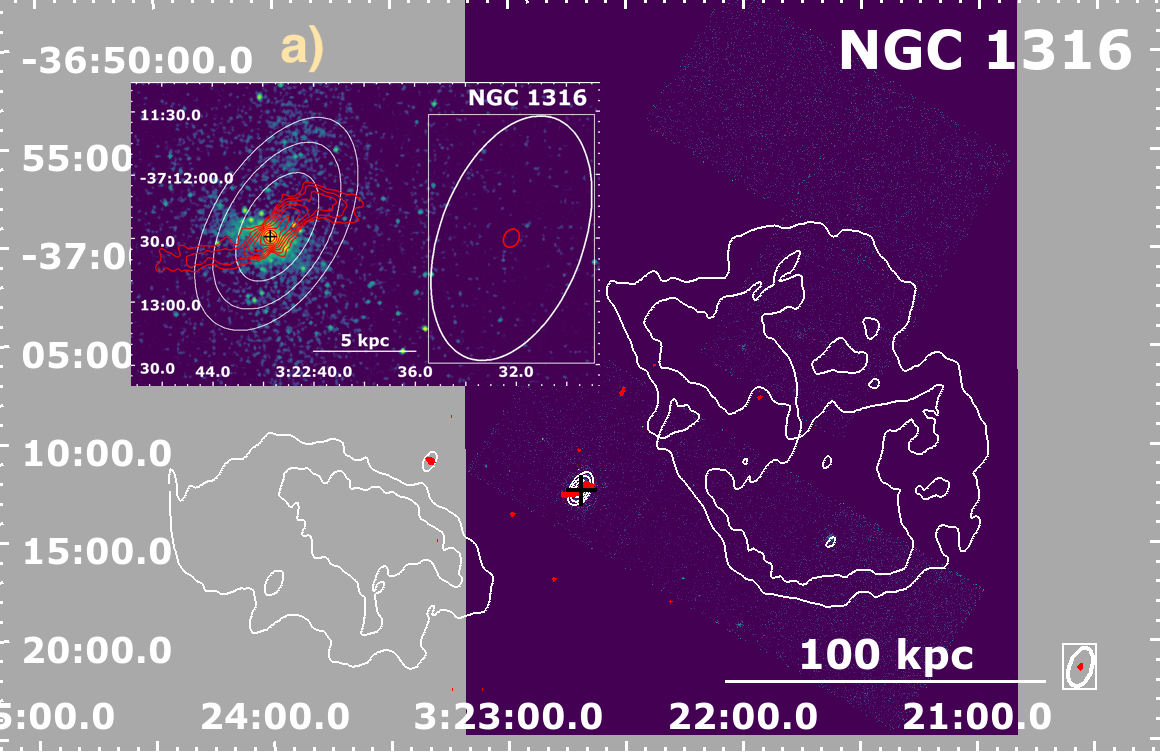}
    \includegraphics[width=250pt]{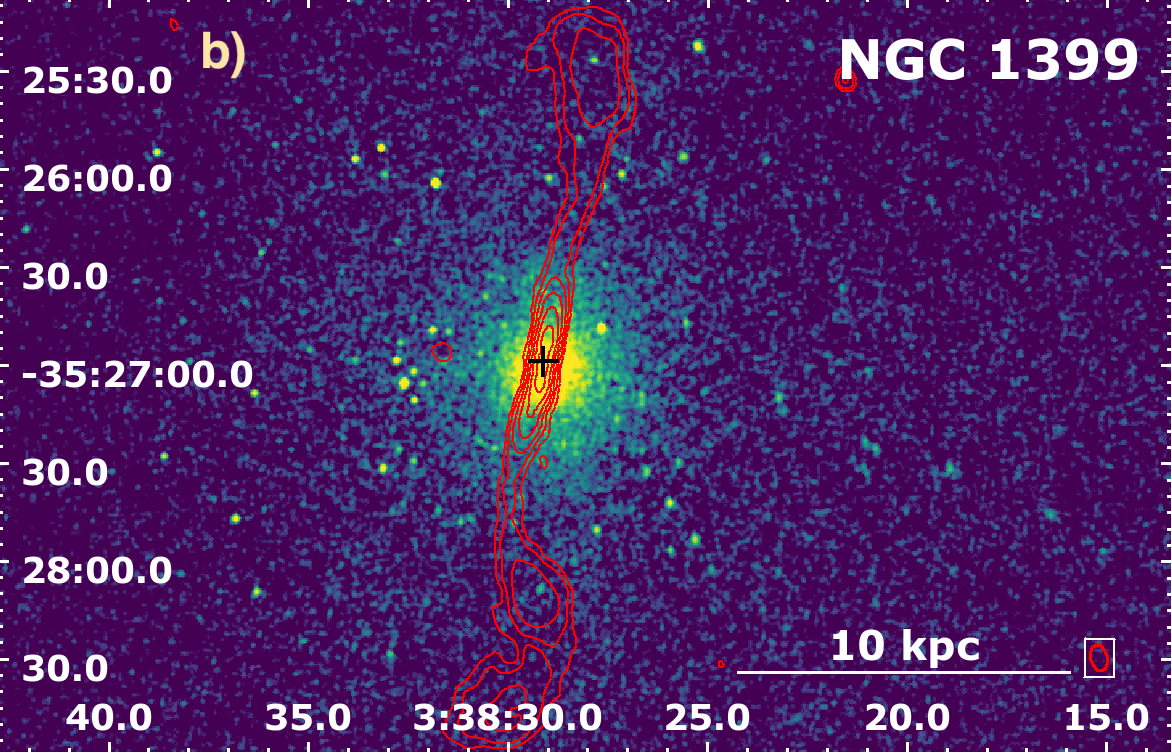}
    \includegraphics[width=250pt]{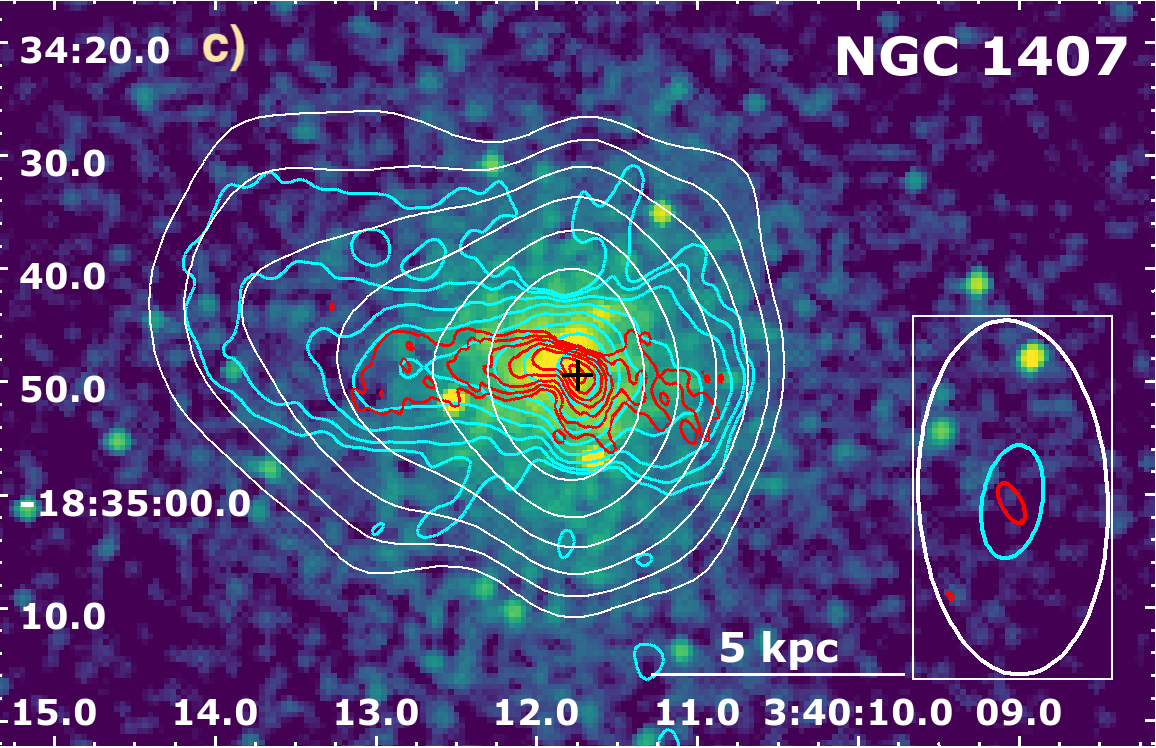}
    \includegraphics[width=250pt]{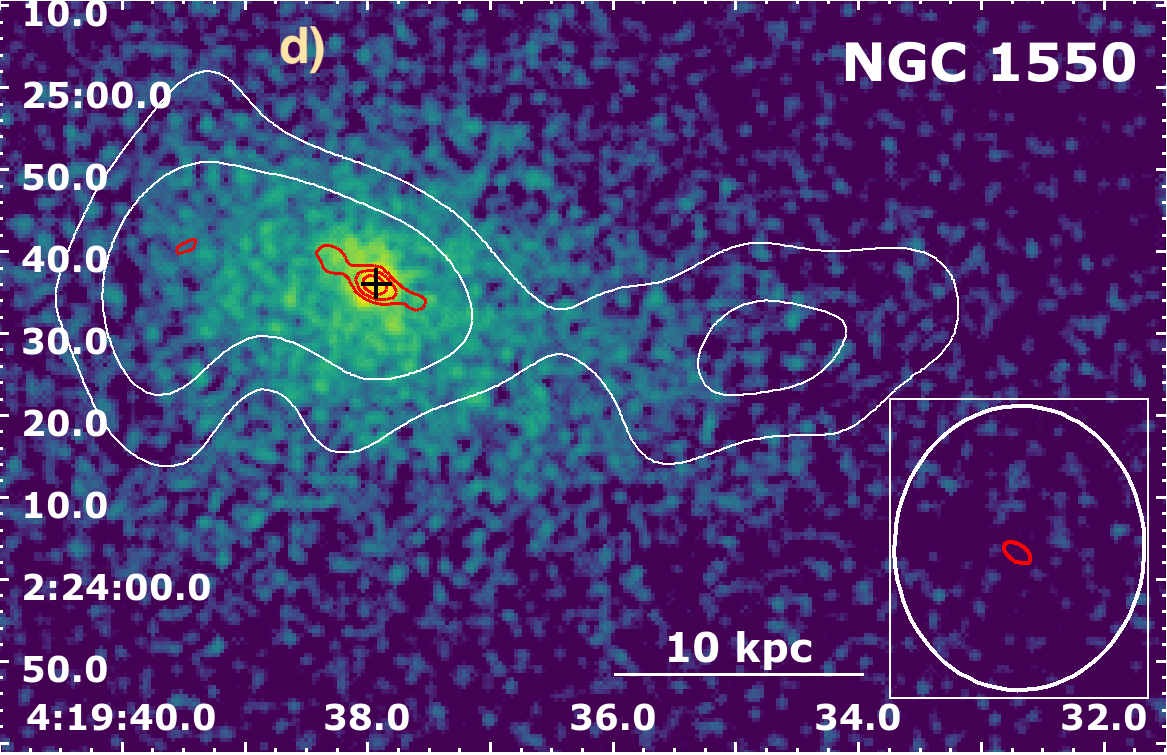}
    \includegraphics[width=250pt]{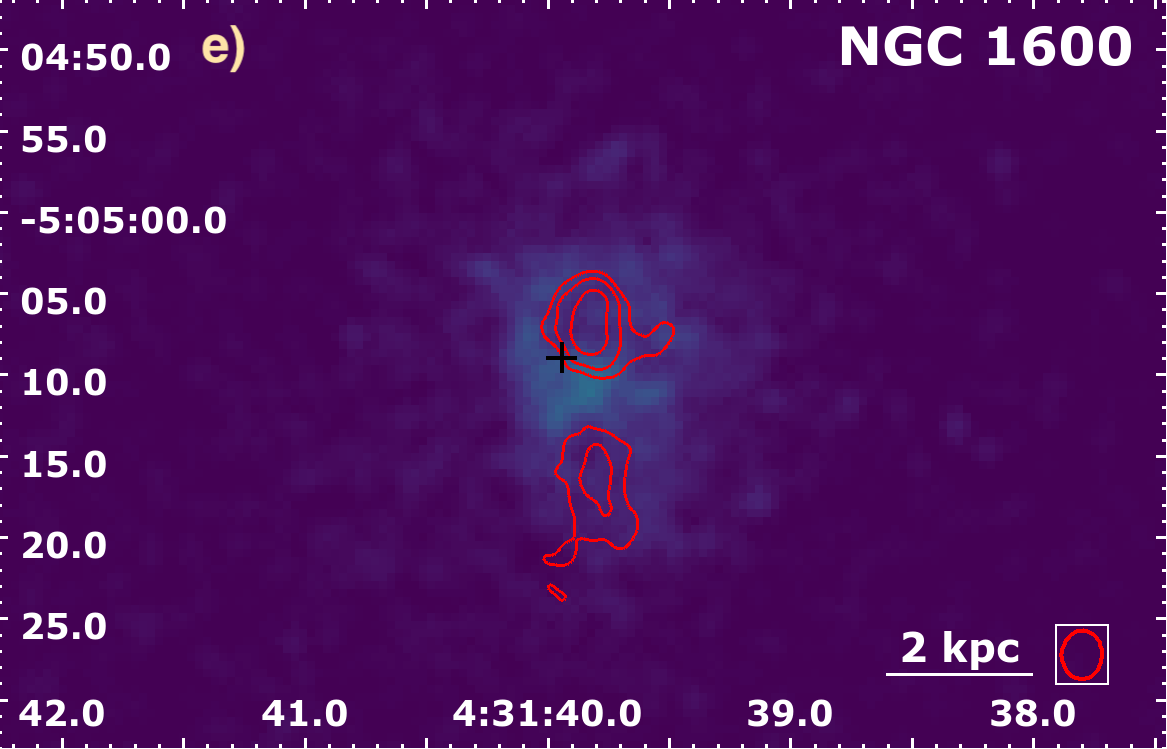}
    \includegraphics[width=250pt]{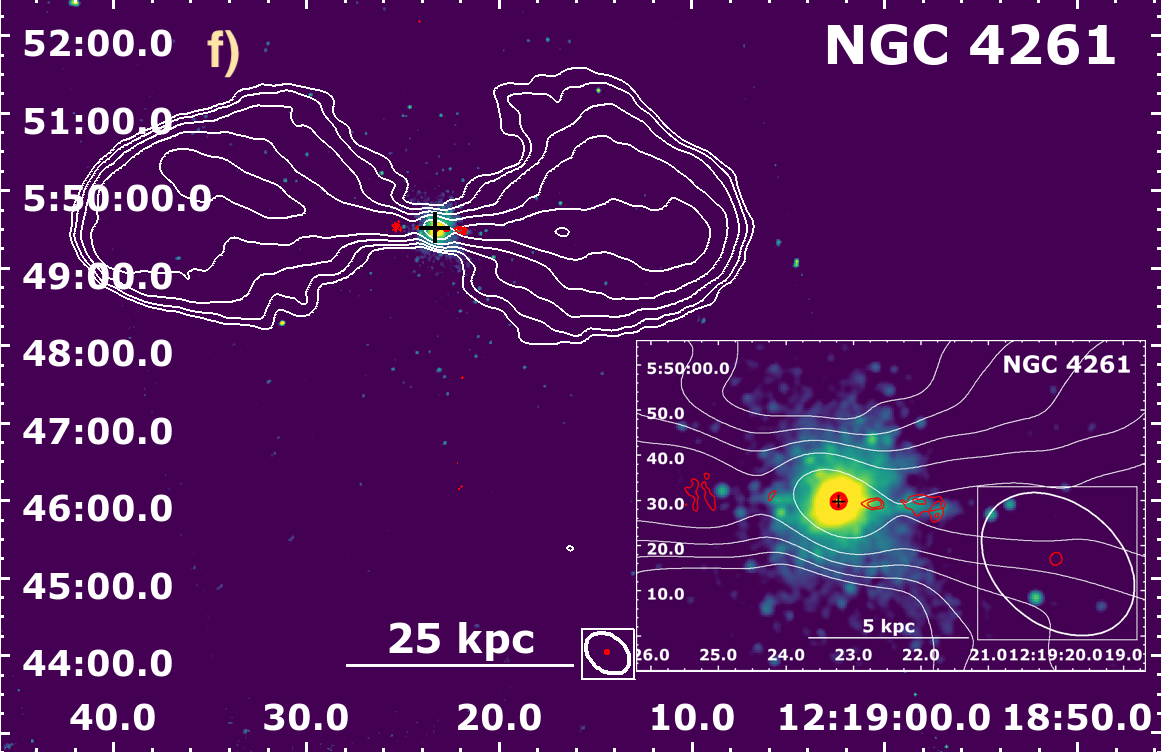}
    \includegraphics[width=250pt]{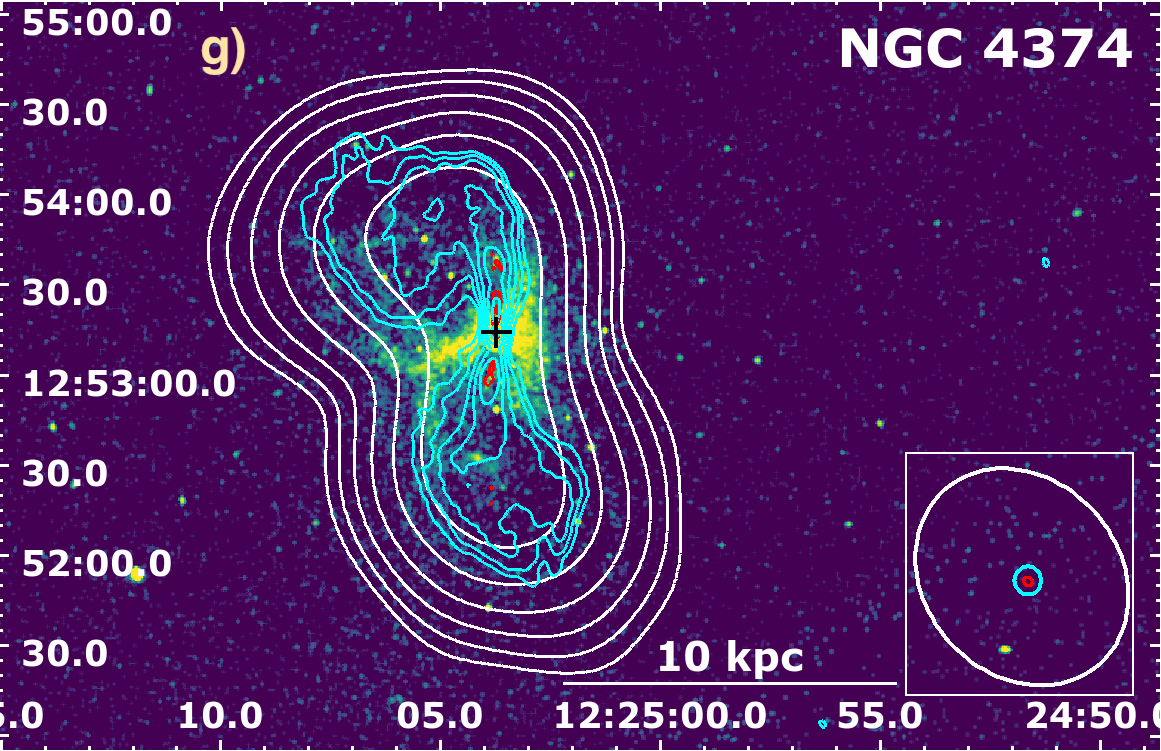}
    \includegraphics[width=250pt]{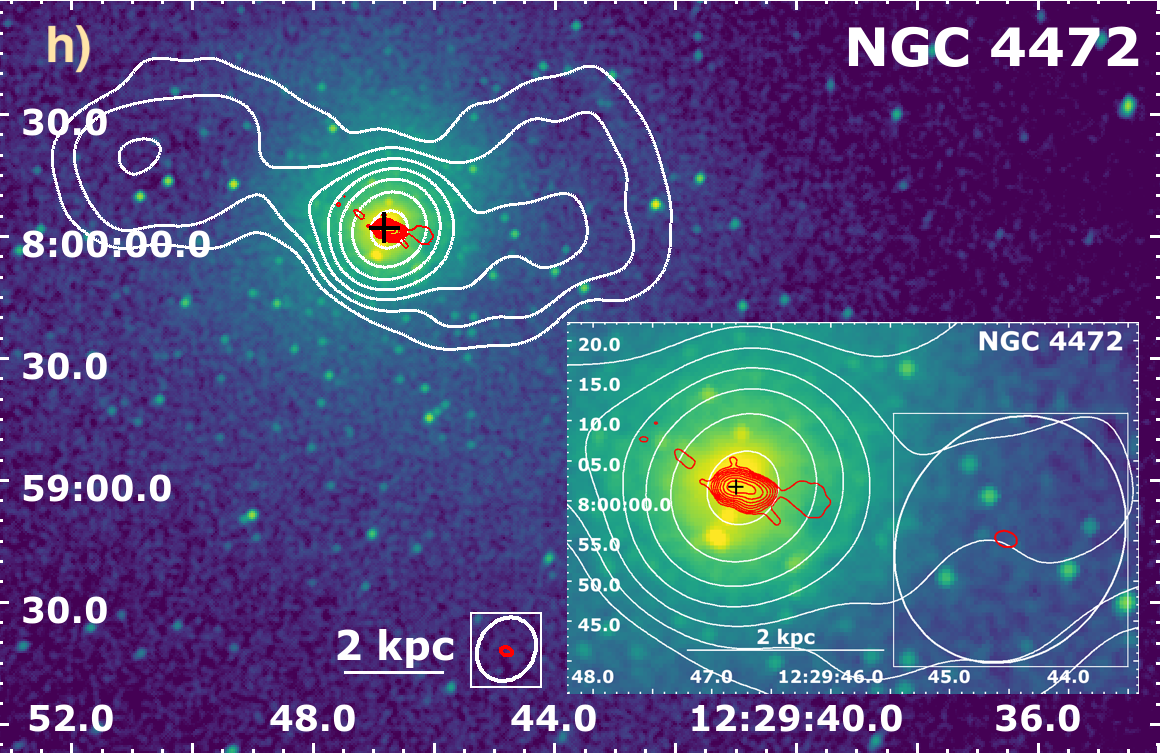}
     \caption{The X-ray {\it Chandra} smoothed (with a 2\,arcseconds Gaussian kernel)  images overlaid by 1.5\,GHz VLA A, AB/B and C/D configuration contours in the red, cyan and white, respectively. In all cases, the contours are created at $5\,\times\,\sigma_{\rm RMS}$ and increase by the power of 2 up to the peak intensity. The center of the galaxy is represented by a black `+' sign. The RMS noise and peak intensity values are given in Table\,\ref{tab:results}.}
    \label{fig:chandra_vla_b}
   \centering
\end{figure}

\renewcommand{\thefigure}{B2.3}
\begin{figure}
    \centering
    \includegraphics[width=250pt]{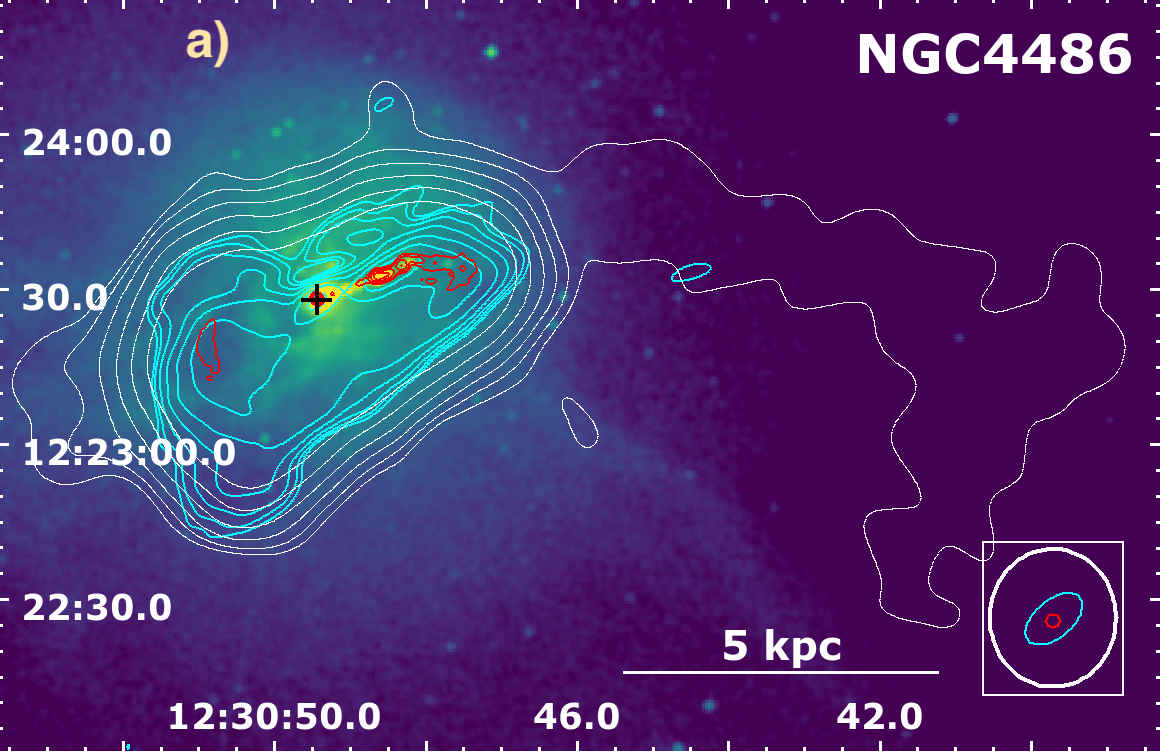}
    \includegraphics[width=250pt]{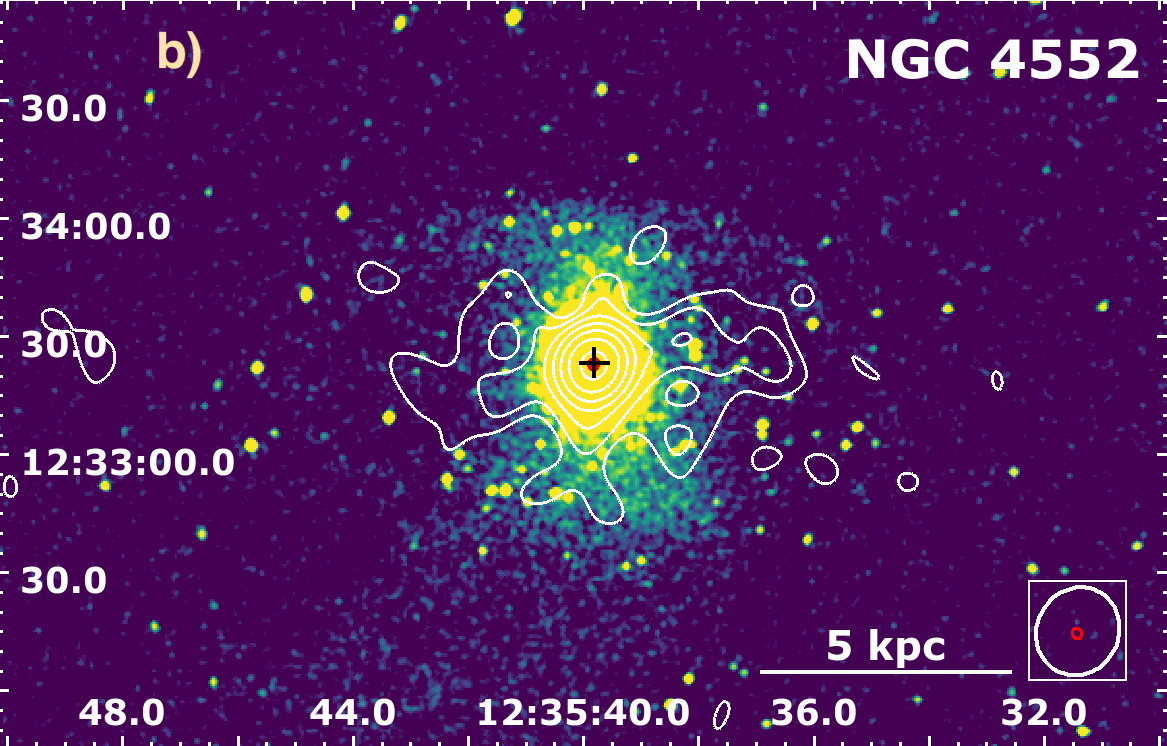}
    \includegraphics[width=250pt]{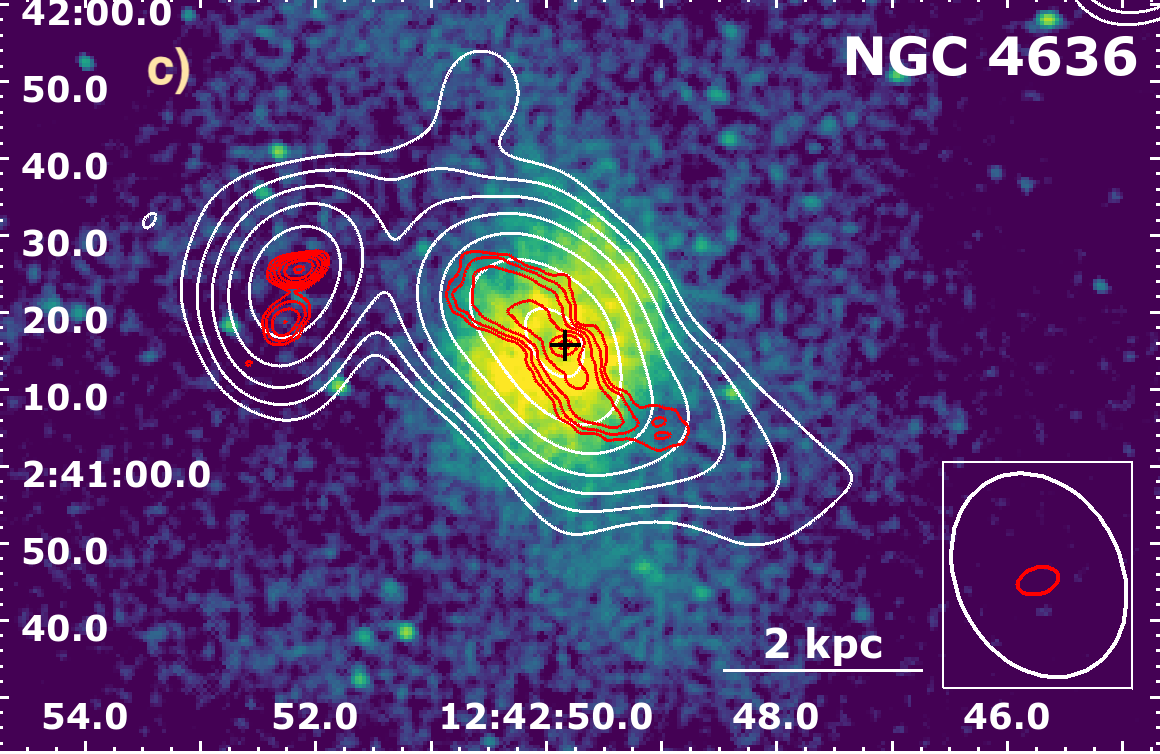}
    \includegraphics[width=250pt]{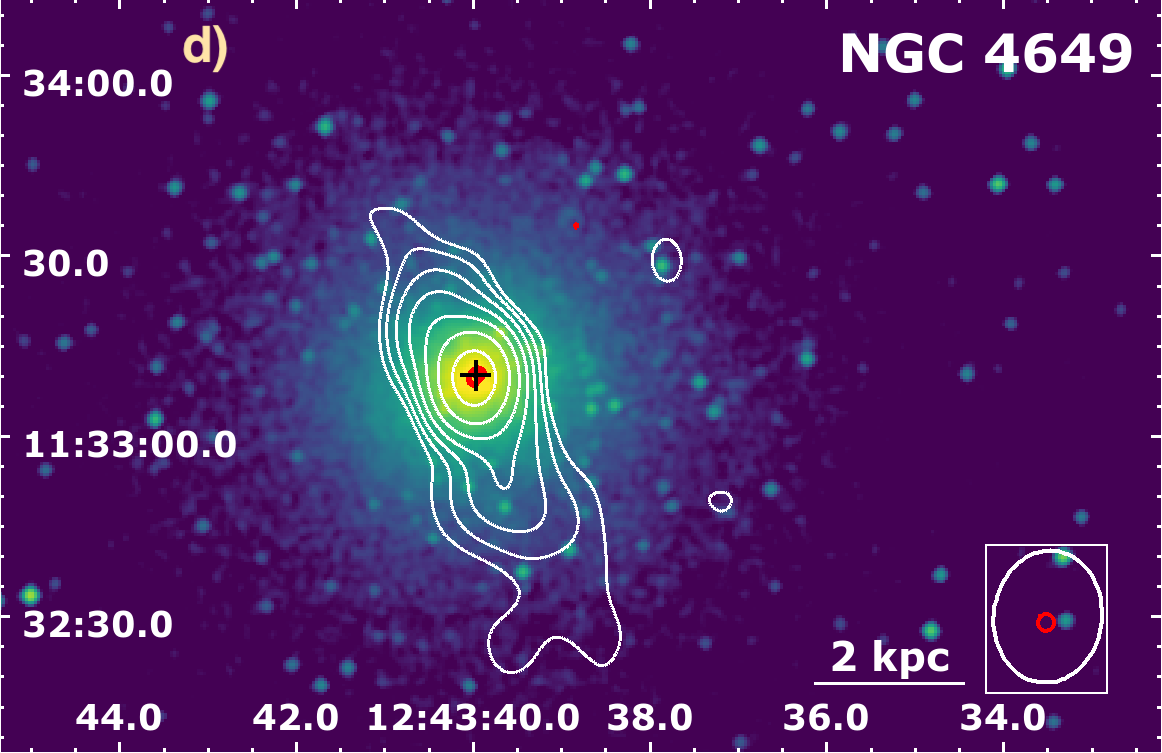}
    \includegraphics[width=250pt]{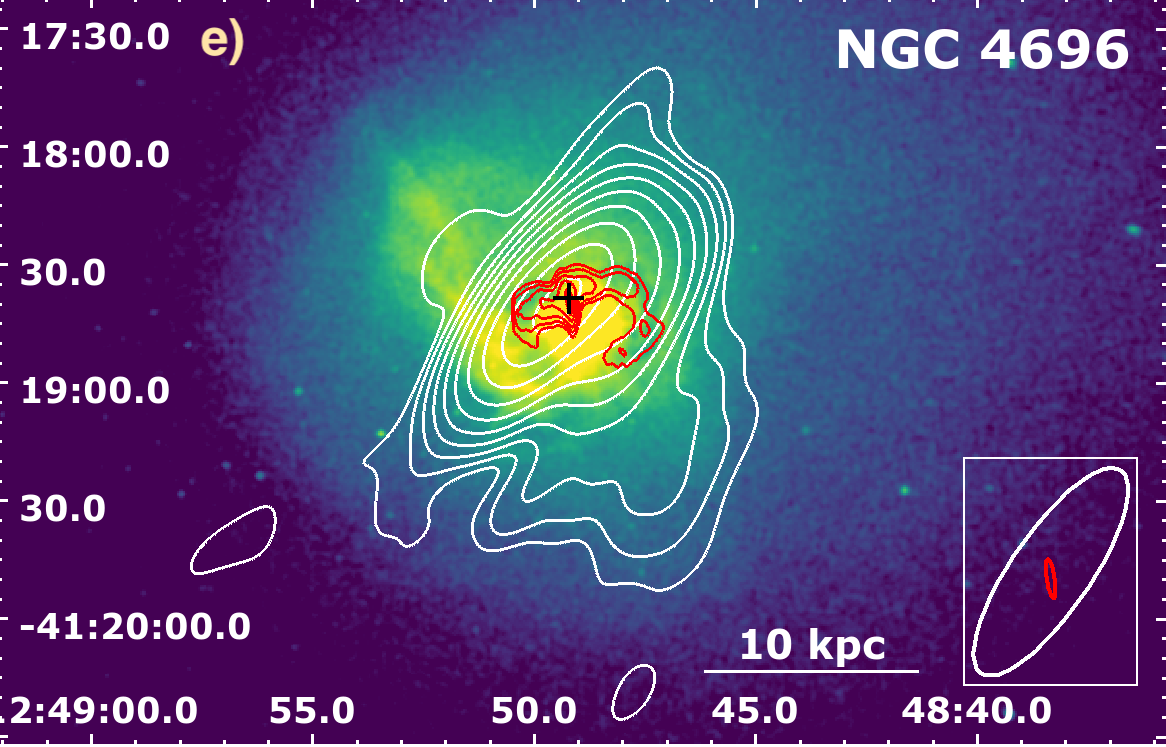}
    \includegraphics[width=250pt]{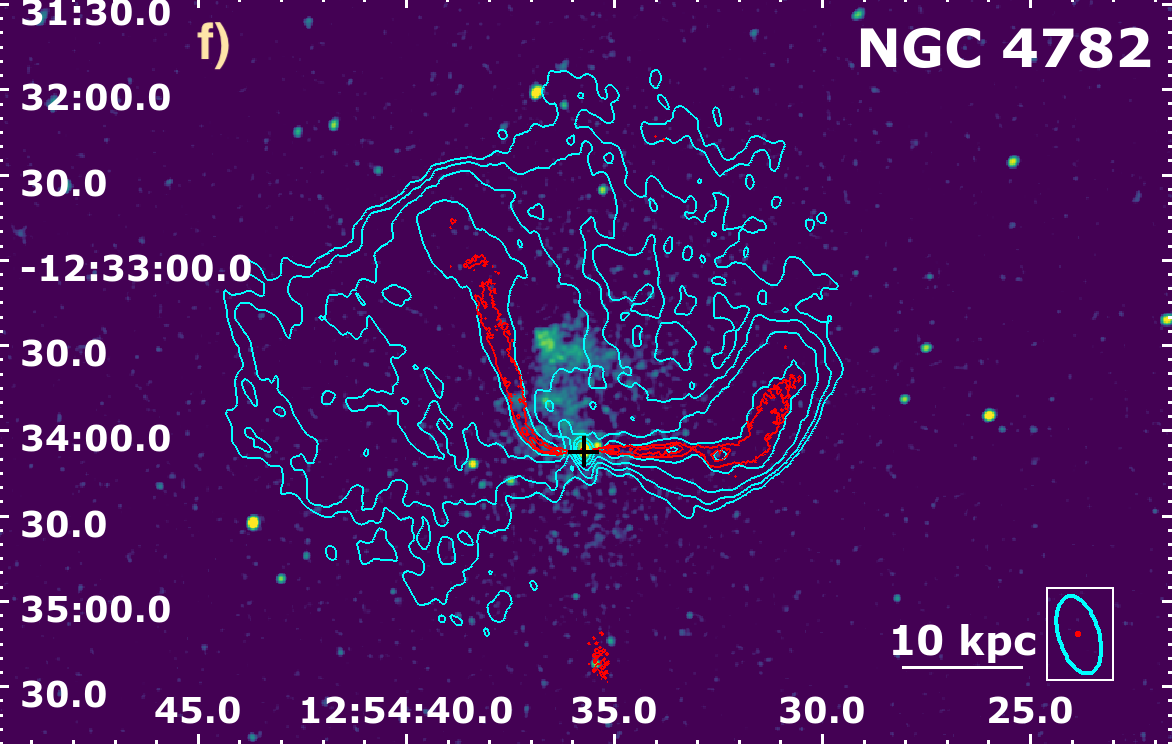}
    \includegraphics[width=250pt]{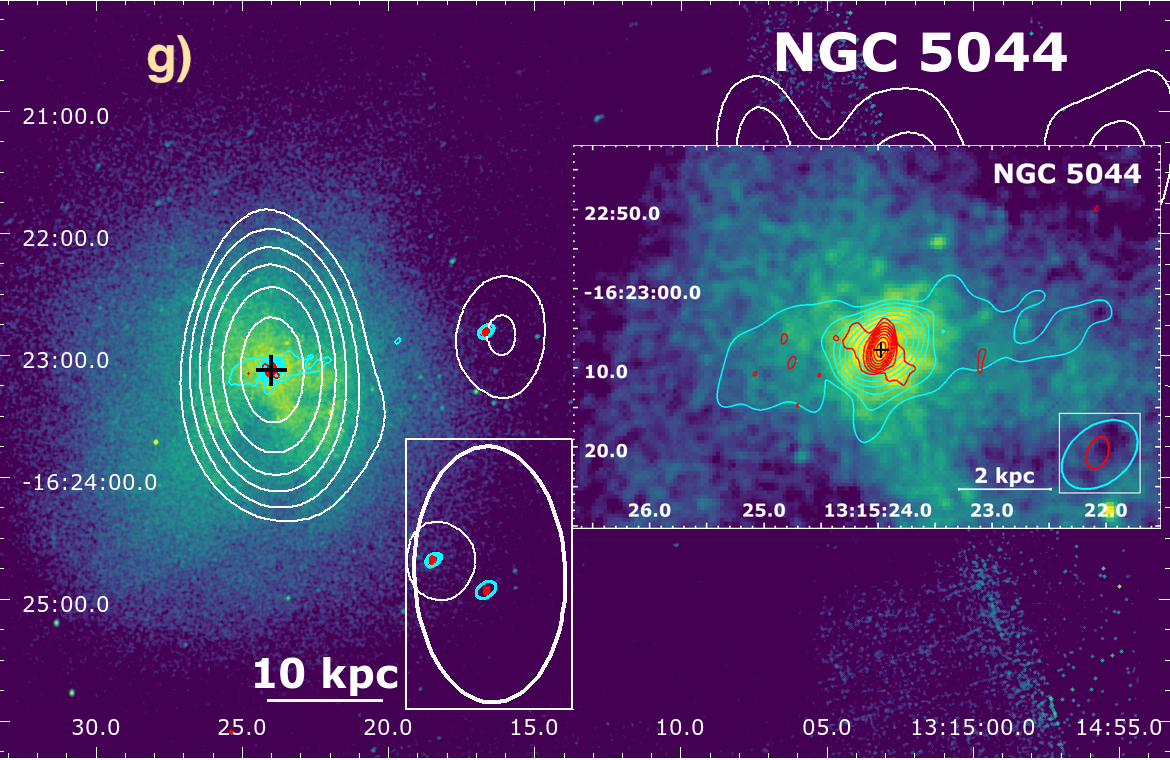}
    \includegraphics[width=250pt]{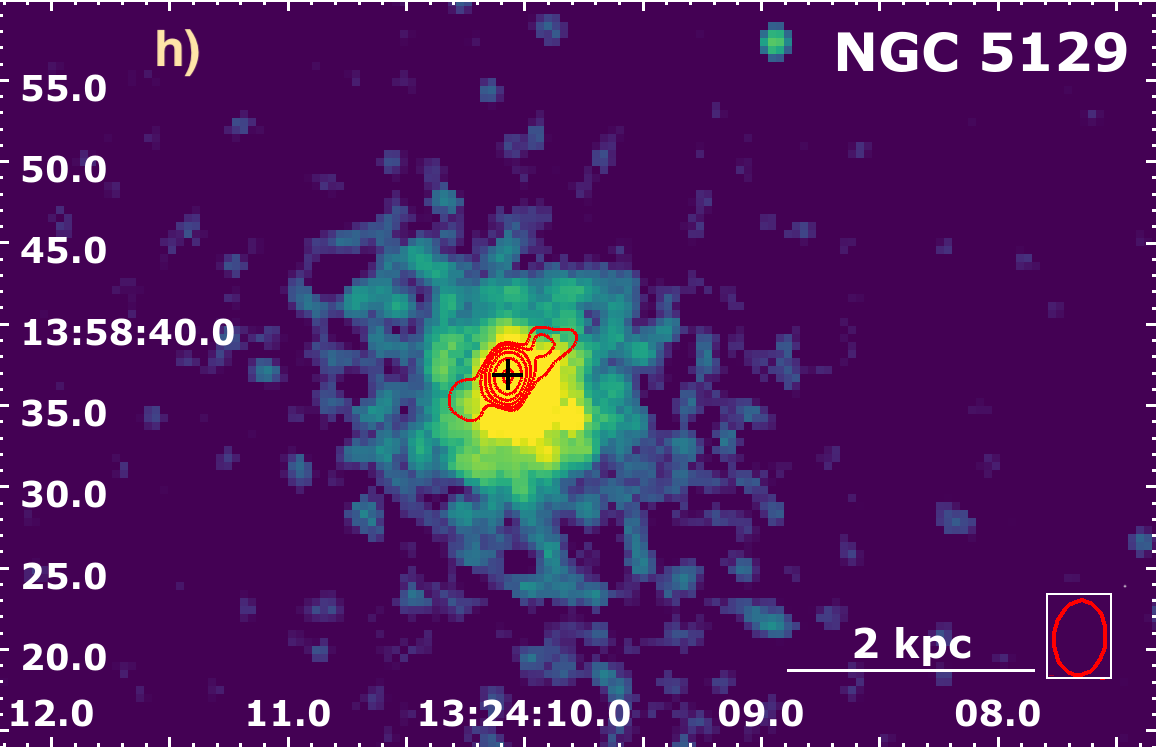}
     \caption{The X-ray {\it Chandra} smoothed (with a 2\,arcseconds Gaussian kernel)  images overlaid by 1.5\,GHz VLA A, AB/B and C/D configuration contours in the red, cyan and white, respectively. In all cases, the contours are created at $5\,\times\,\sigma_{\rm RMS}$ and increase by the power of 2 up to the peak intensity. The center of the galaxy is represented by a black `+' sign. The RMS noise and peak intensity values are given in Table\,\ref{tab:results}.}
    \label{fig:chandra_vla_c}
   \centering
\end{figure}

\renewcommand{\thefigure}{B2.4}
\begin{figure}
    \centering
    \includegraphics[width=250pt]{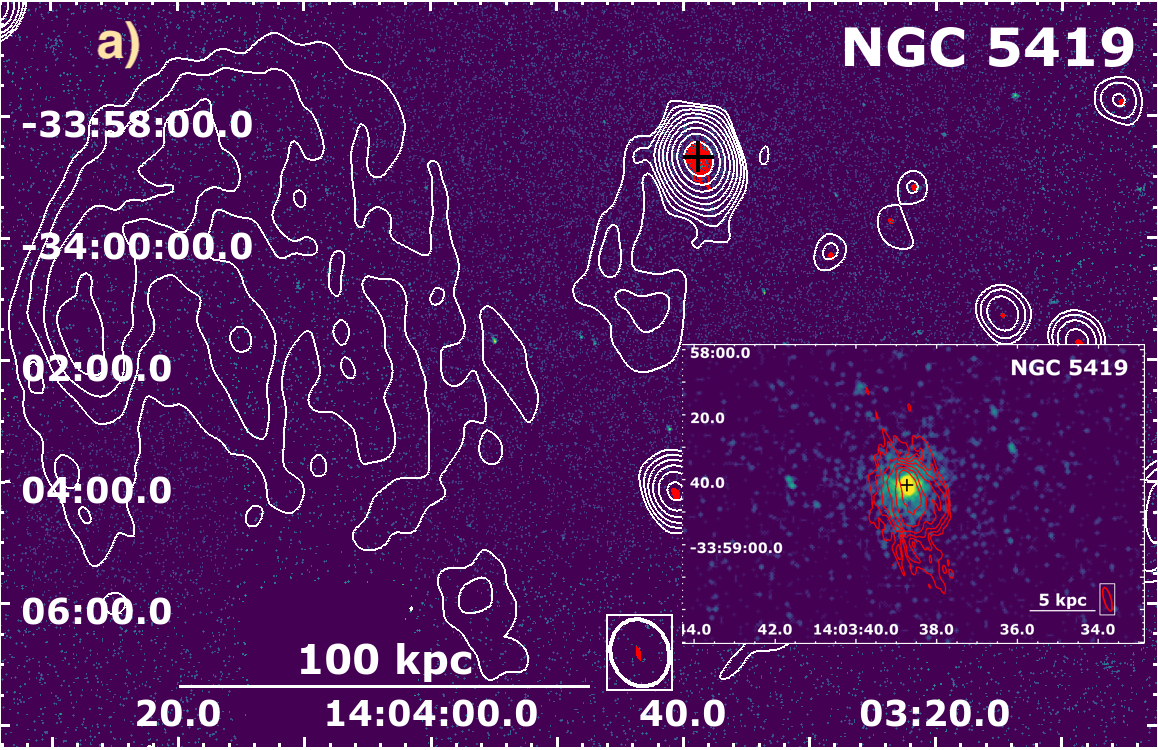}
    \includegraphics[width=250pt]{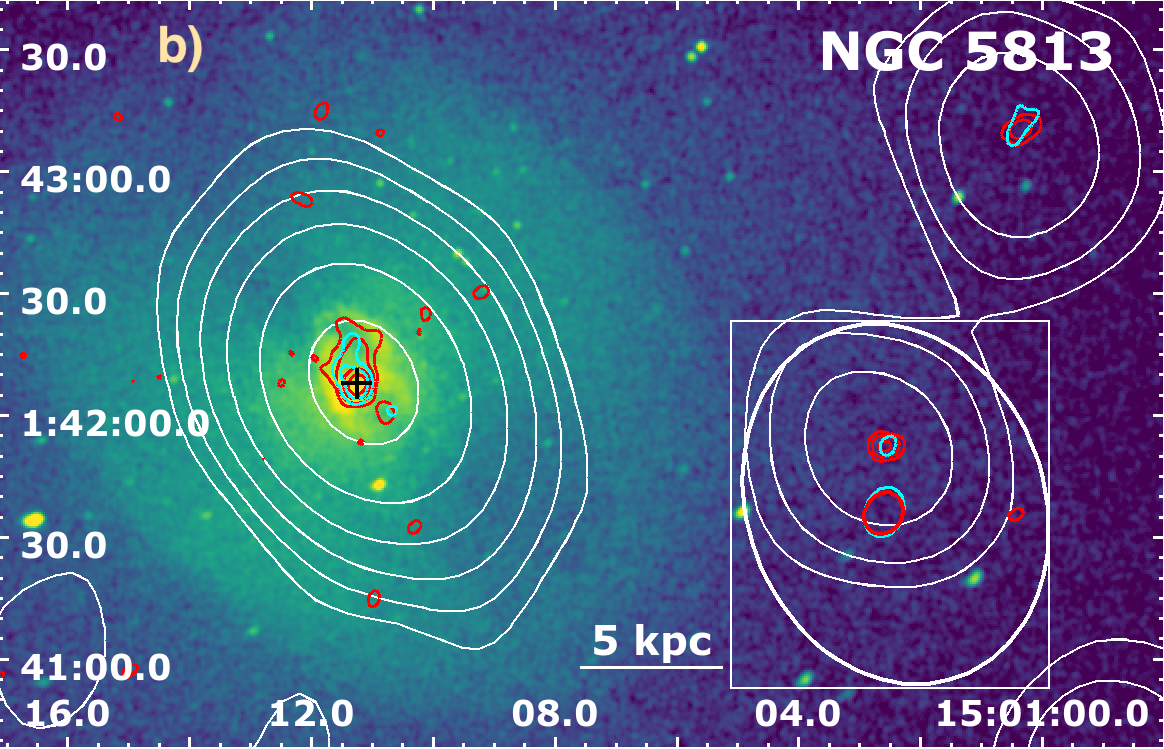}
    \includegraphics[width=250pt]{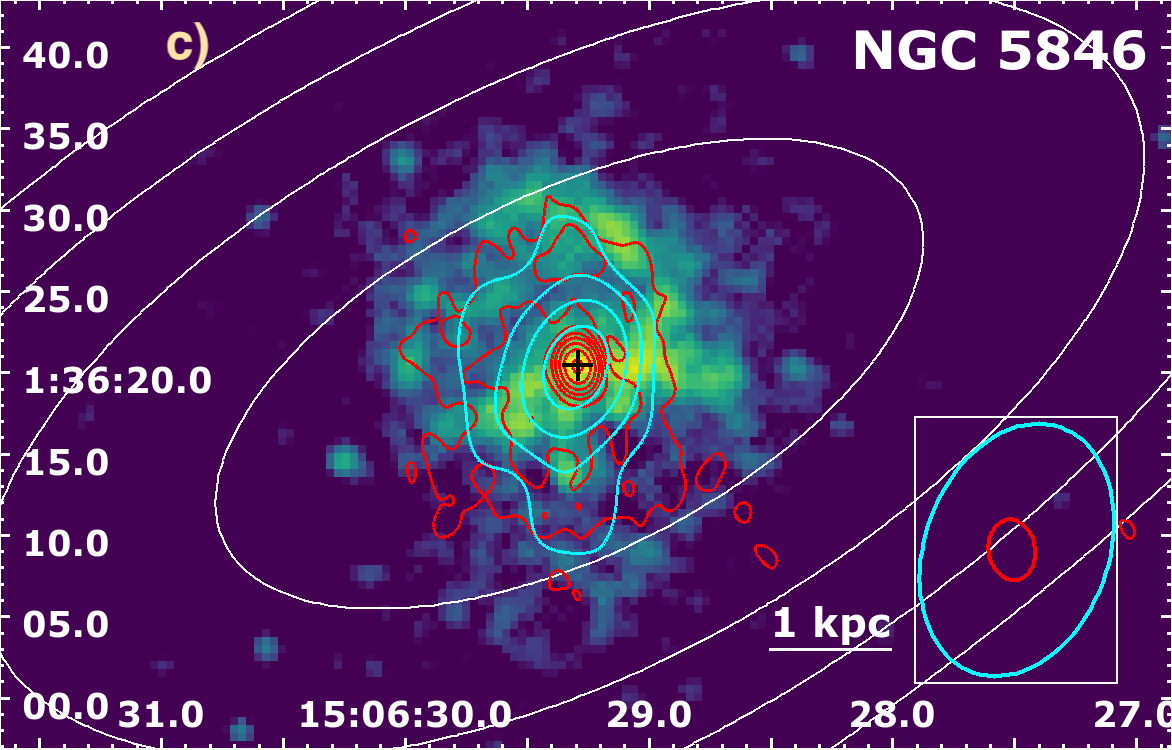}
     \caption{The X-ray {\it Chandra} smoothed (with a 2\,arcseconds Gaussian kernel)  images overlaid by 1.5\,GHz VLA A, AB/B and C/D configuration contours in the red, cyan and white, respectively. In all cases, the contours are created at $5\,\times\,\sigma_{\rm RMS}$ and increase by the power of 2 up to the peak intensity. The center of the galaxy is represented by a black `+' sign. The RMS noise and peak intensity values can be found in Table\,\ref{tab:results}.}
    \label{fig:chandra_vla_d}
   \centering
\end{figure}
\newpage
\onecolumngrid

\section{Appendix: Table}
\label{app:tab}

\subsection{Radio observations details}
\label{app:app_tab_radio}
\begin{startlongtable}
\begin{deluxetable*}{lcccc}
\tablenum{C1}
\tablecaption{\normalfont The main VLA observational information about the early-type galaxies in our sample between 1--2\,GHz. The columns are ordered as follows: (1) the source name; (2) the VLA configuration; (3) the NRAO VLA archive project code; (4) the date (year-month-day; some sources have been observed during on multiple days or within multiple VLA projects); (5)the VLA flux density calibrators from the NRAO VLA archive: details can be found at \href{https://science.nrao.edu/facilities/vla/observing/callist}{https://science.nrao.edu/facilities/vla/observing/callist}.}
\label{tab:obs}
\tablehead{
\colhead{Source} & \colhead{Config} & \colhead{Project} & \colhead{Date} &
\colhead{Flux} \\
\colhead{Name} & \colhead{VLA} & \colhead{Code} & \colhead{(year-month-day)} &
\colhead{Calibrator}
}
\decimalcolnumbers
\startdata
3C\,449 & A & AB920 & 1999-Jul-18 & 3C\,84 \\
3C\,449 & C & AM0740 & 2002-Dec-21 & 3C\,84 \\
IC\,1860 & A & AG0748 & 2009-Jun-19 & 3C\,84 \\
IC\,4296* & A & 15A-305 & 2015-Jul-11 & 3C\,286 \\
IC\,4296 & D & 18A-317 & 2018-Oct-26 & 3C\,286 \\
NGC\,57* & A & 15A-305 & 2015-Jul-03 & 3C\,48 \\
NGC\,315 & A & AM670/AB920 & 1999-Jul-18/2000-Dec-04 & 3C\,286 \\
NGC\,315 & B & AL538 & 2001-Mar-10 & 3C\,286 \\
NGC\,315 & C & AL538 & 2001-Jul-17 & 3C\,48\\
NGC\,410* & A & 15A-305 & 2015-Jun-28 & 3C\,48 \\
NGC\,499 & A & AG748 & 2007-Jun-09/14 & 3C\,84\\
NGC\,507* & A & 15A-305 & 2015-Jun-19 & 3C\,48 \\
NGC\,507 & C & AC0785 & 2005-Aug-17 & 3C\,84 \\
NGC\,533* & A & 15A-305 & 2015-Jun-27 & 3C\,48 \\
NGC\,533 & B & AC0849 & 2006-Sep-02/04/10 & 3C\,84 \\
NGC\,708 & AB & AC470 & 1987-Nov-06 & 3C\,84 \\
NGC\,708 & C & 13A-131 & 2013-Aug-18 & 3C\,286 \\
NGC\,741 & A & AB920 & 1999-Jul-18 & 3C\,84 \\
NGC\,741 & C & 13A-387 & 2013-Jul-26 & 3C\,48 \\
NGC\,777* & A & 15A-305 & 2015-Jun-21 & 3C\,48 \\
NGC\,777 & C & AC0488 & 1997-Sep-18 & 3C\,286 \\
NGC\,1132 & C & AC0488 & 1997-Sep-18 & 3C\,286 \\
NGC\,1316 & BA & AH787 & 2002-May-31/Jun-01 & 3C\,84 \\
NGC\,1316 & CD & AW0110 & 1985-Nov-03 & 3C\,48 \\
NGC\,1399* & A & 15A-305 & 2015-Jun-20 & 3C\,138 \\
NGC\,1404* & A & 15A-305 & 2015-Jun-20 & 3C\,138 \\
NGC\,1404 & CD & AT0285/6 & 2003-Jan-31/Feb-01 & 3C\,84 \\
NGC\,1407* & A & 15A-305 & 2015-Jun-28 & 3C\,138 \\
NGC\,1407 & B & 12A-139 & 2012-Jun-10 & 3C\,147 \\
NGC\,1407 & C & 12A-139 & 2012-Apr-22 & 3C\,147 \\
NGC\,1550* & A & 15A-305 & 2015-Jun-24 & 3C\,138 \\
NGC\,1550 & C & AC0488 & 1997-Sep-18 & 3C\,286 \\
NGC\,1600 & A & AB0289 & 1984-Dec-15 & 3C\,48 \\
NGC\,2300* & A & 15A-305 & 2015-Jun-18 & 3C\,147 \\
NGC\,2300 & D & AS0552 & 1995-Mar-27 & 3C\,48 \\
NGC\,3091* & A & 15A-305 & 2015-Jun-19 & 3C\,286 \\
NGC\,3091 & AB & AM0344 & 1991-Dec-15 & 3C\,286 \\
NGC\,3923* & A & 15A-305 & 2015-Jul-01 & 3C\,286\\
NGC\,3923 & C & AW0110 & 1984-Jun-10 & 3C\,286 \\
NGC\,4073* & A & 15A-305 & 2015-Jun-24 & 3C\,286 \\
NGC\,4125 & D & 14B-396 & 2015-Oct-17 & 3C\,286\\
NGC\,4261 & A & AL0693 & 2007-Jun-08/09 & 3C\,286 \\
NGC\,4261 & C & AL0693 & 2008-May-24/25 & 3C\,286	 \\
NGC\,4374* & A & BH210 & 2015-Jul-17 & 3C\,286 \\
NGC\,4374 & B & BW0003 & 1994-Aug-04 & 3C\,286\\
NGC\,4374 & C & 14A-468 & 2014-Dec-24 & 3C\,286 \\
NGC\,4406* & A & 15A-305 & 2015-Jul-02 & 3C\,286 \\
NGC\,4406 & D & AS623 & 1997-Nov-29 & 3C\,286 \\
NGC\,4472* & A & 15A-305 & 2015-Jun-27 & 3C\,286 \\
NGC\,4472 & C & AB0412 & 1986-Nov-26 & 3C\,286\\
NGC\,4486 & A & AB0920 & 1999-Jul-18 & 3C\,286 \\
NGC\,4486 & B & 16A-245 & 2016-Sep-17 & 3C\,286 \\
NGC\,4486 & C & AO0149 & 2000-May-15 & 3C\,286 \\
NGC\,4552 & A & AC301 & 1991-Aug-24 & 3C\,286 \\
NGC\,4552 & C & 16A-275 & 2016-Apr-04 & 3C\,286 \\
NGC\,4636 & A & AF389 & 2002-Mar-12 & 3C\,286 \\
NGC\,4636 & C & 17A-073 & 2017-May-25 & 3C\,286 \\
NGC\,4649 & A & AC301 & 1991-Aug-24 &	3C\,48 \\
NGC\,4649 & D & 17A-073 & 2017-Jun-01 & 3C\,286 \\
NGC\,4696 & A & AT211 & 1998-Apr-23 & 3C\,286 \\
NGC\,4696 & BC & AB623 & 1992-Feb-06 & 3C\,286 \\
NGC\,4778 & AB & AL0227 & 1990-Jul-09 & 3C\,286 \\
NGC\,4778 & C & AB1150 & 2005-Jul-30/31 & 3C\,286 \\
NGC\,4782 & A & AC0104 & 1984-Dec-26 & 3C\,286 \\
NGC\,4782 & BA & AL0227 & 1990-Jul-09 & 3C\,286 \\
NGC\,4936* & A & 15A-305 & 2015-Jun-20 & 3C\,295 \\
NGC\,5044 & A & AD294 & 1992-Nov-27 & 3C\,286 \\
NGC\,5044 & BA & 15A-243 & 2015-May-25 & 3C\,286 \\
NGC\,5044 & D & 11B-093 & 2011-Dec-03 & 3C\,286\\
NGC\,5129* & A & 15A-305 & 2015-Jun-30 & 3C\,286 \\
NGC\,5419* & A & 15A-305 & 2015-Jun-18 & 3C\,295 \\
NGC\,5419 & B & AE31 & 1984-Feb-26 & 3C\,286\\
NGC\,5419 & CD & AE31 & 1984-Jul-07/8 & 3C\,286	 \\
NGC\,5813 & A & AF188 & 1990-Apr-19 & 3C\,48 \\
NGC\,5813 & B & AW202 & 1988-Jan-30 & 3C\,286 \\
NGC\,5813 & C & AC0488 & 1997-Sep-17 & 3C\,286 \\
NGC\,5846 & A & AF389 & 2002-Mar-12 & 3C\,286 \\
NGC\,5846 & B & AW0202 & 1988-Jan-30 & 3C\,286\\
NGC\,5846 & CD & AM0197 & 1987-Feb-24 & 3C\,48 \\
NGC\,7619* & A & 15A-305 & 2015-Jul-05 & 3C\,48 \\
NGC\,7619 & C & AC0488 & 1997-Sep-17 & 3C\,286 \\
\enddata
\tablecomments{(1) sources marked with * sign are our new VLA A observations within project 15A-305)}
\end{deluxetable*}
\end{startlongtable}

\setlength\tabcolsep{1pt}
\LTcapwidth=\textwidth

\begin{startlongtable}
\begin{longrotatetable}
\subsection{Multifrequency information}
\label{app:app_tab_multi-fr}
\begin{deluxetable*}{lccccl}
\tablenum{C2}
\tabletypesize{\scriptsize}
 \tablecaption{The multifrequency information of the gas in our sample of early-type galaxies. The column order is following: (1) the 3C, NGC, IC source name; (2)  radio morphology from VLA in the L-band at 1--2\,GHz (centered at 1.5\,GHz), LOFAR or GMRT (3) X-ray morphology from {\it Chandra} at 0.5--7\,keV within the innermost $\sim$\,\,10\,kpc or {\it XMM-Newton}; (4) Optical and infrared morphology of the dust and corresponding diameter of the dusty disk from HST or {\it Spitzer}; (5) Atomic gas observed from SOAR through the presence of a warm ionized nebulae H$\alpha +$[N II]; (6) Molecular gas in form of molecular CO lines observed from ALMA and IRAM.}
\label{tab:multi-f}
\tablehead{
\colhead{Source} & \colhead{Radio} & \colhead{X-rays} & \colhead{Dust/extend} &
\colhead{Atomic gas} & \colhead{Molecular gas} \\
\colhead{Name} & \colhead{VLA/GMRT/LOFAR} & \colhead{{\it Chandra/XMM-Newton}} & \colhead{HST/{\it Spitzer}} &
\colhead{SOAR} & \colhead{ALMA/IRAM}
}
\decimalcolnumbers
\startdata
3C\,449 & FR\,I$^{Fan74}$ & inCav$^{Har98;Cro03}$, CFs$^{Lal13}$ & dustyDisk$^{Fer99;Tre06}$ & unkHa$^{Lak18}$ & CO(1-0)$^{Leo01}$\\
IC\,1860 & PS$^{Dun10}$ & CFs$^{Gas07;Gas13}$ & - & nucHa$^{Lak18}$ & - \\
IC\,4296 & FR\,I/FR\,II$^{Gro19}$ & outCav$^{Gro19}$ & dustyDisk$^{Schm02;Boi17}$ & extHa$^{Lak18}$ & CO(2-1)$^{Boi17;Ruf19}$ \\
NGC\,57 & PS$^{Gro}$ & compact & noDust$^{Gou18}$  & noHa$^{Lak18}$ & - \\
NGC\,315 & FR\,I$^{Cap02}$ & X-jet$^{Wor03;Don04}$ & dustyDisk$^{Ver99;Cap00;Boi21}$  & unkHa$^{Lak18}$ & CO(1-0)$^{Fla10}$/(2-1)$^{Boi21}$ \\
NGC\,410 & PS$^{Con98}$ & compact$^{Lak18}$ & noDust$^{Tan9}$ & nucHa$^{Lak18}$ & uppCO(1-0)/(2-1)$^{O'Sul18}$ \\
NGC\,499 & NS$^{Dun10}$/PS$^{Bir20}$ & gCav(in$^{Pls}$/out$^{Pan24b,Kim19}$) & - & nucHa$^{Lak18}$ & - \\
NGC\,507 & FR\,I$^{Cap02,Par86}$ & inCav$^{Don10;Pls}$; CF$^{Fab02;Kra04}$ & noDust$^{Tem07}$  & noHa$^{Lak18}$ & - \\
NGC\,533 & FR\,II?/young?$^{Gro}$ & inCav$^{Dun10}$ & dust$^{Tem07}$ & extHa$^{Lak18}$ & - \\
NGC\,708 & FR\,I$^{Par86,Bla04}$/young?$^{Gro}$ & inCav$^{Bla04;Pan14b}$ & centDust$^{Sah16}$ & extHa$^{Lak18}$ & CO(1-0)$^{Fla10}$/(2-1)$^{Oli19,Nor21}$ \\
NGC\,741 & radioTail$^{Sche17,Ven94}$ & in$^{Sche17}$/outCav$^{Jet08}$ & noDust$^{Ver05}$ & noHa$^{Lak18}$ & noCO$^{Wik95}$  \\
NGC\,777 & compact$^{Bha14;Kol18}$  & pos-Cav$^{Pan14b}$ & lowDust$^{Pah04}$ & noHa$^{Lak18}$ & uppCO(1-0/2-1)$^{O'sul15}$ \\
NGC\,1132 & PS$^{Kim18}$ & gCav$^{Don10;Pls}$ & dustLanes$^{Ala12}$ & noHa$^{Lak18}$ & noCO$^{Dav16}$\\
NGC\,1316 & FR\,I$^{Fom89;Mac20a}$ & in/out(g)-Cav$^{Lan10}$ & dust$^{Schw80;Gri99;Dua14}$ & extHa$^{Lak18}$ & CO(1-0)$^{Sag93}$/CO(2-1)$^{Hor01}$ \\
NGC\,1399 & FR\,I$^{Dun10;Shu08}$ & inCav$^{Pan14b}$ & dust$^{Tem07}$/noDust$^{Dok95,Pra10}$ & noHa$^{Lak18}$ & CO(2-1)$^{Pra10}$ \\
NGC\,1404 & PS$^{Dun10;Gro}$ & CFs$^{Mach05;Su17}$ & dust$^{Tem07}$ & noHa$^{Lak18}$ & - \\
NGC\,1407 & disturbed$^{Gia12}$ & disturbed$^{For06}$ & dust$^{Kul14;Tem07}$ & noHa$^{Lak18}$ &  uppCO(1-0)$^{Bab19}$ \\
NGC\,1550 & disturbed$^{Dun10;Kol18}$ & Cav$^{Pan14b}$ CFs$^{Kol20}$ & - & noHa$^{Lak18}$ & uppCO(1-0)/(2-1)$^{O'sul18}$ \\
NGC\,1600 & FR\,II?$^{Bir85;Gro}$ & inCav$^{Siv04}$ & dust$^{Ferr99}$ & noHa$^{Lak18}$ & - \\
NGC\,2300 & PS$^{Gro}$ & pot-gCav$^{Pls}$ & noDust$^{Xil04}$ & noHa$^{Lak18}$ & - \\
NGC\,3091 & PS$^{Gro}$ & pot-gCav$^{Pls}$ & noDust$^{Col01}$ & noHa$^{Lak18}$ & - \\
NGC\,3923 & PS$^{Dis77;Gro}$ & pot-gCav$^{Pls}$ & filDust$^{Pen86,Bil16}$ & noHa$^{Lak18}$ & -\\
NGC\,4073 & PS$^{Hog14;Gro}$ & pot-gCav$^{Pls}$ & - & noHa$^{Lak18}$ & - \\
NGC\,4125 & PS$^{Kra02;Gro}$ & pot-gCav$^{Pls}$ & dust$^{Ver05;Tem07;Kul14}$ & noHa$^{Lak18}$ & noCO(1-0)$^{Wel10}$/CO(2-1)$^{Wel10}$ \\
NGC\,4261 & FR\,I$^{Kol15;O'sul11}$ & Xjet$^{Gli03;Wor10}$; outCav$^{Cro08;O'su11}$ & dustyDisk & nucHa$^{Lak18}$ & uppCO(1-0)$^{O'sul18}$/CO(2-1)/(3-2)$^{Boi21}$ \\
NGC\,4374 & FR\,I$^{Lai87}$ & outCav$^{Fin01,Dev10}$ & dustLanes$^{Ver99;Boi17}$ & nucHa$^{Lak18}$ & CO(1-0)/(2-1)$^{Fla10,Boi17}$   \\
NGC\,4406 & PS$^{Dun10;Gro}$ & X-ray tail/plume$^{For79,Kim19}$ & dust$^{Smi12}$ & extHa$^{Ken08;Lak18}$ & noCO$^{Wik95;You11}$ \\
NGC\,4472 & FR\,I$^{Con88;Gro}$ & inCav$^{Bil04;Su19}$ & dust$^{Tem07}$  & noHa$^{Lak18}$ & CO(1-0)/(2-1)$^{Huch88,Huch99}$ \\
NGC\,4486 & FR\,I$^{Fan74}$ & in/outCav$^{You02;For05;For07}$; Xjet$^{Mar02}$ & dustyDisk & extHa$^{Lak18}$ & CO(1-0)$^{Fla10}$/(2-1)$^{Sim18;Oli19}$ \\
NGC\,4552 & FR\,I$^{Fil04;Gro}$ & outCav$^{Mach06;All06}$ , CF$^{Mach06;Kra17}$& dust$^{Tem07;Kul14}$ & noHa$^{Lak18}$ & uppCO(1-0)/(2-1)$^{Com07}$ \\
NGC\,4636 & FR\,I$^{Dun10;Gia11}$ & inCav$^{Sta86;Bal09}$ & dust$^{Tem07}$  & nucHa$^{Lak18}$ & uppCO(1-0)$^{O'sul18}$/CO(2-1)$^{O'sul18;Tem18}$ \\
NGC\,4649 & FR\,I?$^{Shu08;Dun10}$ & Cav$^{Shu08;Dun10}$ & dust$^{Kul14;Tem07}$  & noHa$^{Lak18}$ & CO(1-0)$^{Sag89;You11}$ \\
NGC\,4696 & FR\,I$^{Tay02}$  & inCav$^{Tay06;San16}$ & filDust$^{Tem07;Fab16}$  & extHa$^{Lak18}$ & CO(1-0)$^{Bab19,Oli19}$/(2-1)$^{Fab16}$ \\
NGC\,4778 & PS$^{Vrt02;Gro}$ & gCav$^{Mor06,Pan14b}$ & - & nucHa$^{Lak18}$ & - \\
NGC\,4782 & FR\,I$^{Bor96;Mach06}$ & in/outCav$^{Bor96;Mach06}$ & - & nucHa$^{Lak18}$ & - \\
NGC\,4936 & PS$^{Bha14;Gro}$ & disturbed$^{Bha14;Lak18}$ &  & extHa$^{Lak18}$ & - \\
NGC\,5044 & jets$^{Dun10;Sche20a}$ & in/outCav$^{Gas09;Sche20a}$; CFs$^{Gas09}$ & dust$^{Tem07}$  & extHa$^{Lak18}$ & CO(2-1)$^{Lau14;Dav14;Tem18}$ \\
NGC\,5129 & jets$^{Con98;Bha14;Gro}$ & disturbed$^{Bha14}$ & - & nucHa$^{Lak18}$ & - \\
NGC\,5419 & relic-like$^{Gos87;Sub03}$ & central$^{Bal06}$ & - & noHa$^{Lak18}$ & - \\
NGC\,5813 & jets$^{Ran11;Ran15}$ & in/outCav$^{Ran11}$ & dust$^{Tem07}$  & extHa$^{Lak18}$ & uppCO(1-0)/(2-1)$^{O'sul18}$ \\
NGC\,5846 & disturbed$^{Dun10;Bir20}$ & in/outCav$^{Tri02;All06}$, CF$^{Mach11}$ & dust$^{Tem07}$ & nucHa$^{Lak18}$ & CO(1-0)$^{O'sul18}$/(2-1)$^{O'sul18;Tem18}$ \\
NGC\,7619 & PS$^{Con98;Gia11}$ & RPS tails$^{Ran09}$ & dust$^{Tem07}$ & noHa$^{Lak18}$ & -\\
\enddata
  \tablecomments{ {\bf (2)} the Fanaroff-Riley \citep{fanaroff1974} (FR) radio morphology classification: `FR\,I' = edge-darkened, `FR\,II' = edge-brightened; `FR\,II?': potential FR\,II radio source; `PS': point-source radio morphology ($<$ twice the beam size; unresolved source); NS: no radio source; young: young radio source; {\it References column (2):} Bir85: \cite{birkinshaw1985}; Bir20: \cite{birzan2020}; Bha14: \cite{bharadwaj2014}; Bla04: \cite{blanton2004}; Bor96: \cite{borne1996}; Cap02: \cite{capetti2002a}; Con88L \cite{condon1988}; Con98: \cite{condon1998}; Dun10: \cite{dunn2010}; Dis77: \cite{disney1977}; Fan74: \cite{fanaroff1974}; Fil04: \cite{filho2004}; Fom89: \cite{fomalont1989}; Gia12: \cite{giacintucci2012}; Gia11: \cite{giacintucci2011}; Gos87: \cite{goss1987}; Gro: (submitted); Hog14: \cite{hogan2014}; Kim18: \cite{kim2018}; Kol15: \cite{kolokythas2015}; Kol18: \cite{kolokythas2018}; Kra02: \cite{krajnovic2002}; Lai87: \cite{laing1987}; Mac20a: \cite{maccagni2020a}; Mach06: \cite{machacek2006}; O'sul11: \cite{o'sullivan2011}; Par86: \cite{parma1986}; Ran11: \cite{randall2011}; Ran15: \cite{randall2015}; Sche17: \cite{schellenberger2017}; Sche20a: \cite{schellenberger2020a}; Shu08: \cite{shurkin2008}; Sub03: \cite{subrahmanyan2003}; Tay02: \cite{taylor2002}; Ven94: \cite{venkatesan1994}; Vrt02:\cite{vrtilek2002}; {\bf (3)}`inCav': inner cavities, corresponding to innermost jets/lobes and located within innermost 5\,kpc; `outCav': outer cavities, corresponding to outer jets/cavities; `gCav': ghost cavities, no radio counterpart; `pot-()Cav': potential cavities; `CF': cold fronts as a sign of sloshing;`compact': smooth compact morphology of the hot X-ray atmosphere within the host galaxy (up to 10\,kpc from the center); `RPS tail': ram pressure stripped tail; {\it References column (3):} All06: \cite{allen2006}; Bal09: \cite{baldi2009}; Bal06: \cite{balmaverde2006}; Bha14: \cite{bharadwaj2014}; Bil04; \cite{biller2004}; Bla04: \cite{blanton2004}; Bor96: \cite{borne1996}; Cro03: \cite{croston2003}; Cro08: \cite{croston2008}; Dev10: \cite{devereux2010}; Don10: \cite{dong2010}; Dun10: \cite{dunn2010}; Fab02: \cite{fabbiano2002}; Fin01: \cite{finoguenov2001}; For79: \cite{forman1979}; For05: \cite{forman2005}; For07: \cite{forman2007}; For06: \cite{forbes2006}; Gas07: \cite{gastaldello2007}; Gas09: \cite{gastaldello2009}; Gas13: \cite{gastaldello2013}; Gli03: \cite{gliozzi2003}; Gro19: \cite{grossova2019}; Har98: \cite{hardcastle1998}; Jet08: \cite{jetha2008}; Kol20: \cite{kolokythas2020}; Kra04: \cite{kraft2004}; Kra17: \cite{kraft2017}; Kim19: \cite{kim2019}; Lal13: \cite{lal2013}; Lak18: \cite{lakhchaura2018}; Lan10: \cite{lanz2010}; Mach05: \cite{machacek2005}; Mach06: \cite{machacek2006}; Mar02: \cite{marshall2002}; Mor06: \cite{morita2006}; Pan14b: \cite{panagoulia2014b}; Pls: Plšek et al. (in prep.); Ran09: \cite{randal2009}; Ran11: \cite{randall2011}; San16: \cite{sanders2016}; Sche17: \cite{schellenberger2017}; Sche20: \cite{schellenberger2020a}; Shu08: \cite{shurkin2008}; Siv04: \cite{sivakoff2004}; Sta86: \cite{stanger1986}; Su17:\cite{su2017}; Su19: \cite{su2019}; Tay06: \cite{taylor2006}; Tri02: \cite{trinchieri2002}; Wor10: \cite{worrall2010}; You02: \cite{young2002} {\bf (4)} `dustyDisk': morphology of dust in form of disk; `noDust': no dust observed; `centDust': centrally located dust; dustLanes: dust in form of lanes; `filDust': filamentary dust {\it References column (4):}; Ala12: \cite{alamo-martinez2012}; Bil16: \cite{bilek2016}; Boi17: \cite{boizelle2017}; Cap00: \cite{capetti2000}; Dua14: \cite{duah2014}; Kim19: \cite{kim2019}; Col01: \cite{colbert2001}; Kul14: \cite{kulkarni2014}; Fab16: \cite{fabian2016}; Fer99: \cite{feretti1999}; Ferr99: \cite{ferrari1999}; Gou18: \cite{goullaud2018}; Gri99: \cite{grillmair1999}; O'sul18:\cite{o'sullivan2018}; Pah04: \cite{pahre2004}; Pen86: \cite{pence1986}; Schm02: \cite{schmitt2002}; Smi12: \cite{smith2012}; Schw80: \cite{schweizer1980}; Tan9: \cite{tang2009}; Tem07: \cite{temi2007}; Tre06: \cite{tremblay2006}; Sah16: \cite{sahu2016}; Ver99: \cite{verdoes1999}; Ver05: \cite{verdoes2005}; Xil04: \cite{xilouris2004} {\bf (5)} `noHa': no detection of H$\alpha +$[N II] emission in the narrow-band imaging and confirming long slit spectroscopic observation, `unkHa': the detection could not be confirmed, 'extHa': H$\alpha +$[N II] is extending to $\leq$ 2\,kpc, 'nucHa': nuclear H$\alpha +$[N II] line emission, the extent is smaller than 2\,kpc; {\it Reference column (25):} Ken08: \cite{kenney2008}; Lak18: \cite{lakhchaura2018}; {\bf (6)} `noCO': not dectected CO line emission; uppCO(..): upper limit on the CO line emission {\it References column (6):} Bab19: \cite{babyk2019}; Boi17: \cite{boizelle2017}; Boi21: \cite{boizelle2021}; Com07: \cite{combes2007}; Dav14: \cite{david2014}; Dav16: \cite{davis2016}; Fab16: \cite{fabian2016}; Fla10:\cite{flaquer2010}; Hor01: \cite{horellou2001}; Huch88: \cite{huchtmeier1988}; Lau14: \cite{laurence2014}; Leo01: \cite{leon2001}; Nor21: \cite{north2021};  O'sul15: \cite{o'sullivan2015}; O'sul18: \cite{o'sullivan2018}; Oli19: \cite{olivares2019}; Pra10: \cite{prandoni2010}; Ruf19: \cite{ruffa2019}; Sag89: \cite{sage1989}; Sag93: \cite{sage1993}; Sim18: \cite{simionescu2018}; Tem18: \cite{temi2018}; You11: \cite{young2011}; Wik95: \cite{wiklind1995}; Wel10: \cite{welch2010}; (all) `-': not investigated yet}
\end{deluxetable*}
\end{longrotatetable}
\end{startlongtable}

\normalsize
\onecolumngrid
\section{Appendix: Additional sources}
\label{app:add_sources}
Additionally, we provide a list of giant elliptical galaxies that are omitted from our main sample. The reason for each omission is given in Table\,\ref{tab:add_sources} and varies from missing radio VLA data in the frequency range of 1--2\,GHz or insufficient/missing archival {\it Chandra} data due to strong dominance by the central X-ray point source, thus obscures the extended emission from the hot X-ray atmosphere. Moreover, Table \,\ref{tab:add_sources} also contains the VLA observational details together with information about the properties of the observed radio emission from the VLA in the frequency range between 1--2\,GHz. This information might be used as a foundation for future proposals.

\begin{startlongtable}
\begin{deluxetable*}{lcccccc}
\tablenum{D1}
\tablecaption{The list of early-type galaxies omitted from the main sample. The columns are: (1) the source name; `*': sources observed within our project 15A-305; (2) the VLA configuration; (3) the NRAO VLA archive project code. Missing VLA archival data were supplemented by the NRAO VLA Sky Survey (NVSS) data; (4) the observation date (year-month-day); (5) the VLA total flux flux density of the image; `-': non detection of radio emission at 1.5\,GHz and `$\#$': a weak radio source with an integrated flux density of 2.0\,mJy detected at the lower threshold of 4$\times\,\sigma_{\rm RMS}$; (6) the RMS noise value of the image (7) the reason for the exclusion of a certain source from the main sample.}
\label{tab:add_sources}
\tablewidth{0.5pt}
\tablehead{
\colhead{Source} & \colhead{Config.} & \colhead{VLA Project Code} & \colhead{VLA Obs. Date} & $S_{1.5\,{\rm GHz}}$ $\pm$ e$S_{1.5\,{\rm GHz}}$ & RMS & \colhead{Exclusion Reason}\\
\colhead{Name} & \colhead{VLA} & \colhead{/Survey} & \colhead{(year-month-day)} & [mJy] & [mJy] &
}
\decimalcolnumbers
\startdata
IC2006 & CD & NVSS & 1993-09-20 & - & 0.47 &  no archival VLA at 1--2\,GHz \\
 & & & & & & $\&$ insufficient {\it Chandra} (4.55\,ks) \\
IC310* & A & 15A-305 & 2015-Jun-23 & $330\pm13$ & 0.03 & dominant X-ray point source \\
IC310 & C & AB0412 & 1986-Dec-05 & $460\pm20$ & 1.8 & dominant X-ray point source \\
NGC\,1521 & CD & NVSS & 1993-09-20 & \,-$^{{\small\#}}$ & 0.47 & no archival VLA at 1--2\,GHz \\
NGC\,2329* & A & 15A-305 & 2015-Jun-20 & $220\pm9$ & 0.014 & insufficient {\it Chandra} (2.86\,ks) \\
NGC\,2329 & C & AB0476 & 1988-Apr-14 & $650\pm27$ & 0.6 & insufficient {\it Chandra} (2.86\,ks) \\
NGC\,2340* & A & 15A-305 & 2015-Jun-19 & $2.00\pm0.08$ & 0.002 & insufficient {\it Chandra} (1.92\,ks) \\
NGC\,4203* & A & 15A-305 & 2015-Jul-03 & $7.8\pm0.3$ & 0.027 & dominant X-ray point source  \\
NGC\,4203 & D & AB0506 & 1988-Jul-13 & - & 2.8 & dominant X-ray point source  \\
NGC\,5090 & A & AG0454 & 1995-Jul-23 & $350\pm25$ & 4.5 & no {\it Chandra} \\
NGC\,5090 & BC & AS0225 & 1985-Jul-07/08 & $3300\pm130$ & 0.67 & no {\it Chandra} \\
NGC\,5328 & CD & NVSS & 1993-09-20 & - & 0.47 &  no archival VLA at 1--2\,GHz\\
 & & & & & & $\&$ insufficient {\it Chandra} (2.6\,ks) 
\enddata
\end{deluxetable*}
\end{startlongtable}

\end{document}